\documentclass[fleqn,usenatbib]{mnras}

\usepackage{newtxtext,newtxmath}
\usepackage[T1]{fontenc}

\DeclareRobustCommand{\VAN}[3]{#2}
\let\VANthebibliography\thebibliography
\def\thebibliography{\DeclareRobustCommand{\VAN}[3]{##3}\VANthebibliography}

\usepackage{graphicx}	% Including figure files
\usepackage{amsmath}	% Advanced maths commands
\usepackage{lscape}
\title[DES Quasars]{Broadband reverberation mapping of quasars at cosmic noon}

\author[Hern\'andez Santisteban]{
Juan~V. Hern\'andez Santisteban$^{1}$\thanks{E-mail: jvhs1@st-andrews.ac.uk (JVHS)}\\
\\
$^{1}$SUPA Physics and Astronomy, University of St. Andrews, Fife KY16 9SS, UK\\
}

\date{Accepted XXX. Received YYY; in original form ZZZ}

\pubyear{\the\year{}}

\begin{document}
\label{firstpage}
\pagerange{\pageref{firstpage}--\pageref{lastpage}}
\maketitle

\begin{abstract}
%The structure of the accretion flow around supermassive black holes is still unknown.
Broadband reverberation mapping provides a powerful tool to investigate the structure of the accretion flow around supermassive black holes. The ultraviolet region of the quasar spectrum is of particular interest, as it enables the disentanglement of the relative contributions from the accretion disc and the broad-line region to the observed variability of quasars in the local Universe. Observations of quasars at cosmic noon ($z \sim 1$–$2$) offer the opportunity to probe the UV emission across a diverse population, spanning a broad range of black hole masses and luminosities.
Here we re-analysed the 36 multi-year quasar lightcurves from the Dark Energy Survey (DES) between $z\sim0.7-2$ to retrieve interband delays and analyse their spectral energy distribution. Using {\sc PyROA}, we were able to fit all multiple years of observations and the four optical bands simultaneously, significantly improving the lag detection as this method is able to interpolate over the seasonal gaps. We find robust detections of interband delays in 31/36 quasars, suggesting high-Eddington ratios for a large fraction of objects. Their spectral energy distribution is consistent with a geometrically thin disk $f_\nu\propto\lambda^{-1/3}$, with an ensemble slope of $-0.28\pm0.02$. The full sample of quasars do not show excess around the Balmer jump region, suggesting the impact of the broad-line region to the reverberation signals in high-Eddington sources could be small as documented in some AGN in the local Universe. The prospect of LSST to observe thousands of quasars in this redshift range should motivate the detailed study of the UV-rest frame of high-Eddington quasars to account for the BLR contribution and properly infer disc sizes.
\end{abstract}


%%%%%%%%%%%%%%%%%%%%%%%%%%%%%%%%%%%%%%%%%%%%%%%%%%

%%%%%%%%%%%%%%%%% BODY OF PAPER %%%%%%%%%%%%%%%%%%

\section{Introduction}

Active galactic nuclei (AGN) represent some of the most luminous and energetic phenomena in the Universe, powered by the accretion of gas and dust onto supermassive black holes (SMBHs) residing at the centres of galaxies \citep{LyndenBell1969, Volonteri2010}. As material loses angular momentum and spirals inward, it forms a geometrically thin, optically thick accretion disc that efficiently converts gravitational potential energy into radiation \citep{Shakura1973, Novikov1973}. The resulting bolometric luminosities can reach values of $L_{\rm Bol}\sim10^{45}$ erg s$^{-1}$ for a typical $M_{\rm BH}\sim10^8$ M$_{\odot}$ black hole \citep{Soltan1982, Marconi2004}. The accretion process produces a broad, multi-wavelength spectral energy distribution (SED), extending from hard X-rays emitted by the hot corona \citep{Haardt1991, Haardt1993} to infrared radiation reprocessed by circumnuclear dust in the torus \citep{Antonucci1993, Netzer2015}. Owing to their extreme luminosities, AGN can be detected across cosmological distances, serving as powerful probes of SMBH growth \citep{Alexander2012}, feedback mechanisms \citep{Fabian2012}, and the co-evolution of galaxies and their central black holes \citep{Kormendy2013, Heckman2014}. Among these, quasars—representing the most luminous subclass of AGN—provide crucial laboratories for studying accretion physics and tracing the build-up of massive black holes in the early Universe \citep[e.g.,][]{Mortlock2011,Banados2018,Maiolino2025}.

Despite their significance, the physical structure of the central regions of AGN remains poorly resolved observationally due to the small angular scales involved. Reverberation mapping (RM) offers a powerful method for indirectly probing these regions \citep{Blandford:1982,Peterson2004}, relying on the correlated, variable emission of the AGN at different wavelengths. By measuring time delays between variations at different energy bands, one can effectively reconstruct the spatial geometry and structure of the innermost components, with typical $\sim0.5-10$ light-day timescales for the accretion disc. %This is particularly useful for understanding the accretion disc, where gas inflow produces thermal radiation with a temperature that decreases with radius.

Broadband reverberation mapping (BRM) targets inter-band time delays within the accretion disc continuum emission using broadband photometric filters \citep[see][for a recent review]{Cackett2020}. In the framework of the standard geometrically thin, optically thick accretion disc model \citep{Shakura1973}, shorter-wavelength emission arises from the hotter, inner regions of the disc, while longer wavelengths originate in cooler, outer regions, following a radial temperature profile of  $T(R)\propto R^{-3/4}$. Variations in the innermost regions, often driven by fluctuations in the X-ray corona, irradiate and thermally reprocess in the outer disc, producing correlated variability across wavelengths. Since these reprocessing zones are located at different radii from the central source, the corresponding light-travel-time delays are expected to scale as $\tau=R/c\propto \lambda^{4/3}$, where the proportionality constant depends on the black hole mass, accretion rate, and the luminosity of the illuminating X-ray source \citep{Collier1999, Cackett:2007}.

In the local Universe, this technique has provided some of the most stringent observational constraints on the structure of accretion discs in AGN \citep[e.g.,][]{Edelson2015, Fausnaugh:2016, Edelson:2019, Hernandez2020, Kara2021, Gonzalez2025, Prince2025}. One of the most prominent challenges emerging from these studies is the so-called “too-big disc” problem, wherein the measured disc sizes—derived from inter-band time delays—are typically a factor of $\sim2-3$ larger than predictions from the standard thin-disc model \citep[a discrepancy also observed in micro-lensing studies e.g.,][]{Morgan2010}. This persistent discrepancy has motivated a range of theoretical efforts to reconcile observations with accretion physics. Proposed explanations include enhanced reprocessing of radiation from the inner accretion flow \citep{Gardner:2014}, the contribution of vertically extended disc structures that reprocess the incident ionising continuum \citep{Starkey2023, Hagen2023b}, the inclusion of general relativistic corrections to the disc temperature and light-propagation structure \citep{Kammoun2021, kammoun2024}, and the potential effects of local obscuration along the line of sight \citep{Lewin2025}. Moreover, robust detections of excess time lags at specific wavelengths—particularly around the Balmer jump at 3600~\AA\ (with NGC~4593 being a prime example; \citealt{Cackett:2018})—suggest the presence of an additional, more distant reprocessing component beyond the accretion disc. This excess emission is consistent with diffuse continuum emission (DCE) from hydrogen in the broad-line region and/or from a disc wind \citep{Korista2001, Korista:2019, Netzer2020, Kara2023}.

Large-scale photometric surveys have expanded these studies on two key black hole parameters, redshift and mass. These have been instrumental in understanding the ensemble variability properties of AGN which help identify characteristic timescales and variability amplitudes across the quasar population \citep[e.g.,][]{MacLeod2010,Arevalo2024}. However, the measurement of interband delays in individual quasars at high redshift is often very marginal and studies rely on fitting the ensemble population of quasars to infer average parameters. The Sloan Digital Sky Survey Reverberation Mapping project (SDSS-RM) \citep{Shen2023} collected repeated photometric and spectroscopic monitoring of $\sim800$ quasars over a wide redshift range ($0.1 < z < 4.5$), providing one of the largest and most homogeneous datasets for time-domain AGN studies. Analyses of continuum light curves from this survey \citep{Kinemuchi2020} led to the detection of wavelength-dependent continuum lags and temperature profile scaling consistent with the expectations of disk reverberation by the standard thin disk model, albeit at a low significance \citep[$\sim1.5\sigma$,][]{Homayouni2019}. 

The Dark Energy Survey \citep[DES,][]{Flaugher2015DES,Morganson2018DES,Abbott2021DES} has further contributed to BRM by offering long-baseline, multi-band photometry for high-luminosity quasars at intermediate redshifts ($0.7 \lesssim z \lesssim 1.9$). These grizY-band light curves delivered marginal detection of inter-band delays on individual quasars via  cross-correlation techniques \citep{Mudd2018,Yu2020_DES}. However, by fitting the full sample as an ensemble they were able derive disc sizes, finding larger (factor of $\sim3$) than those predicted from their bolometric luminosity. 
These seemingly contradictory results highlight the need for precise delay measurements, particularly when working with noisier and less temporally resolved datasets compared to those obtained from dedicated, intensive BRM campaigns in the local Universe.

Furthermore, BRM at higher redshifts remains challenging due to cosmological time dilation and the requirement for extended monitoring baselines. Nevertheless, the rapid growth of time-domain survey capabilities is beginning to mitigate these limitations. With the forthcoming Vera C. Rubin Observatory’s Legacy Survey of Space and Time \citep[LSST;][]{Ivezic2019}, the field is poised to enter a new era. LSST will deliver high-cadence, multi-band photometric monitoring over a ten-year baseline for millions of quasars, dramatically expanding the available BRM sample and enabling statistical, population-level investigations of accretion disc structure and variability across cosmic time.

In this paper, we reanalyse the DES quasars from \citet{Mudd2018} and \citet{Yu2020_DES}, which we briefly described in Section ~\ref{sec:obs}. We present a simultaneous multi-year, multi-band fit to the lightcurves to retrieve interband delays in Section~\ref{sec:lags} and analyse their spectral energy distributions (SED) in Section~\ref{sec:sed}. We discuss our results in Section~\ref{sec:results} and present our conclusions in Section~\ref{sec:conclusions}. Throughout the paper we use \citep{Planck2020} cosmology with $H_0=67.7$ km Mpc$^{-1}$ s$^{-1}$ and $\Omega_{m_0}=0.31$\footnote{As implemented in \texttt{astropy.cosmology} \citep{astropy2018,astropy2022}}.

\section{The AGN sample}\label{sec:obs}
\begin{figure}
	\includegraphics[width=\columnwidth]{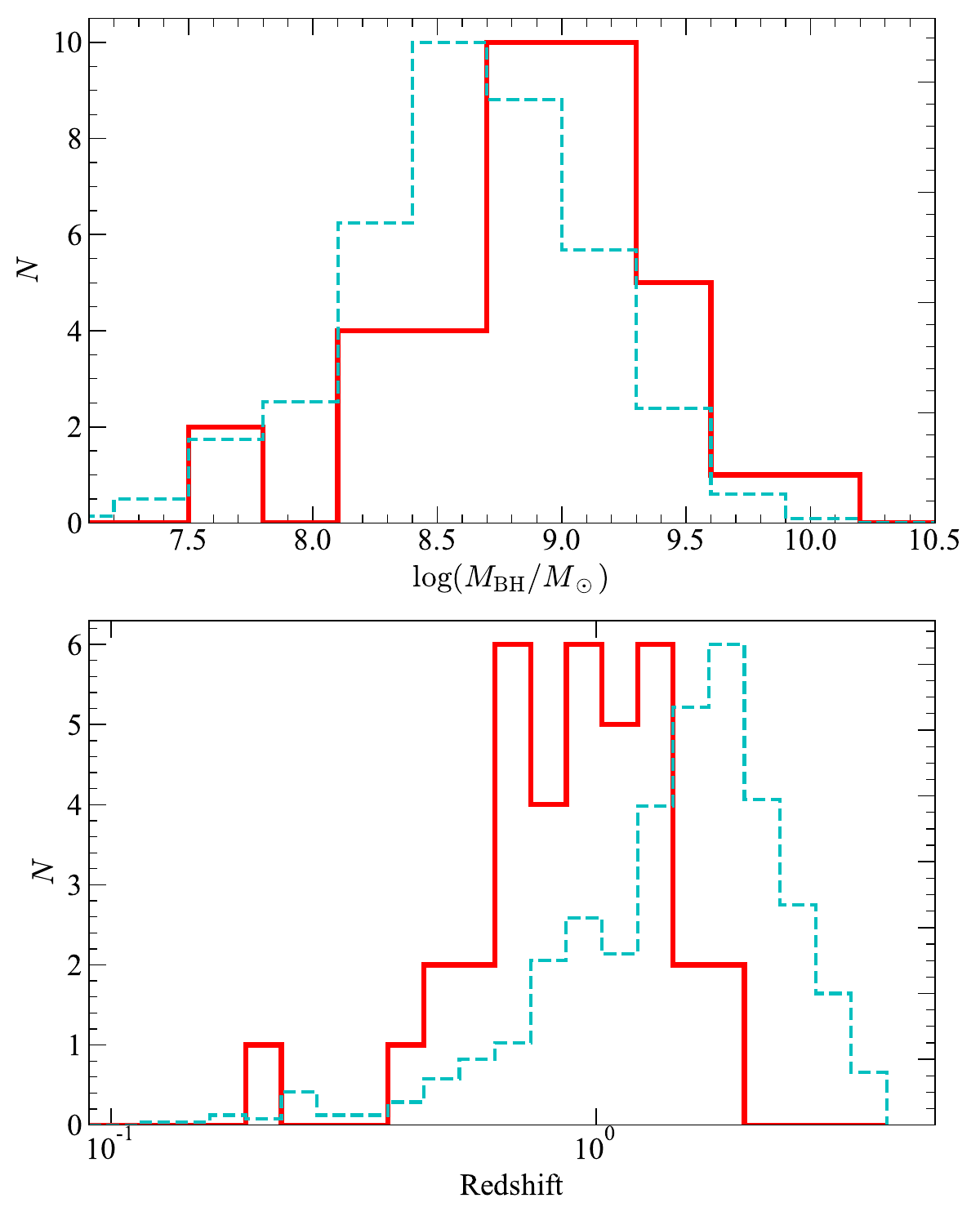}
    \caption{Properties of the DES BRM quasar sample \citep[red,][]{Mudd2018,Yu2020_DES}, showing their black hole mass ({\it top}) and redshift ({\it bottom}) distributions. For comparison, we show the quasar distribution from the SDSS-RM experiment \citep[blue,][]{Shen2023} as a representative sample of the quasar distribution in the deep-drilling fields of LSST. }
    \label{fig:sample}
\end{figure}

In the study, we used the light curves from spectroscopically confirmed quasars from DES \citep{Mudd2018,Yu2020_DES} to re-analyse with novel time-series analysis tools on high redshift quasars. The combined sample consists of 36 luminous and variable quasars with a wide range of redshifts ($z\sim0.7$ to 1.9) and black hole masses ($7.0 \lesssim \log(M_{\rm BH} / M_\odot)\lesssim10$), a shown in Fig.~\ref{fig:sample}. The data contains photometric time-series in the $g$, $r$, $i$, $z$ and $Y$ filters (the latter only available in 3 quasars) over the course of $\sim3-5$~years. We refer to each corresponding paper for details on the data reduction and sample properties. The median cadences for each of the \citet{Mudd2018} and \citet{Yu2020_DES} datasets in the rest-frame of the quasar are $\sim3$ and $\sim2$ days, respectively. Equally, the effective (rest-frame) length of these campaigns are of the order of $\sim1.3$ and $\sim2.2$ years, respectively. Both cadences and lengths of the observations are comparable with intensive BRM campaigns combining space- and ground-based observatories \citep[e.g.,][]{Edelson:2019,Hernandez2020,Gonzalez2025}, making these true analogue to those performed in the local Universe. 

As illustrated in Fig.~\ref{fig:sample}, the DES sample probes a black hole parameter space comparable to that of the SDSS-RM field, which itself serves as a precursor to the deep-drilling fields planned for LSST. Owing to its relatively shallower depth, DES is most sensitive to quasars peaking at lower redshifts ($z\sim1$). Nevertheless, as demonstrated in the following sections, the capability to perform reliable broadband reverberation mapping (BRM) at cosmic noon ($z\sim2$) is now well within reach using current survey datasets.

\section{Time series analysis}\label{sec:lags}
The original analysis in \citet{Mudd2018} and \citep{Yu2020_DES} presented time-series analysis of individual years of data, in search for inter-band delays. This proved challenging in many of the $15+21$ quasars, respectively, in the sample due to the random variability amplitude in each year, as the number of features to anchor the delays can vary widely \citep{Welsh1999}. Also, the choice of a single reference band can strongly affect the accuracy of recovered lags, since the cross-correlation function is heavily shaped by the auto-correlation function of the reference band. Furthermore, a standing issue to take advantage of the multi-year data in a single time-series analysis is the way different techniques deal with the interpolation over the seasonal gaps, which can introduce aliases in the lag distribution. Methods like interpolated cross-correlation functions as implemented by {\sc PyCCF} \citep{pyccf2018} or {\sc javelin} \citep{Sun2020} tend to underestimate the uncertainties on the gaps, producing multiple (often spurious) peaks in the posterior distributions of the lags, \citep[e.g.,][]{Grier17,Shen2024}. In order to circumvent these issues, we analyse the multi-year, multi-band lightcurves simultaneously with PyROA \citep{Donnan2021}.

\subsection{PyROA}
PyROA is a Bayesian time-domain analysis method designed to model AGN light curves by estimating time lags between an observed driving signal (generated by the optimal average of {\it all the data}) and its delayed responses in other wavebands or spectral lines \citep{Donnan2021,Donnan2023}. The core assumption is that observed light curves are generated by convolving a common underlying driving light curve, $X(t)$, with a transfer function that characterises the response of each emission component. In summary, for a simple top-hat transfer function centred at a lag $\tau_i$, the time-dependent lightcurve at every filter, $i$, is modelled as
\begin{equation}\label{eq:pyroa}
F_i(t) = A_i\,\cdot X(t - \tau_i) + B_i \,,
\end{equation}
where $X(t)$ is the running optimal average of the full dataset, with a $\langle X\rangle=0$ and  $\langle X^2\rangle=1$, $A_i$ is a multiplicative amplitude factor (representing the RMS of the variability) and $B_i$ is an additive background. In order to construct $X(t)$, a running optimal average of the combined dataset is measured (after correcting for the lag $\tau_i$) using a Gaussian memory function, with a width $\Delta$. Then, Bayesian inference is performed using Markov Chain Monte Carlo (MCMC) sampling over the parameter space, including the lag parameters $\tau_i$ and the $\Delta$. The posterior probability is maximised over all model components, allowing PyROA to simultaneously fit multiple light curves with rigorous propagation of uncertainties. An additional extra variance per filter, $\sigma_i$, is added in quadrature to each datum to address any systematic uncertainties in the modelling. One main advantage of PyROA is that it that naturally expands the uncertainties at larger gaps (thus down-weighting their importance in the lag estimation), overcoming the challenge presented in other methods.
\begin{figure*}
	\includegraphics[width=18cm,trim=3cm 4.2cm 2cm 5.3cm,clip]{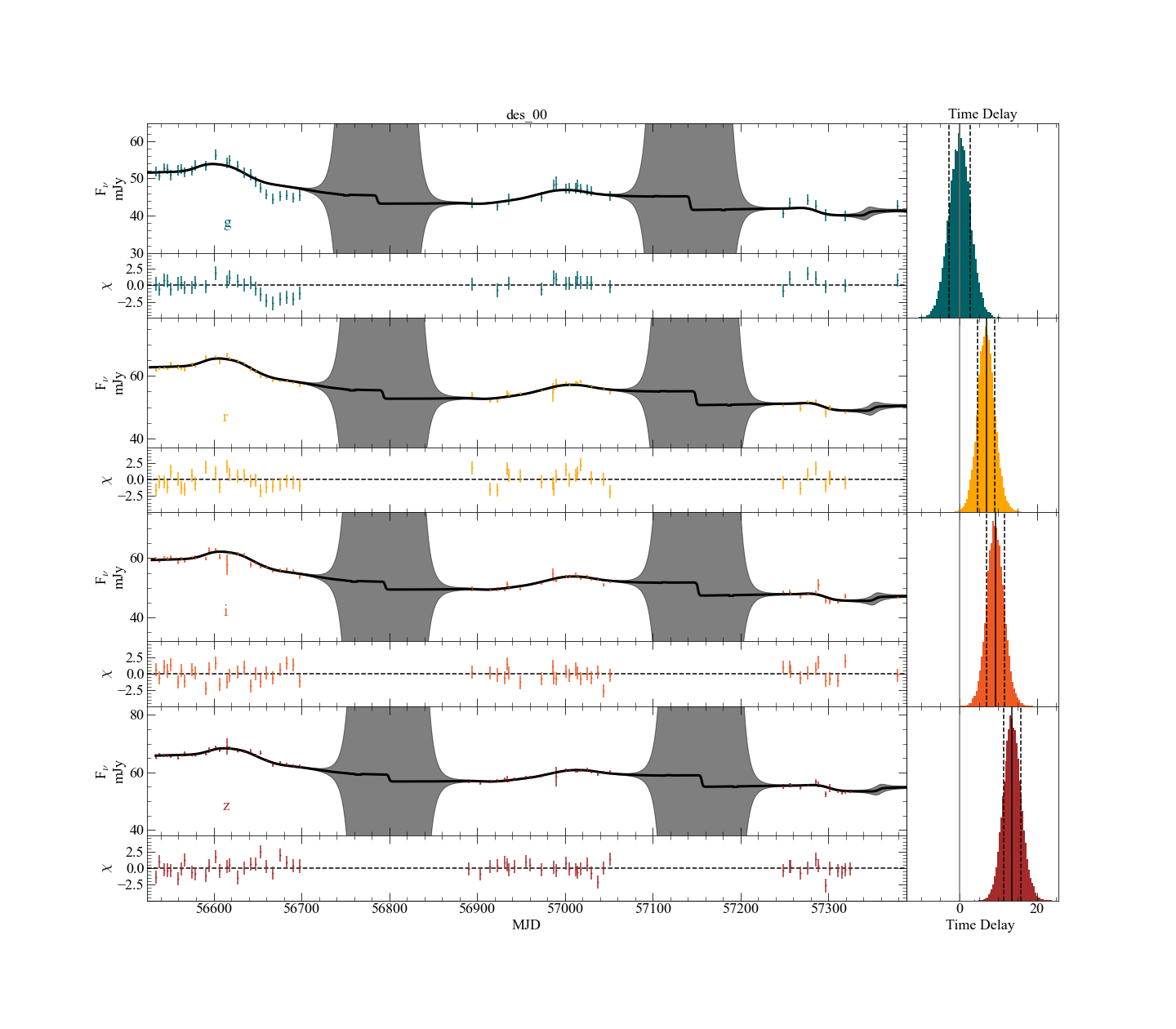}
    \caption{Light curve fits to M00 with {\sc PyROA} in the observed frame. The best fit (black) and its 68\% confidence interval are shown for each filter on every panel. Below, we show the normalised residuals. On the right of every panel, we display the marginalised posterior distribution of the delay in respect to $g$-band (observed frame). The median and its 68\% confidence interval are shown with a solid and dashed line, respectively. A grey line at $\tau=0$ is shown for reference.}
    \label{fig:des_00}
\end{figure*}

We fitted all 36 DES lightcurves from \citet{Mudd2018,Yu2020_DES} with PyROA, using 50,000 iterations in order to ensure convergence. An example of this fitting procedure, with the data and model shown in Fig.~\ref{fig:des_00}, as well as the marginal posterior distributions of the lags. Overall we find a good fit to the lightcurves, as demonstrated by the residuals below every lightcurve\footnote{The fits to each lightcurve is shown in the Supplementary material (online).}. In some instances, there are certain sections of the lightcurve which are not fully captured by the $X(t)$. It is expected, and demonstrated observationally, that longer wavelength lightcurves are smoother than their shorter wavelength counterparts in any single AGN. The transfer function that governs the accretion disc’s response to the driving signal—originating from the X-ray corona in the reprocessing model—broadens with increasing wavelength (due to the $T(R)\propto R^{-3/4}$ temperature radial profile). Consequently, emission from larger radii exhibits a progressively smoother response, leading to an overall wavelength-dependent smoothing of the observed light curves \citep[see][]{Cackett:2007, Starkey:2016}. As a result, the ROA used to construct the driving light curve, 
$X(t)$, is preferentially weighted toward the longer-wavelength bands.
This weighting introduces a bias, producing a smoother $X(t)$ compared to the shorter-wavelength light curves (e.g., the $g$-band). This is further compounded by the natural smoothing effect of short variability by the Gaussian memory function.  These mismatch effect is most apparent in portions of the light curve with sharp variability features at short wavelength, such as at the end of the first observing season in M00 (see Fig.~\ref{fig:des_00}).

\begin{figure*}
	\includegraphics[width=16cm,trim=0cm 0cm 0cm 0cm,clip]{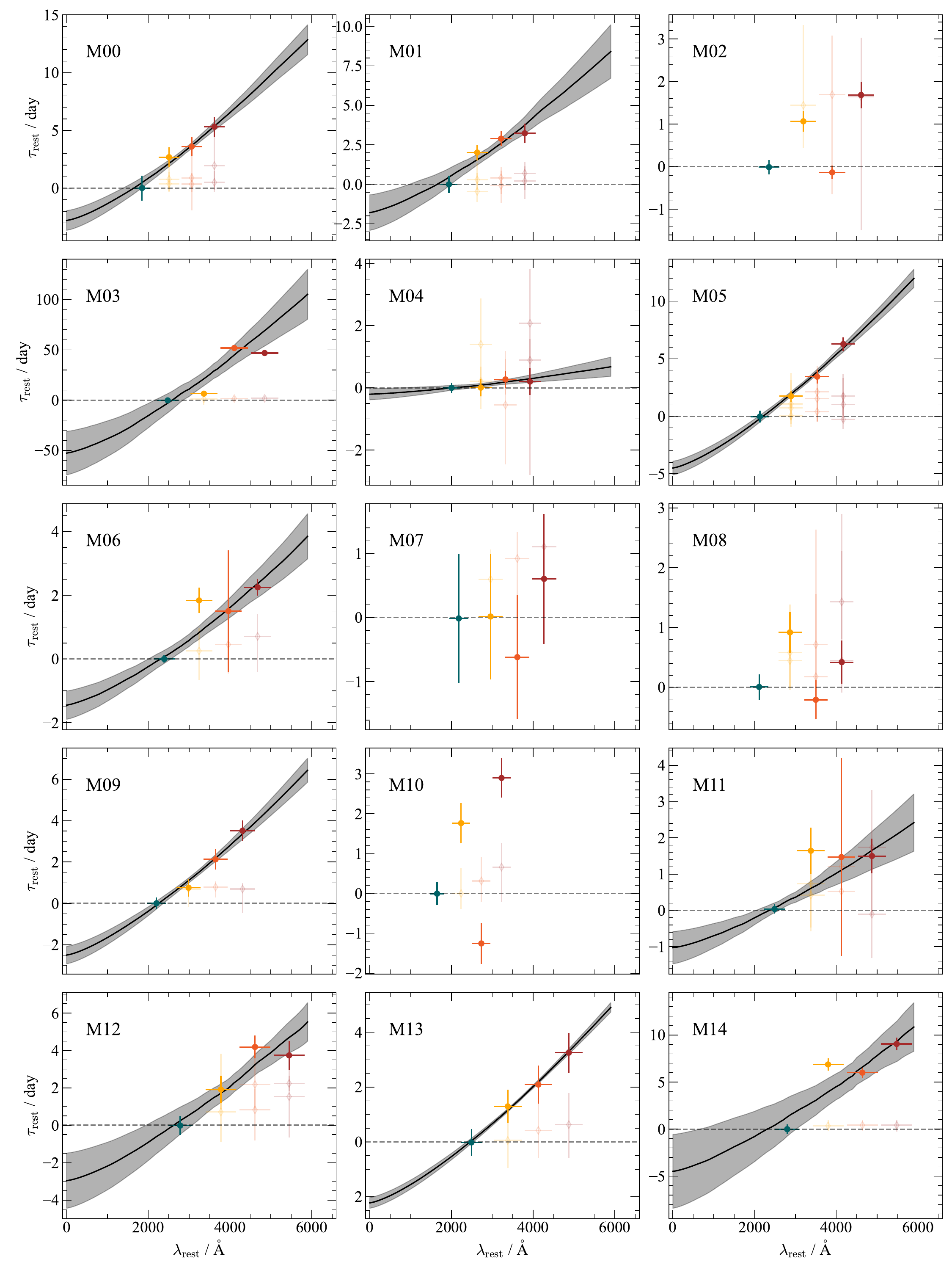}
    \caption{Delay spectrum for all 15 quasars in the \citet{Mudd2018} DES sample. Each panel shows the rest-frame lag measurements for each of the four filters, using the same colour scheme as in Fig.~\ref{fig:des_00}. The best fit power-law is shown in black, with its 68\% confidence interval as a grey band. A dashed line at $\tau=0$ is shown for reference. The lags from \citet{Mudd2018} in diamond for comparison. In some quasars, there are lag detection in different years.}
    \label{fig:des_lag_all}
\end{figure*}

\begin{figure*}
	\includegraphics[width=16cm,trim=0cm 0cm 0cm 0cm,clip]{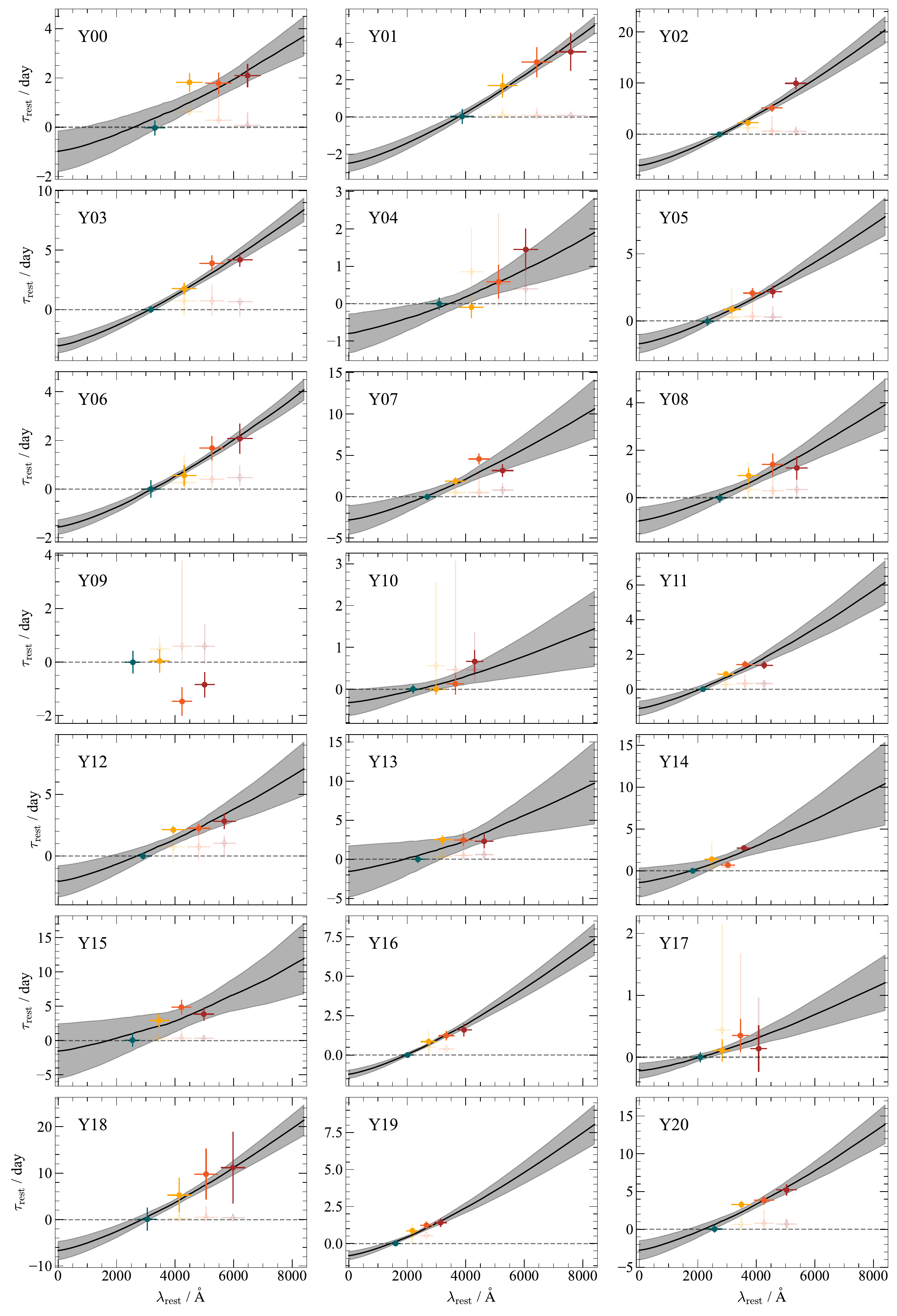}
    \caption{Delay spectrum for all 21 quasars in the \citet{Yu2020_DES} DES sample. Lines have same meanings as described in Fig.~\ref{fig:des_lag_all}.}
    \label{fig:des_lag_all_yu}
\end{figure*}

Including all three years of data significantly enhances the quality of the recovered delay spectra, yielding lags that are both more significant and better constrained than those obtained with {\sc javelin} in previous studies \citep{Mudd2018, Yu2020_DES}. By combining the full multi-year dataset and all available photometric bands to construct the driving light curve, $X(t)$, PyROA successfully recovers robust inter-band continuum delays for 31 out of 36 quasars in our sample. The resulting lag–wavelength relations (i.e., delay spectra) are presented in Fig.~\ref{fig:des_lag_all} and Fig.~\ref{fig:des_lag_all_yu}, with corresponding values listed in Table~\ref{tab:lags}. We also are able to recover lags on the Y-band from \citet{Yu2020_DES} which are not reported allowing to extend the lags spectrum to longer wavelengths on three quasars which had sufficient data available. The original measured lags (often only measured in one year of data) are also shown in the same figure with diamond markers for reference. In all quasars with significant lag detections, we find positive trends of lags increasing with wavelength. In three cases, M02, M08 and M10, we find robust detections of lags except on one/two of the bands. Only in two instances, M07 and Y07, we find no evidence for inter-band delays. These five AGN were not used to measure the size of the accretion disc in Sec.~\ref{sec:discsize} but they still contain information on the variable component, thus are included in the spectral energy distribution analysis in Sec.~\ref{sec:sed}.

It has been noted in recent studies, that the lag measured is a strong function of the timescale at which this is calculated \citep[][Marculewicz et al., in prep.]{Cackett2022,Lewin2023,Lewin2024}. This frequency or timescale resolved lags provide additional constraints on the origin of the reverberating components. We find that overall, the best fit parameter $\Delta$, which controls the timescale at which the lags is measured is comparable for the full sample, with a mean around $\Delta\sim2$ days. This allows us to directly compare the timescale and thus the accretion disc size between all quasars in the combined sample in Sec.~\ref{sec:discsize}.

\section{Spectral energy distribution}\label{sec:sed}
\begin{figure*}
	\includegraphics[width=2\columnwidth]{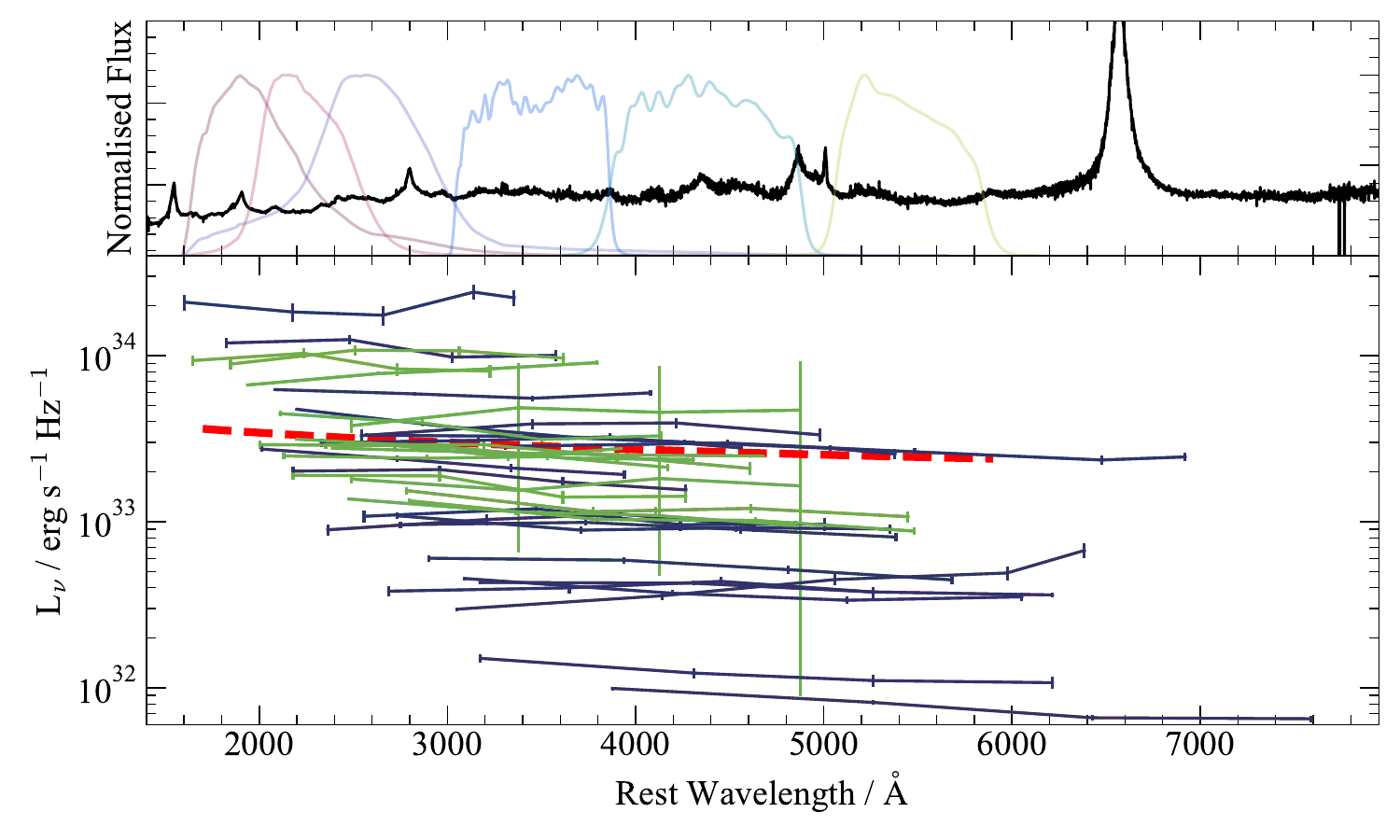}
    \caption{The spectral energy distribution of the variable component of DES quasars. {\it Top:} A quasar template at $z\simeq1-2$ shown for reference \citep{Selsing2016}.  These BRM lightcurves represent UV analogue experiments done in the local Universe where only space-based facilities are capable of obtaining the data e.g., Neil Gehrels {\it Swift} Observatory \citep{Gehrels:2004}. The filter transmission from {\it Swift}/UVOT are shown for reference \citep{Roming:2005}. {\it Bottom:} The Milky Way dust-corrected SEDs for all the quasars. The powerlaw expectation from the plateau of a geometrically thin accretion disc, $F_\nu\propto\lambda^{-1/3}$, is shown for reference as a red dashed line \citep{Shakura1973}.}
    \label{fig:all_seds}
\end{figure*}
An additional independent test of the accretion flow in quasars its their spectral energy distribution (SED). The linear model assumption of PyROA allows to naturally decompose AGN light curves into a variable (AGN) and non-variable components (host galaxy) by exploiting correlations between simultaneous flux measurements in different photometric bands. The method involves plotting the flux in one band against the reference lightcurve $X(t)$ to produce a linear relationship when the variability is dominated by a single spectral component—typically the accretion disc continuum \citep{Winkler1992,winkler1997,Haas2011}. The slope, $A_i$, of this flux–flux diagram reflects the colour of the variable component, while the intercept captures the contribution of constant components such as the host galaxy or non-varying emission lines. By performing a linear regression in flux space, one can extract the SED of the variable component independently of assumptions about its physical origin.  This technique is particularly useful for identifying non-variable contamination and isolating the intrinsic variability signal, which is crucial for accurate reverberation mapping and SED modelling \citep[e.g.,][]{Cackett:2007,Starkey2023}.

The linear model that {\sc PyROA} used to fit the lightcurves, naturally separates these two components (see Eq.~\ref{eq:pyroa}). The SEDs for the DES quasars are displayed in Fig.~\ref{fig:all_seds}. These have been corrected for their line-of-sight Milky Way dust extinction \citep{Schlegel1998} as implemented in {\sc dustmaps} \citep{Green2018}. Overall, most quasars follow a powerlaw behaviour, increasing towards smaller wavelengths. We note that quasars might be affected by additional host galaxy absorption and thus show a redder slope \citep[e.g.,][; we discuss the implications of this in Sec.~\ref{sec:thindisc?}]{Weaver2022}. We have not accounted for intrinsic absorption in the following analysis as it would depend on assuming an underlying model or an independent estimate of the internal redenning (e.g., Balmer decrement).

To quantify the Milky Way dust corrected SED slopes of each quasar, we fitted a simple power-law:
\begin{equation}
    f_\nu = A * \left(\frac{\lambda}{\lambda_{\rm SED,0}}\right)^\beta\,,
\end{equation}
where $\lambda_{\rm SED,0}$ is the arithmetic mean rest-wavelength of all the filters (used to minimise correlations in the fitting), $A$ is the flux at  $\lambda_{\rm SED,0}$ and $\beta$ is the power-law index. We then optimised the parameters by maximising the likelihood:
\begin{equation}
\ln \mathcal{L} = -\frac{1}{2} \sum_{i} \left[ \frac{(y_i - f_\nu(\lambda; \boldsymbol{A,\beta}))^2}{\sigma_i^2 + \sigma_{\text{int}}^2} + \ln\left( \sigma_i^2 + \sigma_{\text{int}}^2 \right) \right]\,,
\end{equation}
where the uncertainties from the flux-flux method, $\sigma_{\text{i}}$, have been modified by an additional extra variance $\sigma_{\text{int}}$, to reflect any deviations from a smooth powerlaw. This is expected as the SED will be modified by emission lines, as shown in the top panel of Fig.~\ref{fig:all_seds}. The results for all individual quasars are shown in Table~\ref{tab:sed_fits_mudd} and \ref{tab:sed_fits_yu}.

\begin{figure}
	\includegraphics[width=1.1\columnwidth,trim=0.3cm 0.2cm 0cm 1.5cm,clip]{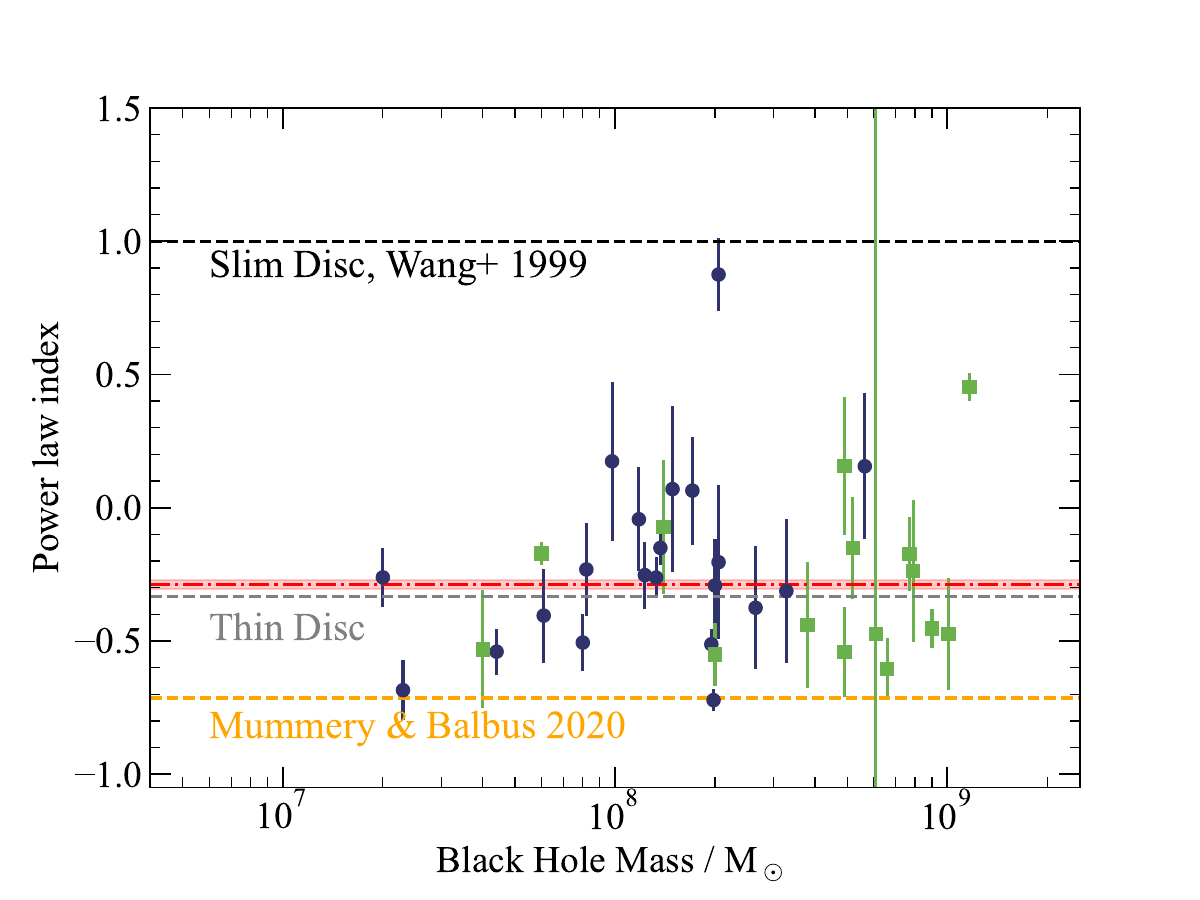}
    \caption{The spectral energy distribution powerlaw index, $\beta$, as a function of black hole mass, where $F_\nu\propto\lambda^\beta$. These show a consistent value around the geometrically thin disc at $\beta-1/3$ (grey line) rather than a slim disc (black line) \citep{Wang1999}. We find an average powerlaw index of $\beta=-0.28\pm0.017$ for the DES sample shown as the red dotted line, with its $1\sigma$ uncertainty as the red band.}
    \label{fig:slope_seds}
\end{figure}

Overall, the powerlaw index recovered have a large scatter around the geometrically thin accretion disc prediction \citep{Shakura1973} of $\beta=-1/3$ at $\sim1\sigma$, as displayed in Fig.~\ref{fig:slope_seds}, albeit with large uncertainties in many cases. An inverse-variance weighted average of all SEDs reveal a $\beta=-0.28\pm0.02$. A good fraction of these quasars show red spectra, with $\beta>0$, which implies additional host-galaxy absorption or a different temperature gradient \citep[e.g.,][; we discuss the implications of the latter in Sec.~\ref{sec:thindisc?}]{Weaver2022} and two AGN with significant steeper SEDs with $\beta\sim-0.7$ -- Y01 and Y10.

\subsection{The host galaxy}
\begin{figure}
	\includegraphics[width=1\columnwidth]{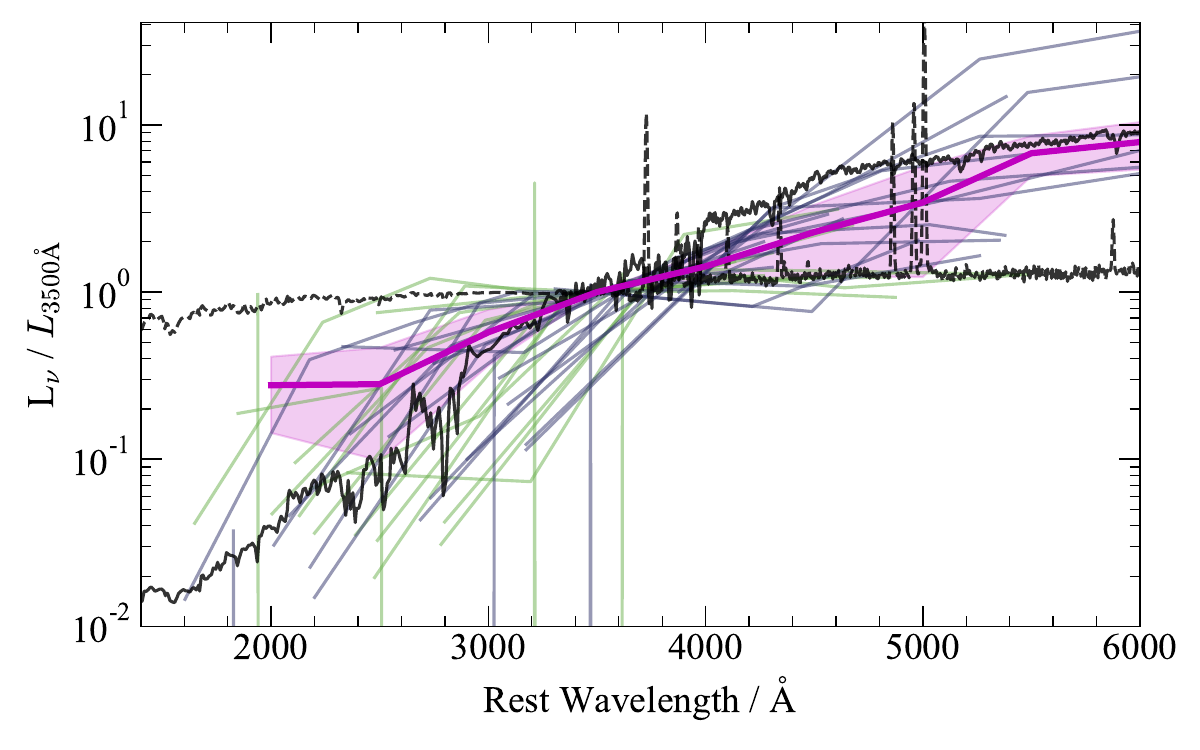}
    \caption{The spectral energy distribution of the constant component/host galaxy of DES quasars. All host galaxies from \citet[][blue]{Mudd2018} and \citet[][green]{Yu2020_DES} have been normalised to a common continuum at $3500$~\AA. For reference, we show a typical red sequence galaxy (NGC 7585, solid line) and a blue-cloud galaxy (compact starburst Mrk 930, dashed line) from \citet{Brown2014}.}
    \label{fig:host}
\end{figure}

The other outcome of the flux-flux analysis is the retrieval of the static component, associated to the host galaxy. In Fig.~\ref{fig:host}, we show all the host galaxy SED for the full sample, normalised to 3500~\AA, for ease of comparison. To make an average host galaxy from our sample, we have taken a median spectra, interpolating linearly between the centre of the filter bandpasses (omitting any extrapolation outside of the data for any given AGN). We show the median SED in Fig.~\ref{fig:host}, as well as the standard deviation around it. On average, the SEDs resemble early-type galaxies, as seen in studies in the local Universe \citep[e.g.,][]{Kauffmann2003} as well as high-redshift quasars in the SDSS sample \citep{Weaver2022}. However, there is a large dispersion around the mean host galaxy spectrum, due to the low number of sources in addition to the uneven sampling at any given wavelength (accentuated at the edges of the energy range). 

The shapes of the host-galaxy SEDs should be interpreted with caution. The shortest-wavelength point in each SED represents only a lower limit to the true constant component (host-galaxy) flux. In most BRM campaigns, the dynamic range of AGN variability is insufficient to probe low-luminosity states where the host galaxy dominates the emission. Consequently, the assumption that a linear extrapolation of the variability model (Eq.~\ref{eq:pyroa}) to $F_\nu = 0$ accurately represents the turn-off of the variable (AGN) component is, at best, an approximation \citep[or model-dependent, in cases where the host-galaxy SED is assumed in order to recover the AGN component; see][]{Gianniotis2022,Donnan2023}. Nevertheless, the derived intersection is less sensitive to the flux at longer wavelengths, which are typically brighter due to their redder spectra, and should therefore yield a relatively robust estimate of the host SED.

An additional complication in isolating the true host-galaxy contribution arises from circumnuclear star-formation rings—structures typically located a few hundred parsecs from galactic centres \citep[e.g.,][]{Wilson1991,Riffel2016}. These compact regions, unresolved by ground-based observatories, likely contribute excess emission at shorter wavelengths, thereby biasing the inferred host-galaxy SED when applied to single templates as shown in Fig.~\ref{fig:host}.

\section{Discussion}\label{sec:results}
\subsection{The size of the accretion disc}\label{sec:discsize}
In studies of AGN in the local Universe, the delay spectrum---i.e., the lag between variations in different wavelength bands---is a powerful diagnostic of the structure and physics of the accretion disc. Observations often reveal that the time delays between continuum emission at different wavelengths scale approximately as a power law with wavelength, typically of the form $\tau \propto \lambda^\beta$. This behaviour is broadly consistent with expectations from standard thin disk models \citep{Shakura1973}, where longer wavelengths are emitted from cooler, outer regions of the disk, thus exhibiting larger light-travel-time delays. Empirically, the exponent $\beta$ is often found to be close to $4/3$ \citep{Edelson2015, Fausnaugh:2016}, matching theoretical predictions under assumptions of a steady-state, optically thick, geometrically thin disk. Deviations from the expected power-law index can indicate additional physical processes such as disk inhomogeneities \citep{Starkey2017,Starkey2023}, temperature fluctuations \citep{Neustadt2024} or contributions from diffuse continuum emission from the BLR \citep{Korista2001,Cackett:2018,Korista:2019,Netzer2020}. Fitting the delay spectrum with a power-law model thus provides a simplistic yet informative phenomenological constraint on the accretion disc size and the reprocessing geometry in AGN. 

Here, we follow a similar procedure by performing a fit using the parametrised power-law:
\begin{equation}
\tau(\lambda) = \tau_0 \left[\left( \frac{\lambda}{\lambda_0} \right)^\beta - y_0\right]\,,
\end{equation}
where $\tau(\lambda)$ is the time delay measured at wavelength $\lambda$,
$\tau_0$ is the normalisation constant, representing the delay at the reference wavelength $\lambda_0$, $\lambda_0$ is the reference wavelength, $y_0$ as a normalisation constant and $\beta$ is the power-law index, with a theoretically expected value of $\beta = 4/3$, which was kept fixed, for a standard geometrically thin, optically thick accretion disk \citep{Collier1999,Cackett:2007}. Thus, we fitted for both $y_0$ and $\tau_0$. These fits are shown in Fig.~\ref{fig:des_lag_all} and Fig.~\ref{fig:des_lag_all_yu} with their 68\% confidence interval sampled from $10^5$ Monte Carlo simulations accounting for the covariance between the parameters. The best fit parameters and uncertainties are shown in each panel in Figures and presented in Table~\ref{tab:lags}.

For the most part, the {\sc PyROA} analyses of the DES quasars reveal an improvement on the disc size, $\tau_0$, from using individual years only and providing robust evidence for lag spectra detection up to $z\lesssim2$. This improvement arises from 
using all the multi-year data and the four/five bands simultaneously to create a `driving lightcurve' ($X(t)$) rather than relying exclusively on one of the lightcurves to measure the delays, provides significant detection power. Furthermore, the larger inter-band delays in the observed frame (due to the expansion of the Universe, a factor of $(1+z)$) result in a better sampling of $X(t)$, as it naturally fills the gaps when all shifted into a common frame within PyROA. This enables to constrain the lag spectrum even when in individual years are not detected with other methods. Overall, the shape of the lag spectra are well described by a power-law,  as expected from accretion disc theory \citep{Collier1999,Cackett:2007} and comparable with results in the local Universe \citep[e.g.,][]{Fausnaugh:2016,Cackett:2018,Hernandez2020,Vincentelli2021}. Given that we find robust detections on the individual quasars, we analysed them separately rather than as an ensemble \citep{Mudd2018,Homayouni2019}.
\begin{figure}
	\includegraphics[width=1.1\columnwidth]{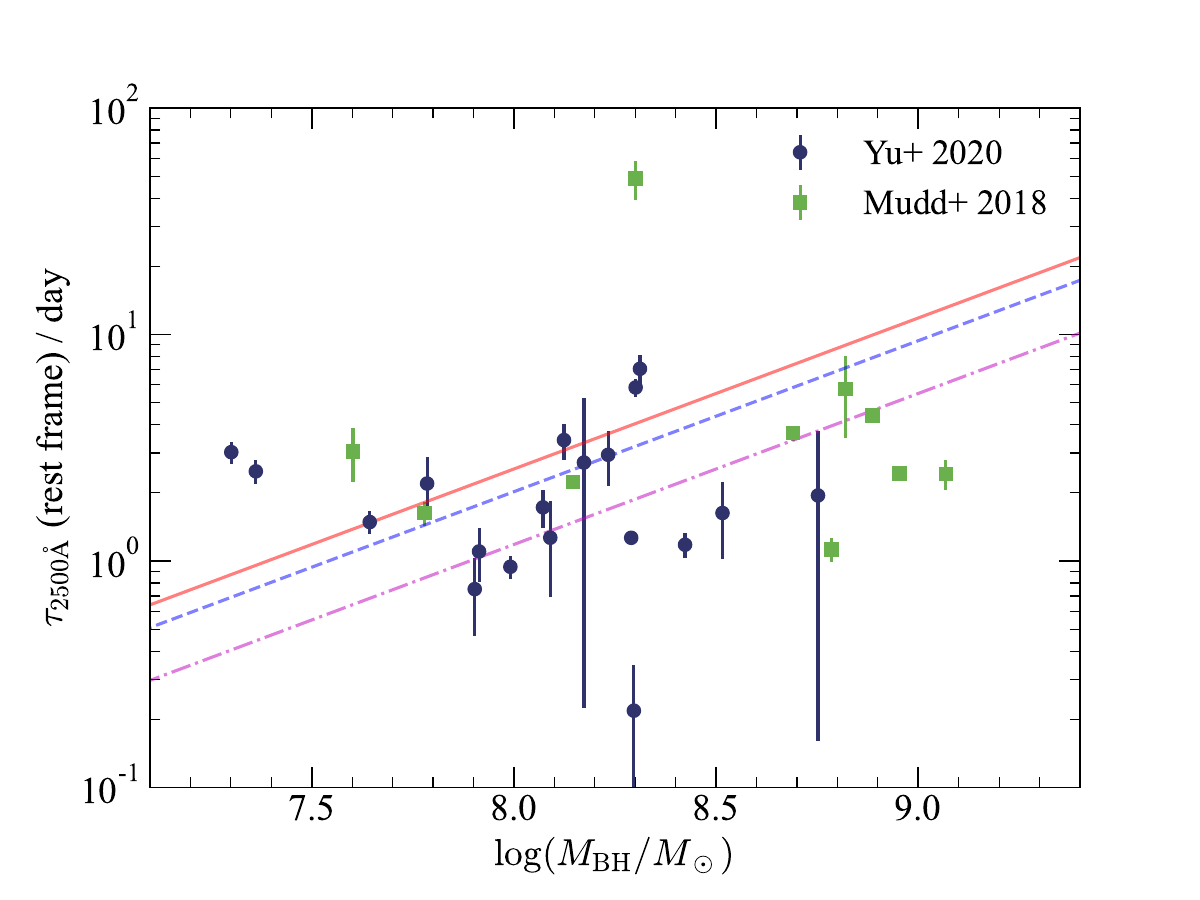}
    \caption{The accretion disc size at a reference wavelength of 2500~\AA as a function of black hole mass. These are taken from the extrapolations of the delay spectra fits in Sec.~\ref{sec:lags}. The lines display the expected accretion disc size for a quasar with an Eddington ratio of 1, 0.5 and 0.1 shown in red, blue and magenta, respectively. We note that the expected uncertainties in the black hole are mass are of order 0.4 dex \citep{Mudd2018}, which would introduce additional scatter to this relations.}
    \label{fig:scaling}
\end{figure}

To compare the sizes of these discs, as they are all measured at different redshifts, we extrapolate from the best power-law fit the size at 2500~\AA\ i.e., $\tau_{2500}$, as it lies close the observed rest-frame bands in all quasars (minimising any extrapolation outside of the data range). Using the inferred black hole masses from single-epoch spectroscopy derived from \ion{Mg}{ii} emission line \citep{Mudd2018}, we show the predicted size normalisation $\tau_0$ for a standard accretion disc \citep{Shakura1973,Fausnaugh:2016}:
\begin{equation}
    \tau_0 = \frac{1}{c}\left(X\frac{k\lambda_0}{hc}\right)^{4/3}\left[\left(\frac{GM_{\rm BH}}{8\pi\sigma}\right)\left(\frac{L_{\rm Edd}}{\eta c^2}\right)\left(3+\kappa\right)\dot{m}_{\rm Edd}\right]^{1/3}\,
\end{equation}
where $X = 4.9$ is the conversion constant that accounts for the conversion of temperature to lag at a given radius, $\lambda_0$ is the reference rest-wavelength,  $L_{\rm Edd}$ is the Eddington luminosity, $\eta=0.1$ is the radiative efficiency, $\kappa = 1$ is the local ratio of external to internal heating (assumed constant throughout the disc), $\dot{m}_{\rm Edd} = L_{\rm Bol}/L_{\rm Edd}$ is the Eddington ratio.
In Fig.~\ref{fig:scaling}, we show the expected $\tau_{2500}$ as a function of black hole mass, at a fixed Eddington ratio as the dashed lines. The measured sizes, assuming no contribution from additional reprocessors (e.g., diffuse continuum of the BLR) suggest that most quasars in this sample are highly accreting SMBHs at $\dot{m}_{\rm Edd}>0.5$ at lower black hole mass ($\log(M_{\rm BH}/M_\odot) \lesssim 8$). At higher values, we find a larger diversity in Eddington ratio. This effect is likely a selection bias, as this sample is flux-limited, we expect to pick-up the highest $\dot{m}_{\rm Edd}$ AGN, particularly at low black hole mass. In a similar way, the photometric cadence of DES also favours longer lags which arise either from large  $M_{\rm BH}$ and/or $\dot{m}_{\rm Edd}$.

\subsection{Slim or thin disc?}\label{sec:thindisc?}

Theoretical models of disc accretion predict distinct spectral energy distributions (SEDs) and temperature profiles depending on the accretion regime. In the standard geometrically thin, optically thick accretion disk model formulated by \citet{Shakura1973}, the disk temperature scales as \( T(R) \propto R^{-3/4} \), leading to a multicolour blackbody spectrum with a characteristic UV/optical spectral slope of \( F_\nu \propto \lambda^{-1/3} \). However, at higher accretion rates—near or above the Eddington limit—accretion flows are expected to transition into a slim disk regime \citep{Abramowicz1988}, where advection becomes significant and the radial temperature profile flattens, altering the emergent spectrum.

Observations of AGN in the local Universe have provided a critical test-bed for distinguishing between these accretion regimes. Spectral slope measurements from high-quality UV/optical spectra, e.g., SDSS, show a diversity of slopes that often deviate from the \( \nu^{1/3} \) prediction. For instance, \citet{Davis2007} examined a sample of local AGN and found that the optical/UV slopes were generally redder than the standard thin disk prediction, suggesting either non-standard disk structures or the effects of extinction and host galaxy contamination. 
At high Eddington ratios, particularly in narrow-line Seyfert 1 galaxies, the observed spectral slopes tend to be flatter or even inverted, indicating significant deviations from standard thin disk theory. Slim disk models predict a flatter radial temperature profile, such as \( T(R) \propto R^{-1/2} \), which results in a broader, less peaked spectrum with a characteristic slope of $F_\nu\propto\lambda$. Observationally, sources with high accretion rates (\( L_{\text{bol}}/L_{\text{Edd}} \gtrsim 0.3 \)) often display SEDs consistent with such modified structures, supporting the presence of slim disks \citep{Wang1999,Done2012D}. 

However, intensive BRM campaigns on nearby ($z\lesssim0.1$) high-Eddington ratio sources show mixed inferences on the SED shape of the variable component. Both Mrk 142 \citep{Cackett2020} and PG 1119+102 \citep{Donnan2023} show that the variable SED shape is well described by a simple powerlaw, which slope is consistent with the plateau of a geometrically thin disc $f_\nu\propto\lambda^{-1/3}$. \citet{Donnan2023} showed that the slow-varying component in PG~1119+102 had not only longer timescale delay spectrum and but its SED showed an $u$-band excess as well, suggestive of hydrogen Balmer edge emission. This supports the scenario found in most intensive BRM campaigns that require at least two distinct reprocessors to reproduce the UV/optical lags, hinting at the BLR as the possible culprit \citep{Korista2001,Korista:2019,Netzer2020,Lewin2024}.

While the average slope of the full sample is consistent with the geometrically thin prediction, there is significant scatter above and below the expected $\beta=-1/3$. Intrinsic host galaxy extinction, which should be present in all quasars to some degree, is likely responsible for the flatter or even positive slopes observed. An alternative is that the inner disc temperature is cool enough to observe the turn-over in the spectrum. As noted in Sec.~\ref{sec:discsize}, the disc sizes alone suggest higher Eddington ratio sources for which the inner temperature be high enough to produce the turn-over in the far-ultraviolet. 

Interestingly, we find the slopes of several quasars which are steeper than the geometrically thin prediction. Strong emission lines could potentially bias the SED continuum estimates, as they will be stronger that expected at particular bands. However, given the broad-wavelength coverage of the DES filters, their influence would be greatly diminished. Another possibility is that the radial profile is different from the \citet{Shakura1973} prediction. Steep slopes are mostly set by the temperature gradient across the accretion disc. An alternative gradient can be introduced by finite stress boundary condition at the ISCO, which modifies the slope of the SED to $\beta=-0.7$ \citep{Mummery2020}, precisely where these two AGN lie. We note that \citet{Weaver2022} also found in a larger sample of quasars, that a steeper slope (in addition to host-galaxy extinction) was a better match to the data.

\subsection{The impact of the DCE in high-redshift quasar RM}\label{sec:lsst}
\begin{figure}
    \includegraphics[width=1.1\columnwidth,trim=1cm 0.2cm 0cm 0cm,clip]{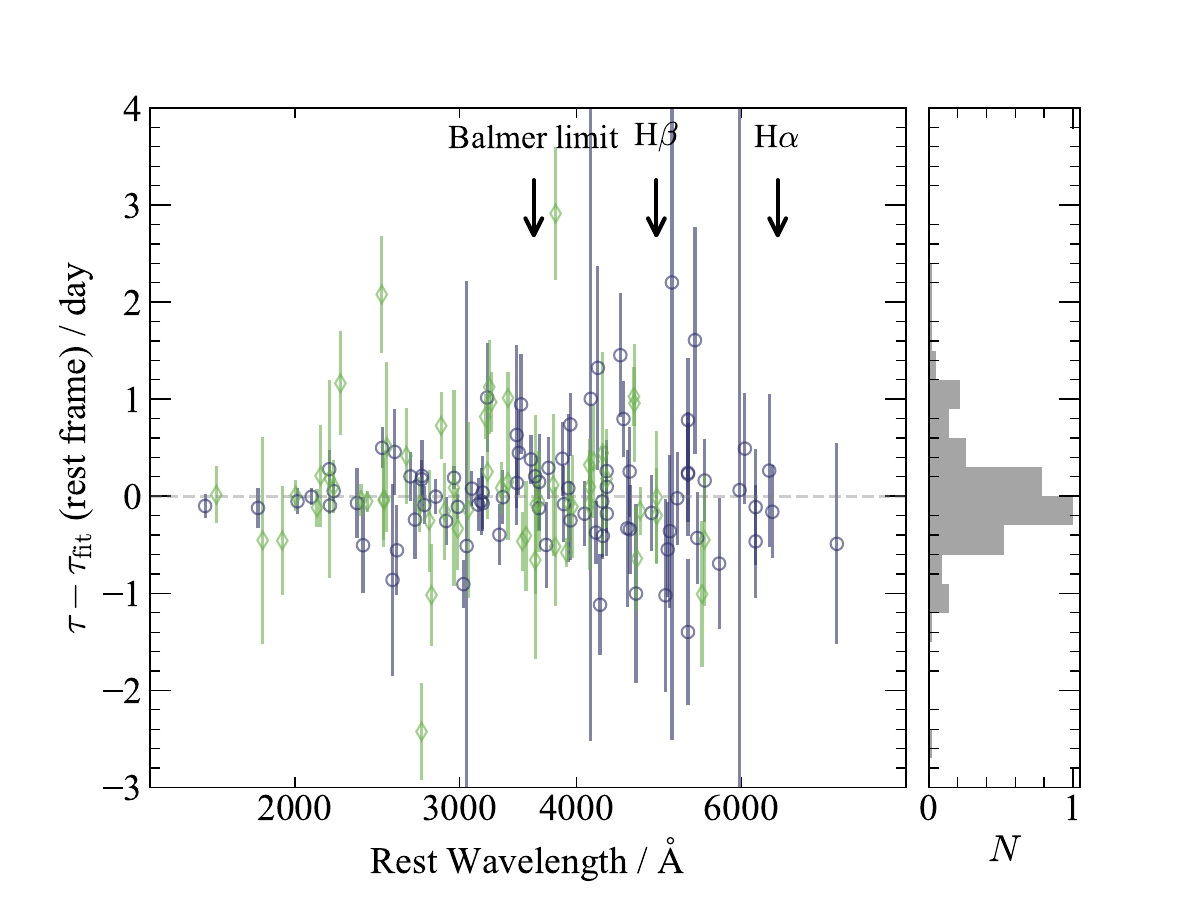}
	\includegraphics[width=1.1\columnwidth,trim=1cm 0.2cm 0cm 2cm,clip]{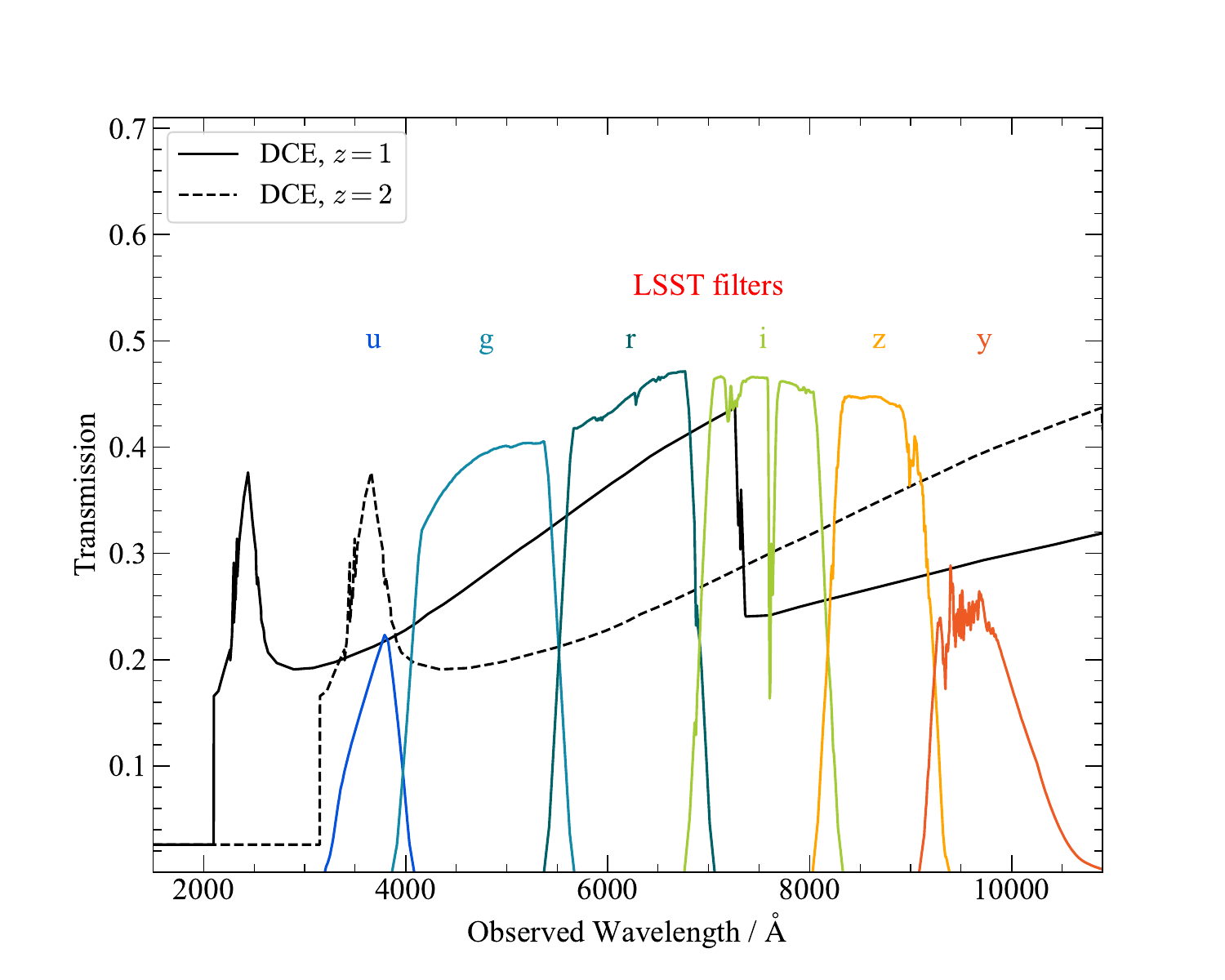}
    \caption{The high-redshift quasars in DES and LSST probe the ultraviolet range of the AGN (redshifted to the optical bands). {\it Top:} Residual lag from the best power-law fit as a function of wavelength. There is no evidence for an excess around the Balmer limit. The histogram on the right reflect the distribution of the data around the best fit. {\it Bottom: }We show the delay spectrum which includes the DCE from the BLR from \citet{Korista:2019} in two redshifts: $z=1$ (solid line) and $z=2$ (dashed line). This exemplifies the importance of the Balmer jump in emission feature for $z\sim1$ quasars as it lies between the $r$- and $i$-bands, with an expected drop in the $z$-band.}
    \label{fig:lsst}
\end{figure}

Given the evidence of the DCE in both the lag spectrum and SED in the local Universe across a large sample of AGN mass and luminosity, it is at least plausible that high-redshift quasar will also be contaminated from variations of their own BLR. Given the redshift at which these quasars lie, the optical bands we measure from DES or the upcoming Legacy Survey of Space and Time (LSST) \citep{Brandt:2015} are mostly sensitive to the UV part of the quasar spectrum. This is explicitly shown in the lower panel of Fig.~\ref{fig:lsst}. For $z\sim1$, the Balmer jump lies around the $i$-band, which would create an excess and/or flattening in the lag when compared to their adjacent bands. This is the same redshifted $u$-band excess ubiquitously observed in low-redshift AGN \citep[e.g.,][]{Fausnaugh:2016,Cackett:2018,Edelson:2019}. %

In order to search for any excess lag or deviation from a simple power-law, we plotted the residuals of the delay spectrum against the best fit in the top panel of Fig.~\ref{fig:lsst}. There is no obvious trend in the residuals nor excess lag in particular regions. In fact, most of the lags detected lie very close to the fit, as demonstrated in the histogram on the the right panel. Given that most fits to the delay spectra have only two degrees of freedom, it is not surprising that strong deviations are difficult to capture in these fits. The prospect of quantifying the DCE contribution will likely be possible with LSST, as it will enable the retrieval of time delays on six bands. The extended wavelength coverage should provide a better constraint on the power-law shape at both ends of the Balmer jump and determine any excess far better than DES. These constraints do not live in isolation, as they should be present in the SED as well. The broad continuum contribution of the DCE naturally produces flatter SED \citep[accretion disc $+$ DCE, see figure 9 in][]{Korista:2019,Netzer2022}, thus affecting any inferred slope attributed to the accretion disc only. This additional component, currently not considered, might contribute towards the flatter and large dispersion observed in the SED slopes (e.g., Fig.~\ref{fig:slope_seds}). In order to properly quantify the contribution, it will be necessary to either model both components simultaneously, or remove the DCE contribution (e.g., by estimating its broad-line luminosities from the single-epoch spectra and compute the corresponding DCE continuum emission) to retreive unbiased measurements of the accretion disc properties (e.g., mass accretion rate).

Recently, the detailed study of I Zw 1 (a nearby narrow-line Seyfert 1, Drewes et al, in prep) show that an additional reprocessor is required to modelled the timescale dependent lags in a high-Eddington AGN. This is interpreted as evidence of the DCE artificially boosting the lags despite a weak U-band excess \citep[e.g.,][]{Edelson:2019,Cackett:2018,Hernandez2020,Netzer2022,Hagen2023b}. %Furthermore, there is evidence of this DCE in the lag spectrum as the well-documented excess in the U-band attributed to hydrogen Balmer emission . 
The solution to determine the true influence of the DCE to the delay spectra in these high-redshift, high-Eddington sources will lies on the long-term monitoring and analysis of the delay spectrum and SED as a function of timescales \citep[or better known as frequency-resolved lags][]{Uttley2014, Cackett2020,Lewin2023,Lewin2024,Edelson2024} across all bands in order to search for the low-frequency delays (associated to reprocessing at large distances). 
%These long-term variations are often removed by detrending the lightcurves before the time-series analysis \citep[e.g.,][]{Hernandez2020,McHardy2022} or modelled simultaneously with the high-frequency variations \citep{Donnan2023}.
The full ten-year survey of LSST should provide sufficient coverage on the rest-frame of the quasar, $10\,(1+z)^{-1}$~years, to retrieve these long-timescale delays. Even at $z\sim2$, it should be possible to uncover the true size-scale of the additional reprocessor and its SED, as it would represent three years of data on the quasar rest frame, similar to nearby studies which have successfully characterise this component in analogue local high-Eddington AGN (e.g., Drewes et al., in prep).

\section{Conclusions}\label{sec:conclusions}
We have re-analysed the DES quasar lightcurves from \citet{Mudd2018} and \citet{Yu2020_DES} to perform broadband reverberation mapping at cosmic noon ($0.7\lesssim z\lesssim2.0$). We used {\sc PyROA} \citep{Donnan2021,Donnan2023} to successfully model all the multi-epoch and multi-band data which revealed:
\begin{itemize}
    \item Robust detections of inter-band delays in 31/36 quasars in the sample. In general, these delays scale with increasing wavelength, consistent with the prediction for a geometrically thin, optically thick disc $\tau\propto\lambda^{4/3}$. In three quasars, lags are detected with a one of the bands being consistent with zero. Due to the redshift probe by the DES survey, this sample represents the largest UV-rest frame BRM detections, analogues to the space-based experiments done in the local Universe \citep[e.g.,][]{Edelson:2019}.
    \item The size of the accretion discs suggest they are highly-accreting black holes with Eddington ratios $\dot{m}_{\rm Edd}> 0.5$
    \item The spectral energy distribution of the quasars have slopes in the UV range consistent with the predicted plateau of $f_\nu\propto\lambda^{-1/3}$ from accretion disc theory.
    \item There is no evidence for a u-band excess, often attributed to diffuse continuum emission from the BLR, as often observed in AGN in the local Universe.
\end{itemize}

LSST will deliver wide-field, high-cadence observations probing the rest-frame ultraviolet emission of quasars at cosmic noon over the coming decade. This capability will enable large-scale, reverberation mapping studies across a broad range of supermassive black hole (SMBH) properties, including black hole mass and accretion rate. To fully exploit this potential, it is crucial to quantify the impact of the broad-line region and diffuse continuum emission (BLR/DCE) on the observed UV variability \citep{Edelson:2019,Hagen2025,Lewin2025} and to develop robust methodologies to disentangle their respective observational signatures.

\section*{Acknowledgements}

JVHS acknowledges Keith Horne, Roberta Vieliute, Nikko Juengsophonvitavas, Aditya Sinha, Farah Wallauer and Reinosuke Kusano for fruitful discussions and comments that improved the paper.

This project used public archival data from the Dark Energy Survey (DES). Funding for the DES Projects has been provided by the U.S. Department of Energy, the U.S. National Science Foundation, the Ministry of Science and Education of Spain, the Science and Technology FacilitiesCouncil of the United Kingdom, the Higher Education Funding Council for England, the National Center for Supercomputing Applications at the University of Illinois at Urbana-Champaign, the Kavli Institute of Cosmological Physics at the University of Chicago, the Center for Cosmology and Astro-Particle Physics at the Ohio State University, the Mitchell Institute for Fundamental Physics and Astronomy at Texas A\&M University, Financiadora de Estudos e Projetos, Funda{\c c}{\~a}o Carlos Chagas Filho de Amparo {\`a} Pesquisa do Estado do Rio de Janeiro, Conselho Nacional de Desenvolvimento Cient{\'i}fico e Tecnol{\'o}gico and the Minist{\'e}rio da Ci{\^e}ncia, Tecnologia e Inova{\c c}{\~a}o, the Deutsche Forschungsgemeinschaft, and the Collaborating Institutions in the Dark Energy Survey.
The Collaborating Institutions are Argonne National Laboratory, the University of California at Santa Cruz, the University of Cambridge, Centro de Investigaciones Energ{\'e}ticas, Medioambientales y Tecnol{\'o}gicas-Madrid, the University of Chicago, University College London, the DES-Brazil Consortium, the University of Edinburgh, the Eidgen{\"o}ssische Technische Hochschule (ETH) Z{\"u}rich,  Fermi National Accelerator Laboratory, the University of Illinois at Urbana-Champaign, the Institut de Ci{\`e}ncies de l'Espai (IEEC/CSIC), the Institut de F{\'i}sica d'Altes Energies, Lawrence Berkeley National Laboratory, the Ludwig-Maximilians Universit{\"a}t M{\"u}nchen and the associated Excellence Cluster Universe, the University of Michigan, the National Optical Astronomy Observatory, the University of Nottingham, The Ohio State University, the OzDES Membership Consortium, the University of Pennsylvania, the University of Portsmouth, SLAC National Accelerator Laboratory, Stanford University, the University of Sussex, and Texas A\&M University.
Based in part on observations at Cerro Tololo Inter-American Observatory, National Optical Astronomy Observatory, which is operated by the Association of Universities for Research in Astronomy (AURA) under a cooperative agreement with the National Science Foundation.
%%%%%%%%%%%%%%%%%%%%%%%%%%%%%%%%%%%%%%%%%%%%%%%%%%
\section*{Data Availability}

The data used in this paper was taken from the original studies of \citet{Mudd2018} and \citet{Yu2020_DES}.



\bibliographystyle{mnras}
\bibliography{bibliography} % if your bibtex file is called example.bib

@ARTICLE{Starkey:2016,
       author = {{Starkey}, D.~A. and {Horne}, Keith and {Villforth}, C.},
        title = "{Accretion disc time lag distributions: applying CREAM to simulated AGN light curves}",
      journal = {\mnras},
         year = "2016",
        month = "Feb",
       volume = {456},
       number = {2},
        pages = {1960-1973},
          doi = {10.1093/mnras/stv2744},
archivePrefix = {arXiv},
       eprint = {1511.06162},
 primaryClass = {astro-ph.GA},
       adsurl = {https://ui.adsabs.harvard.edu/abs/2016MNRAS.456.1960S}
}

@ARTICLE{Blandford:1982,
       author = {{Blandford}, R.~D. and {McKee}, C.~F.},
        title = "{Reverberation mapping of the emission line regions of Seyfert galaxies and quasars.}",
      journal = {\apj},
         year = "1982",
        month = "Apr",
       volume = {255},
        pages = {419-439},
          doi = {10.1086/159843},
       adsurl = {https://ui.adsabs.harvard.edu/abs/1982ApJ...255..419B}
}

@ARTICLE{Cackett2020,
       author = {{Cackett}, Edward M. and {Gelbord}, Jonathan and {Li}, Yan-Rong and {Horne}, Keith and {Wang}, Jian-Min and {Barth}, Aaron J. and {Bai}, Jin-Ming and {Bian}, Wei-Hao and {Carroll}, Russell W. and {Du}, Pu and {Edelson}, Rick and {Goad}, Michael R. and {Ho}, Luis C. and {Hu}, Chen and {Khatu}, Viraja C. and {Luo}, Bin and {Miller}, Jake and {Yuan}, Ye-Fei},
        title = "{Supermassive Black Holes with High Accretion Rates in Active Galactic Nuclei. XI. Accretion Disk Reverberation Mapping of Mrk 142}",
      journal = {\apj},
         year = 2020,
        month = jun,
       volume = {896},
       number = {1},
          eid = {1},
        pages = {1},
          doi = {10.3847/1538-4357/ab91b5},
archivePrefix = {arXiv},
       eprint = {2005.03685},
 primaryClass = {astro-ph.HE},
       adsurl = {https://ui.adsabs.harvard.edu/abs/2020ApJ...896....1C}
}

@ARTICLE{Cackett:2018,
       author = {{Cackett}, Edward M. and {Chiang}, Chia-Ying and {McHardy}, Ian and
         {Edelson}, Rick and {Goad}, Michael R. and {Horne}, Keith and
         {Korista}, Kirk T.},
        title = "{Accretion Disk Reverberation with Hubble Space Telescope Observations of NGC 4593: Evidence for Diffuse Continuum Lags}",
      journal = {\apj},
         year = "2018",
        month = "Apr",
       volume = {857},
       number = {1},
          eid = {53},
        pages = {53},
          doi = {10.3847/1538-4357/aab4f7},
archivePrefix = {arXiv},
       eprint = {1712.04025},
 primaryClass = {astro-ph.HE},
       adsurl = {https://ui.adsabs.harvard.edu/abs/2018ApJ...857...53C}
}

@ARTICLE{Cackett:2007,
       author = {{Cackett}, Edward M. and {Horne}, Keith and {Winkler}, Hartmut},
        title = "{Testing thermal reprocessing in active galactic nuclei accretion discs}",
      journal = {\mnras},
         year = "2007",
        month = "Sep",
       volume = {380},
       number = {2},
        pages = {669-682},
          doi = {10.1111/j.1365-2966.2007.12098.x},
archivePrefix = {arXiv},
       eprint = {0706.1464},
 primaryClass = {astro-ph},
       adsurl = {https://ui.adsabs.harvard.edu/abs/2007MNRAS.380..669C}
}

@ARTICLE{Edelson:2019,
   author = {{Edelson}, R. and {Gelbord}, J. and {Cackett}, E. and {Peterson}, B.~M. and 
	{Horne}, K. and {Barth}, A.~J. and {Starkey}, D.~A. and {Bentz}, M. and 
	{Brandt}, W.~N. and {Goad}, M. and {Joner}, M. and {Korista}, K. and 
	{Netzer}, H. and {Page}, K. and {Uttley}, P. and {Vaughan}, S. and 
	{Breeveld}, A. and {Cenko}, S.~B. and {Done}, C. and {Evans}, P. and 
	{Fausnaugh}, M. and {Ferland}, G. and {Gonzalez-Buitrago}, D. and 
	{Gropp}, J. and {Grupe}, D. and {Kaastra}, J. and {Kennea}, J. and 
	{Kriss}, G. and {Mathur}, S. and {Mehdipour}, M. and {Mudd}, D. and 
	{Nousek}, J. and {Schmidt}, T. and {Vestergaard}, M. and {Villforth}, C.
	},
    title = "{The First Swift Intensive AGN Accretion Disk Reverberation Mapping Survey}",
  journal = {\apj},
archivePrefix = "arXiv",
   eprint = {1811.07956},
 primaryClass = "astro-ph.HE",
     year = 2019,
    month = jan,
   volume = 870,
      eid = {123},
    pages = {123},
      doi = {10.3847/1538-4357/aaf3b4},
   adsurl = {https://ui.adsabs.harvard.edu/abs/2019ApJ...870..123E}
}

@ARTICLE{Fausnaugh:2016,
       author = {{Fausnaugh}, M.~M. and {Denney}, K.~D. and {Barth}, A.~J. and
         {Bentz}, M.~C. and {Bottorff}, M.~C. and {Carini}, M.~T. and
         {Croxall}, K.~V. and {De Rosa}, G. and {Goad}, M.~R. and
         {Horne}, Keith and {Joner}, M.~D. and {Kaspi}, S. and {Kim}, M. and
         {Klimanov}, S.~A. and {Kochanek}, C.~S. and {Leonard}, D.~C. and
         {Netzer}, H. and {Peterson}, B.~M. and {Schn{\"u}lle}, K. and
         {Sergeev}, S.~G. and {Vestergaard}, M. and {Zheng}, W. -K. and
         {Zu}, Y. and {Anderson}, M.~D. and {Ar{\'e}valo}, P. and {Bazhaw}, C. and
         {Borman}, G.~A. and {Boroson}, T.~A. and {Brandt}, W.~N. and
         {Breeveld}, A.~A. and {Brewer}, B.~J. and {Cackett}, E.~M. and
         {Crenshaw}, D.~M. and {Dalla Bont{\`a}}, E. and
         {De Lorenzo-C{\'a}ceres}, A. and {Dietrich}, M. and {Edelson}, R. and
         {Efimova}, N.~V. and {Ely}, J. and {Evans}, P.~A. and
         {Filippenko}, A.~V. and {Flatland}, K. and {Gehrels}, N. and
         {Geier}, S. and {Gelbord}, J.~M. and {Gonzalez}, L. and {Gorjian}, V. and
         {Grier}, C.~J. and {Grupe}, D. and {Hall}, P.~B. and {Hicks}, S. and
         {Horenstein}, D. and {Hutchison}, T. and {Im}, M. and {Jensen}, J.~J. and
         {Jones}, J. and {Kaastra}, J. and {Kelly}, B.~C. and {Kennea}, J.~A. and
         {Kim}, S.~C. and {Korista}, K.~T. and {Kriss}, G.~A. and {Lee}, J.~C. and
         {Lira}, P. and {MacInnis}, F. and {Manne-Nicholas}, E.~R. and
         {Mathur}, S. and {McHardy}, I.~M. and {Montouri}, C. and {Musso}, R. and
         {Nazarov}, S.~V. and {Norris}, R.~P. and {Nousek}, J.~A. and
         {Okhmat}, D.~N. and {Pancoast}, A. and {Papadakis}, I. and
         {Parks}, J.~R. and {Pei}, L. and {Pogge}, R.~W. and {Pott}, J. -U. and
         {Rafter}, S.~E. and {Rix}, H. -W. and {Saylor}, D.~A. and
         {Schimoia}, J.~S. and {Siegel}, M. and {Spencer}, M. and {Starkey}, D. and
         {Sung}, H. -I. and {Teems}, K.~G. and {Treu}, T. and {Turner}, C.~S. and
         {Uttley}, P. and {Villforth}, C. and {Weiss}, Y. and {Woo}, J. -H. and
         {Yan}, H. and {Young}, S.},
        title = "{Space Telescope and Optical Reverberation Mapping Project. III. Optical Continuum Emission and Broadband Time Delays in NGC 5548}",
      journal = {\apj},
         year = "2016",
        month = "Apr",
       volume = {821},
       number = {1},
          eid = {56},
        pages = {56},
          doi = {10.3847/0004-637X/821/1/56},
archivePrefix = {arXiv},
       eprint = {1510.05648},
 primaryClass = {astro-ph.GA},
       adsurl = {https://ui.adsabs.harvard.edu/abs/2016ApJ...821...56F}
}

@ARTICLE{Gardner:2014,
       author = {{Gardner}, Emma and {Done}, Chris},
        title = "{A physical model for the X-ray time lags of narrow-line Seyfert type 1 active galactic nuclei}",
      journal = {\mnras},
         year = "2014",
        month = "Aug",
       volume = {442},
       number = {3},
        pages = {2456-2473},
          doi = {10.1093/mnras/stu1026},
archivePrefix = {arXiv},
       eprint = {1403.2929},
 primaryClass = {astro-ph.HE},
       adsurl = {https://ui.adsabs.harvard.edu/abs/2014MNRAS.442.2456G}
}

@ARTICLE{Korista:2019,
       author = {{Korista}, K.~T. and {Goad}, M.~R.},
        title = "{Quantifying the impact of variable BLR diffuse continuum contributions on measured continuum interband delays}",
      journal = {\mnras},
         year = "2019",
        month = "Nov",
       volume = {489},
       number = {4},
        pages = {5284-5300},
          doi = {10.1093/mnras/stz2330},
archivePrefix = {arXiv},
       eprint = {1908.07757},
 primaryClass = {astro-ph.GA},
       adsurl = {https://ui.adsabs.harvard.edu/abs/2019MNRAS.489.5284K}
}

@ARTICLE{Hernandez2020,
       author = {{Hern{\'a}ndez Santisteban}, J.~V. and {Edelson}, R. and {Horne}, K. and {Gelbord}, J.~M. and {Barth}, A.~J. and {Cackett}, E.~M. and {Goad}, M.~R. and {Netzer}, H. and {Starkey}, D. and {Uttley}, P. and {Brandt}, W.~N. and {Korista}, K. and {Lohfink}, A.~M. and {Onken}, C.~A. and {Page}, K.~L. and {Siegel}, M. and {Vestergaard}, M. and {Bisogni}, S. and {Breeveld}, A.~A. and {Cenko}, S.~B. and {Dalla Bont{\`a}}, E. and {Evans}, P.~A. and {Ferland}, G. and {Gonzalez-Buitrago}, D.~H. and {Grupe}, D. and {Joner}, M.~D. and {Kriss}, G. and {LaPorte}, S.~J. and {Mathur}, S. and {Marshall}, F. and {Mehdipour}, M. and {Mudd}, D. and {Peterson}, B.~M. and {Schmidt}, T. and {Vaughan}, S. and {Valenti}, S.},
        title = "{Intensive disc-reverberation mapping of Fairall 9: first year of Swift and LCO monitoring}",
      journal = {\mnras},
         year = 2020,
        month = nov,
       volume = {498},
       number = {4},
        pages = {5399-5416},
          doi = {10.1093/mnras/staa2365},
archivePrefix = {arXiv},
       eprint = {2008.02134},
 primaryClass = {astro-ph.GA},
       adsurl = {https://ui.adsabs.harvard.edu/abs/2020MNRAS.498.5399H}
}

@ARTICLE{Kara2021,
       author = {{Kara}, Erin and {Mehdipour}, Missagh and {Kriss}, Gerard A. and {Cackett}, Edward M. and {Arav}, Nahum and {Barth}, Aaron J. and {Byun}, Doyee and {Brotherton}, Michael S. and {De Rosa}, Gisella and {Gelbord}, Jonathan and {Hern{\'a}ndez Santisteban}, Juan V. and {Hu}, Chen and {Kaastra}, Jelle and {Landt}, Hermine and {Li}, Yan-Rong and {Miller}, Jake A. and {Montano}, John and {Partington}, Ethan and {Aceituno}, Jes{\'u}s and {Bai}, Jin-Ming and {Bao}, Dongwei and {Bentz}, Misty C. and {Brink}, Thomas G. and {Chelouche}, Doron and {Chen}, Yong-Jie and {Colmenero}, Encarni Romero and {Dalla Bont{\`a}}, Elena and {Dehghanian}, Maryam and {Du}, Pu and {Edelson}, Rick and {Ferland}, Gary J. and {Ferrarese}, Laura and {Fian}, Carina and {Filippenko}, Alexei V. and {Fischer}, Travis and {Goad}, Michael R. and {Gonz{\'a}lez Buitrago}, Diego H. and {Gorjian}, Varoujan and {Grier}, Catherine J. and {Guo}, Wei-Jian and {Hall}, Patrick B. and {Ho}, Luis C. and {Homayouni}, Y. and {Horne}, Keith and {Ili{\'c}}, Dragana and {Jiang}, Bo-Wei and {Joner}, Michael D. and {Kaspi}, Shai and {Kochanek}, Christopher S. and {Korista}, Kirk T. and {Kynoch}, Daniel and {Li}, Sha-Sha and {Liu}, Jun-Rong and {McHardy}, Ian M. and {McLane}, Jacob N. and {Mitchell}, Jake A.~J. and {Netzer}, Hagai and {Olson}, Kianna A. and {Pogge}, Richard W. and {Popovi{\'c}}, Luka {\v{C}}. and {Proga}, Daniel and {Storchi-Bergmann}, Thaisa and {Strasburger}, Erika and {Treu}, Tommaso and {Vestergaard}, Marianne and {Wang}, Jian-Min and {Ward}, Martin J. and {Waters}, Tim and {Williams}, Peter R. and {Yang}, Sen and {Yao}, Zhu-Heng and {Zastrocky}, Theodora E. and {Zhai}, Shuo and {Zu}, Ying},
        title = "{AGN STORM 2. I. First results: A Change in the Weather of Mrk 817}",
      journal = {\apj},
         year = 2021,
        month = dec,
       volume = {922},
       number = {2},
          eid = {151},
        pages = {151},
          doi = {10.3847/1538-4357/ac2159},
archivePrefix = {arXiv},
       eprint = {2105.05840},
 primaryClass = {astro-ph.HE},
       adsurl = {https://ui.adsabs.harvard.edu/abs/2021ApJ...922..151K}
}

@ARTICLE{Vincentelli2021,
       author = {{Vincentelli}, F.~M. and {McHardy}, I. and {Cackett}, E.~M. and {Barth}, A.~J. and {Horne}, K. and {Goad}, M. and {Korista}, K. and {Gelbord}, J. and {Brandt}, W. and {Edelson}, R. and {Miller}, J.~A. and {Pahari}, M. and {Peterson}, B.~M. and {Schmidt}, T. and {Baldi}, R.~D. and {Breedt}, E. and {Hern{\'a}ndez Santisteban}, J.~V. and {Romero-Colmenero}, E. and {Ward}, M. and {Williams}, D.~R.~A.},
        title = "{On the multiwavelength variability of Mrk 110: two components acting at different time-scales}",
      journal = {\mnras},
         year = 2021,
        month = jul,
       volume = {504},
       number = {3},
        pages = {4337-4353},
          doi = {10.1093/mnras/stab1033},
archivePrefix = {arXiv},
       eprint = {2104.04530},
 primaryClass = {astro-ph.HE},
       adsurl = {https://ui.adsabs.harvard.edu/abs/2021MNRAS.504.4337V}
}

@ARTICLE{Netzer2020,
       author = {{Netzer}, Hagai},
        title = "{Testing broad-line region models with reverberation mapping}",
      journal = {\mnras},
         year = 2020,
        month = may,
       volume = {494},
       number = {2},
        pages = {1611-1621},
          doi = {10.1093/mnras/staa767},
archivePrefix = {arXiv},
       eprint = {2003.07660},
 primaryClass = {astro-ph.GA},
       adsurl = {https://ui.adsabs.harvard.edu/abs/2020MNRAS.494.1611N}
}

@ARTICLE{Netzer2022,
       author = {{Netzer}, Hagai},
        title = "{Continuum reverberation mapping and a new lag-luminosity relationship for AGN}",
      journal = {\mnras},
         year = 2022,
        month = jan,
       volume = {509},
       number = {2},
        pages = {2637-2646},
          doi = {10.1093/mnras/stab3133},
archivePrefix = {arXiv},
       eprint = {2110.05512},
 primaryClass = {astro-ph.GA},
       adsurl = {https://ui.adsabs.harvard.edu/abs/2022MNRAS.509.2637N}
}

@ARTICLE{Cackett2022,
       author = {{Cackett}, Edward M. and {Zoghbi}, Abderahmen and {Ulrich}, Otho},
        title = "{Frequency-resolved Lags in UV/Optical Continuum Reverberation Mapping}",
      journal = {\apj},
         year = 2022,
        month = jan,
       volume = {925},
       number = {1},
          eid = {29},
        pages = {29},
          doi = {10.3847/1538-4357/ac3913},
archivePrefix = {arXiv},
       eprint = {2109.02155},
 primaryClass = {astro-ph.GA},
       adsurl = {https://ui.adsabs.harvard.edu/abs/2022ApJ...925...29C}
}

@ARTICLE{Shakura1973,
       author = {{Shakura}, N.~I. and {Sunyaev}, R.~A.},
        title = "{Black holes in binary systems. Observational appearance.}",
      journal = {\aap},
         year = 1973,
        month = jan,
       volume = {24},
        pages = {337-355},
       adsurl = {https://ui.adsabs.harvard.edu/abs/1973A&A....24..337S}
}

@ARTICLE{Haardt1991,
       author = {{Haardt}, F. and {Maraschi}, L.},
        title = "{A Two-Phase Model for the X-Ray Emission from Seyfert Galaxies}",
      journal = {\apjl},
         year = 1991,
        month = oct,
       volume = {380},
        pages = {L51},
          doi = {10.1086/186171},
       adsurl = {https://ui.adsabs.harvard.edu/abs/1991ApJ...380L..51H}
}

@ARTICLE{Edelson2015,
       author = {{Edelson}, R. and {Gelbord}, J.~M. and {Horne}, K. and {McHardy}, I.~M. and {Peterson}, B.~M. and {Ar{\'e}valo}, P. and {Breeveld}, A.~A. and {De Rosa}, G. and {Evans}, P.~A. and {Goad}, M.~R. and {Kriss}, G.~A. and {Brandt}, W.~N. and {Gehrels}, N. and {Grupe}, D. and {Kennea}, J.~A. and {Kochanek}, C.~S. and {Nousek}, J.~A. and {Papadakis}, I. and {Siegel}, M. and {Starkey}, D. and {Uttley}, P. and {Vaughan}, S. and {Young}, S. and {Barth}, A.~J. and {Bentz}, M.~C. and {Brewer}, B.~J. and {Crenshaw}, D.~M. and {Dalla Bont{\`a}}, E. and {De Lorenzo-C{\'a}ceres}, A. and {Denney}, K.~D. and {Dietrich}, M. and {Ely}, J. and {Fausnaugh}, M.~M. and {Grier}, C.~J. and {Hall}, P.~B. and {Kaastra}, J. and {Kelly}, B.~C. and {Korista}, K.~T. and {Lira}, P. and {Mathur}, S. and {Netzer}, H. and {Pancoast}, A. and {Pei}, L. and {Pogge}, R.~W. and {Schimoia}, J.~S. and {Treu}, T. and {Vestergaard}, M. and {Villforth}, C. and {Yan}, H. and {Zu}, Y.},
        title = "{Space Telescope and Optical Reverberation Mapping Project. II. Swift and HST Reverberation Mapping of the Accretion Disk of NGC 5548}",
      journal = {\apj},
         year = 2015,
        month = jun,
       volume = {806},
       number = {1},
          eid = {129},
        pages = {129},
          doi = {10.1088/0004-637X/806/1/129},
archivePrefix = {arXiv},
       eprint = {1501.05951},
 primaryClass = {astro-ph.GA},
       adsurl = {https://ui.adsabs.harvard.edu/abs/2015ApJ...806..129E}
}

@ARTICLE{Korista2001,
       author = {{Korista}, Kirk T. and {Goad}, Michael R.},
        title = "{The Variable Diffuse Continuum Emission of Broad-Line Clouds}",
      journal = {\apj},
         year = 2001,
        month = jun,
       volume = {553},
       number = {2},
        pages = {695-708},
          doi = {10.1086/320964},
archivePrefix = {arXiv},
       eprint = {astro-ph/0101117},
 primaryClass = {astro-ph},
       adsurl = {https://ui.adsabs.harvard.edu/abs/2001ApJ...553..695K}
}

@ARTICLE{Kormendy2013,
       author = {{Kormendy}, John and {Ho}, Luis C.},
        title = "{Coevolution (Or Not) of Supermassive Black Holes and Host Galaxies}",
      journal = {\araa},
         year = 2013,
        month = aug,
       volume = {51},
       number = {1},
        pages = {511-653},
          doi = {10.1146/annurev-astro-082708-101811},
archivePrefix = {arXiv},
       eprint = {1304.7762},
 primaryClass = {astro-ph.CO},
       adsurl = {https://ui.adsabs.harvard.edu/abs/2013ARA&A..51..511K}
}

@ARTICLE{Peterson2004,
       author = {{Peterson}, B.~M. and {Ferrarese}, L. and {Gilbert}, K.~M. and {Kaspi}, S. and {Malkan}, M.~A. and {Maoz}, D. and {Merritt}, D. and {Netzer}, H. and {Onken}, C.~A. and {Pogge}, R.~W. and {Vestergaard}, M. and {Wandel}, A.},
        title = "{Central Masses and Broad-Line Region Sizes of Active Galactic Nuclei. II. A Homogeneous Analysis of a Large Reverberation-Mapping Database}",
      journal = {\apj},
         year = 2004,
        month = oct,
       volume = {613},
       number = {2},
        pages = {682-699},
          doi = {10.1086/423269},
archivePrefix = {arXiv},
       eprint = {astro-ph/0407299},
 primaryClass = {astro-ph},
       adsurl = {https://ui.adsabs.harvard.edu/abs/2004ApJ...613..682P}
}

@ARTICLE{Morgan2010,
       author = {{Morgan}, Christopher W. and {Kochanek}, C.~S. and {Morgan}, Nicholas D. and {Falco}, Emilio E.},
        title = "{The Quasar Accretion Disk Size-Black Hole Mass Relation}",
      journal = {\apj},
         year = 2010,
        month = apr,
       volume = {712},
       number = {2},
        pages = {1129-1136},
          doi = {10.1088/0004-637X/712/2/1129},
archivePrefix = {arXiv},
       eprint = {1002.4160},
 primaryClass = {astro-ph.CO},
       adsurl = {https://ui.adsabs.harvard.edu/abs/2010ApJ...712.1129M}
}

@ARTICLE{Kammoun2021,
       author = {{Kammoun}, E.~S. and {Dov{\v{c}}iak}, M. and {Papadakis}, I.~E. and {Caballero-Garc{\'\i}a}, M.~D. and {Karas}, V.},
        title = "{UV/Optical Disk Thermal Reverberation in Active Galactic Nuclei: An In-depth Study with an Analytic Prescription for Time-lag Spectra}",
      journal = {\apj},
         year = 2021,
        month = jan,
       volume = {907},
       number = {1},
          eid = {20},
        pages = {20},
          doi = {10.3847/1538-4357/abcb93},
archivePrefix = {arXiv},
       eprint = {2011.08563},
 primaryClass = {astro-ph.HE},
       adsurl = {https://ui.adsabs.harvard.edu/abs/2021ApJ...907...20K}
}

@ARTICLE{Donnan2021,
       author = {{Donnan}, Fergus R. and {Horne}, Keith and {Hern{\'a}ndez Santisteban}, Juan V.},
        title = "{Bayesian analysis of quasar light curves with a running optimal average: new time delay measurements of COSMOGRAIL gravitationally lensed quasars}",
      journal = {\mnras},
         year = 2021,
        month = dec,
       volume = {508},
       number = {4},
        pages = {5449-5467},
          doi = {10.1093/mnras/stab2832},
archivePrefix = {arXiv},
       eprint = {2107.12318},
 primaryClass = {astro-ph.IM},
       adsurl = {https://ui.adsabs.harvard.edu/abs/2021MNRAS.508.5449D}
}

@ARTICLE{Netzer2015,
   author = {{Netzer}, H.},
    title = "{Revisiting the Unified Model of Active Galactic Nuclei}",
  journal = {\araa},
archivePrefix = "arXiv",
   eprint = {1505.00811},
     year = 2015,
    month = aug,
   volume = 53,
    pages = {365-408},
      doi = {10.1146/annurev-astro-082214-122302},
   adsurl = {http://adsabs.harvard.edu/abs/2015ARA\&A..53..365N}
}

@ARTICLE{Ivezic2019,
       author = {{Ivezi{\'c}}, {\v{Z}}eljko and {Kahn}, Steven M. and {Tyson}, J. Anthony and {Abel}, Bob and {Acosta}, Emily and {Allsman}, Robyn and {Alonso}, David and {AlSayyad}, Yusra and {Anderson}, Scott F. and {Andrew}, John and {Angel}, James Roger P. and {Angeli}, George Z. and {Ansari}, Reza and {Antilogus}, Pierre and {Araujo}, Constanza and {Armstrong}, Robert and {Arndt}, Kirk T. and {Astier}, Pierre and {Aubourg}, {\'E}ric and {Auza}, Nicole and {Axelrod}, Tim S. and {Bard}, Deborah J. and {Barr}, Jeff D. and {Barrau}, Aurelian and {Bartlett}, James G. and {Bauer}, Amanda E. and {Bauman}, Brian J. and {Baumont}, Sylvain and {Bechtol}, Ellen and {Bechtol}, Keith and {Becker}, Andrew C. and {Becla}, Jacek and {Beldica}, Cristina and {Bellavia}, Steve and {Bianco}, Federica B. and {Biswas}, Rahul and {Blanc}, Guillaume and {Blazek}, Jonathan and {Blandford}, Roger D. and {Bloom}, Josh S. and {Bogart}, Joanne and {Bond}, Tim W. and {Booth}, Michael T. and {Borgland}, Anders W. and {Borne}, Kirk and {Bosch}, James F. and {Boutigny}, Dominique and {Brackett}, Craig A. and {Bradshaw}, Andrew and {Brandt}, William Nielsen and {Brown}, Michael E. and {Bullock}, James S. and {Burchat}, Patricia and {Burke}, David L. and {Cagnoli}, Gianpietro and {Calabrese}, Daniel and {Callahan}, Shawn and {Callen}, Alice L. and {Carlin}, Jeffrey L. and {Carlson}, Erin L. and {Chandrasekharan}, Srinivasan and {Charles-Emerson}, Glenaver and {Chesley}, Steve and {Cheu}, Elliott C. and {Chiang}, Hsin-Fang and {Chiang}, James and {Chirino}, Carol and {Chow}, Derek and {Ciardi}, David R. and {Claver}, Charles F. and {Cohen-Tanugi}, Johann and {Cockrum}, Joseph J. and {Coles}, Rebecca and {Connolly}, Andrew J. and {Cook}, Kem H. and {Cooray}, Asantha and {Covey}, Kevin R. and {Cribbs}, Chris and {Cui}, Wei and {Cutri}, Roc and {Daly}, Philip N. and {Daniel}, Scott F. and {Daruich}, Felipe and {Daubard}, Guillaume and {Daues}, Greg and {Dawson}, William and {Delgado}, Francisco and {Dellapenna}, Alfred and {de Peyster}, Robert and {de Val-Borro}, Miguel and {Digel}, Seth W. and {Doherty}, Peter and {Dubois}, Richard and {Dubois-Felsmann}, Gregory P. and {Durech}, Josef and {Economou}, Frossie and {Eifler}, Tim and {Eracleous}, Michael and {Emmons}, Benjamin L. and {Fausti Neto}, Angelo and {Ferguson}, Henry and {Figueroa}, Enrique and {Fisher-Levine}, Merlin and {Focke}, Warren and {Foss}, Michael D. and {Frank}, James and {Freemon}, Michael D. and {Gangler}, Emmanuel and {Gawiser}, Eric and {Geary}, John C. and {Gee}, Perry and {Geha}, Marla and {Gessner}, Charles J.~B. and {Gibson}, Robert R. and {Gilmore}, D. Kirk and {Glanzman}, Thomas and {Glick}, William and {Goldina}, Tatiana and {Goldstein}, Daniel A. and {Goodenow}, Iain and {Graham}, Melissa L. and {Gressler}, William J. and {Gris}, Philippe and {Guy}, Leanne P. and {Guyonnet}, Augustin and {Haller}, Gunther and {Harris}, Ron and {Hascall}, Patrick A. and {Haupt}, Justine and {Hernandez}, Fabio and {Herrmann}, Sven and {Hileman}, Edward and {Hoblitt}, Joshua and {Hodgson}, John A. and {Hogan}, Craig and {Howard}, James D. and {Huang}, Dajun and {Huffer}, Michael E. and {Ingraham}, Patrick and {Innes}, Walter R. and {Jacoby}, Suzanne H. and {Jain}, Bhuvnesh and {Jammes}, Fabrice and {Jee}, M. James and {Jenness}, Tim and {Jernigan}, Garrett and {Jevremovi{\'c}}, Darko and {Johns}, Kenneth and {Johnson}, Anthony S. and {Johnson}, Margaret W.~G. and {Jones}, R. Lynne and {Juramy-Gilles}, Claire and {Juri{\'c}}, Mario and {Kalirai}, Jason S. and {Kallivayalil}, Nitya J. and {Kalmbach}, Bryce and {Kantor}, Jeffrey P. and {Karst}, Pierre and {Kasliwal}, Mansi M. and {Kelly}, Heather and {Kessler}, Richard and {Kinnison}, Veronica and {Kirkby}, David and {Knox}, Lloyd and {Kotov}, Ivan V. and {Krabbendam}, Victor L. and {Krughoff}, K. Simon and {Kub{\'a}nek}, Petr and {Kuczewski}, John and {Kulkarni}, Shri and {Ku}, John and {Kurita}, Nadine R. and {Lage}, Craig S. and {Lambert}, Ron and {Lange}, Travis and {Langton}, J. Brian and {Le Guillou}, Laurent and {Levine}, Deborah and {Liang}, Ming and {Lim}, Kian-Tat and {Lintott}, Chris J. and {Long}, Kevin E. and {Lopez}, Margaux and {Lotz}, Paul J. and {Lupton}, Robert H. and {Lust}, Nate B. and {MacArthur}, Lauren A. and {Mahabal}, Ashish and {Mandelbaum}, Rachel and {Markiewicz}, Thomas W. and {Marsh}, Darren S. and {Marshall}, Philip J. and {Marshall}, Stuart and {May}, Morgan and {McKercher}, Robert and {McQueen}, Michelle and {Meyers}, Joshua and {Migliore}, Myriam and {Miller}, Michelle and {Mills}, David J. and {Miraval}, Connor and {Moeyens}, Joachim and {Moolekamp}, Fred E. and {Monet}, David G. and {Moniez}, Marc and {Monkewitz}, Serge and {Montgomery}, Christopher and {Morrison}, Christopher B. and {Mueller}, Fritz and {Muller}, Gary P. and {Mu{\~n}oz Arancibia}, Freddy and {Neill}, Douglas R. and {Newbry}, Scott P. and {Nief}, Jean-Yves and {Nomerotski}, Andrei and {Nordby}, Martin and {O'Connor}, Paul and {Oliver}, John and {Olivier}, Scot S. and {Olsen}, Knut and {O'Mullane}, William and {Ortiz}, Sandra and {Osier}, Shawn and {Owen}, Russell E. and {Pain}, Reynald and {Palecek}, Paul E. and {Parejko}, John K. and {Parsons}, James B. and {Pease}, Nathan M. and {Peterson}, J. Matt and {Peterson}, John R. and {Petravick}, Donald L. and {Libby Petrick}, M.~E. and {Petry}, Cathy E. and {Pierfederici}, Francesco and {Pietrowicz}, Stephen and {Pike}, Rob and {Pinto}, Philip A. and {Plante}, Raymond and {Plate}, Stephen and {Plutchak}, Joel P. and {Price}, Paul A. and {Prouza}, Michael and {Radeka}, Veljko and {Rajagopal}, Jayadev and {Rasmussen}, Andrew P. and {Regnault}, Nicolas and {Reil}, Kevin A. and {Reiss}, David J. and {Reuter}, Michael A. and {Ridgway}, Stephen T. and {Riot}, Vincent J. and {Ritz}, Steve and {Robinson}, Sean and {Roby}, William and {Roodman}, Aaron and {Rosing}, Wayne and {Roucelle}, Cecille and {Rumore}, Matthew R. and {Russo}, Stefano and {Saha}, Abhijit and {Sassolas}, Benoit and {Schalk}, Terry L. and {Schellart}, Pim and {Schindler}, Rafe H. and {Schmidt}, Samuel and {Schneider}, Donald P. and {Schneider}, Michael D. and {Schoening}, William and {Schumacher}, German and {Schwamb}, Megan E. and {Sebag}, Jacques and {Selvy}, Brian and {Sembroski}, Glenn H. and {Seppala}, Lynn G. and {Serio}, Andrew and {Serrano}, Eduardo and {Shaw}, Richard A. and {Shipsey}, Ian and {Sick}, Jonathan and {Silvestri}, Nicole and {Slater}, Colin T. and {Smith}, J. Allyn and {Smith}, R. Chris and {Sobhani}, Shahram and {Soldahl}, Christine and {Storrie-Lombardi}, Lisa and {Stover}, Edward and {Strauss}, Michael A. and {Street}, Rachel A. and {Stubbs}, Christopher W. and {Sullivan}, Ian S. and {Sweeney}, Donald and {Swinbank}, John D. and {Szalay}, Alexander and {Takacs}, Peter and {Tether}, Stephen A. and {Thaler}, Jon J. and {Thayer}, John Gregg and {Thomas}, Sandrine and {Thornton}, Adam J. and {Thukral}, Vaikunth and {Tice}, Jeffrey and {Trilling}, David E. and {Turri}, Max and {Van Berg}, Richard and {Vanden Berk}, Daniel and {Vetter}, Kurt and {Virieux}, Francoise and {Vucina}, Tomislav and {Wahl}, William and {Walkowicz}, Lucianne and {Walsh}, Brian and {Walter}, Christopher W. and {Wang}, Daniel L. and {Wang}, Shin-Yawn and {Warner}, Michael and {Wiecha}, Oliver and {Willman}, Beth and {Winters}, Scott E. and {Wittman}, David and {Wolff}, Sidney C. and {Wood-Vasey}, W. Michael and {Wu}, Xiuqin and {Xin}, Bo and {Yoachim}, Peter and {Zhan}, Hu},
        title = "{LSST: From Science Drivers to Reference Design and Anticipated Data Products}",
      journal = {\apj},
         year = 2019,
        month = mar,
       volume = {873},
       number = {2},
          eid = {111},
        pages = {111},
          doi = {10.3847/1538-4357/ab042c},
archivePrefix = {arXiv},
       eprint = {0805.2366},
 primaryClass = {astro-ph},
       adsurl = {https://ui.adsabs.harvard.edu/abs/2019ApJ...873..111I}
}

@ARTICLE{Sun2020,
       author = {{Sun}, Mouyuan and {Xue}, Yongquan and {Brandt}, W.~N. and {Gu}, Wei-Min and {Trump}, Jonathan R. and {Cai}, Zhenyi and {He}, Zhicheng and {Lin}, Da-bin and {Liu}, Tong and {Wang}, Junxian},
        title = "{Corona-heated Accretion-disk Reprocessing: A Physical Model to Decipher the Melody of AGN UV/Optical Twinkling}",
      journal = {\apj},
         year = 2020,
        month = mar,
       volume = {891},
       number = {2},
          eid = {178},
        pages = {178},
          doi = {10.3847/1538-4357/ab789e},
archivePrefix = {arXiv},
       eprint = {2002.08564},
 primaryClass = {astro-ph.HE},
       adsurl = {https://ui.adsabs.harvard.edu/abs/2020ApJ...891..178S}
}

@ARTICLE{Starkey2017,
       author = {{Starkey}, D. and {Horne}, Keith and {Fausnaugh}, M.~M. and {Peterson}, B.~M. and {Bentz}, M.~C. and {Kochanek}, C.~S. and {Denney}, K.~D. and {Edelson}, R. and {Goad}, M.~R. and {De Rosa}, G. and {Anderson}, M.~D. and {Ar{\'e}valo}, P. and {Barth}, A.~J. and {Bazhaw}, C. and {Borman}, G.~A. and {Boroson}, T.~A. and {Bottorff}, M.~C. and {Brandt}, W.~N. and {Breeveld}, A.~A. and {Cackett}, E.~M. and {Carini}, M.~T. and {Croxall}, K.~V. and {Crenshaw}, D.~M. and {Dalla Bont{\`a}}, E. and {De Lorenzo-C{\'a}ceres}, A. and {Dietrich}, M. and {Efimova}, N.~V. and {Ely}, J. and {Evans}, P.~A. and {Filippenko}, A.~V. and {Flatland}, K. and {Gehrels}, N. and {Geier}, S. and {Gelbord}, J.~M. and {Gonzalez}, L. and {Gorjian}, V. and {Grier}, C.~J. and {Grupe}, D. and {Hall}, P.~B. and {Hicks}, S. and {Horenstein}, D. and {Hutchison}, T. and {Im}, M. and {Jensen}, J.~J. and {Joner}, M.~D. and {Jones}, J. and {Kaastra}, J. and {Kaspi}, S. and {Kelly}, B.~C. and {Kennea}, J.~A. and {Kim}, S.~C. and {Kim}, M. and {Klimanov}, S.~A. and {Korista}, K.~T. and {Kriss}, G.~A. and {Lee}, J.~C. and {Leonard}, D.~C. and {Lira}, P. and {MacInnis}, F. and {Manne-Nicholas}, E.~R. and {Mathur}, S. and {McHardy}, I.~M. and {Montouri}, C. and {Musso}, R. and {Nazarov}, S.~V. and {Norris}, R.~P. and {Nousek}, J.~A. and {Okhmat}, D.~N. and {Pancoast}, A. and {Parks}, J.~R. and {Pei}, L. and {Pogge}, R.~W. and {Pott}, J. -U. and {Rafter}, S.~E. and {Rix}, H. -W. and {Saylor}, D.~A. and {Schimoia}, J.~S. and {Schn{\"u}lle}, K. and {Sergeev}, S.~G. and {Siegel}, M.~H. and {Spencer}, M. and {Sung}, H. -I. and {Teems}, K.~G. and {Turner}, C.~S. and {Uttley}, P. and {Vestergaard}, M. and {Villforth}, C. and {Weiss}, Y. and {Woo}, J. -H. and {Yan}, H. and {Young}, S. and {Zheng}, W. and {Zu}, Y.},
        title = "{Space Telescope and Optical Reverberation Mapping Project.VI. Reverberating Disk Models for NGC 5548}",
      journal = {\apj},
         year = 2017,
        month = jan,
       volume = {835},
       number = {1},
          eid = {65},
        pages = {65},
          doi = {10.3847/1538-4357/835/1/65},
archivePrefix = {arXiv},
       eprint = {1611.06051},
 primaryClass = {astro-ph.GA},
       adsurl = {https://ui.adsabs.harvard.edu/abs/2017ApJ...835...65S}
}

@ARTICLE{Starkey2023,
       author = {{Starkey}, D.~A. and {Huang}, Jiamu and {Horne}, Keith and {Lin}, Douglas N.~C.},
        title = "{Rimmed and rippled accretion disc models to explain AGN continuum lags}",
      journal = {\mnras},
         year = 2023,
        month = feb,
       volume = {519},
       number = {2},
        pages = {2754-2768},
          doi = {10.1093/mnras/stac3579},
archivePrefix = {arXiv},
       eprint = {2212.01379},
 primaryClass = {astro-ph.GA},
       adsurl = {https://ui.adsabs.harvard.edu/abs/2023MNRAS.519.2754S}
}

@ARTICLE{Donnan2023,
       author = {{Donnan}, Fergus R. and {Hern{\'a}ndez Santisteban}, Juan V. and {Horne}, Keith and {Hu}, Chen and {Du}, Pu and {Li}, Yan-Rong and {Xiao}, Ming and {Ho}, Luis C. and {Aceituno}, Jes{\'u}s and {Wang}, Jian-Min and {Guo}, Wei-Jian and {Yang}, Sen and {Jiang}, Bo-Wei and {Yao}, Zhu-Heng},
        title = "{Testing Super-Eddington Accretion onto a Supermassive Black Hole: Reverberation Mapping of PG 1119+120}",
      journal = {arXiv e-prints},
         year = 2023,
        month = feb,
          eid = {arXiv:2302.09370},
        pages = {arXiv:2302.09370},
          doi = {10.48550/arXiv.2302.09370},
archivePrefix = {arXiv},
       eprint = {2302.09370},
 primaryClass = {astro-ph.GA},
       adsurl = {https://ui.adsabs.harvard.edu/abs/2023arXiv230209370D}
}

@ARTICLE{Homayouni2019,
       author = {{Homayouni}, Y. and {Trump}, Jonathan R. and {Grier}, C.~J. and {Shen}, Yue and {Starkey}, D.~A. and {Brandt}, W.~N. and {Fonseca Alvarez}, G. and {Hall}, P.~B. and {Horne}, Keith and {Kinemuchi}, Karen and {I-Hsiu Li}, Jennifer and {McGreer}, Ian D. and {Sun}, Mouyuan and {Ho}, L.~C. and {Schneider}, D.~P.},
        title = "{The Sloan Digital Sky Survey Reverberation Mapping Project: Accretion Disk Sizes from Continuum Lags}",
      journal = {\apj},
         year = 2019,
        month = aug,
       volume = {880},
       number = {2},
          eid = {126},
        pages = {126},
          doi = {10.3847/1538-4357/ab2638},
archivePrefix = {arXiv},
       eprint = {1806.08360},
 primaryClass = {astro-ph.GA},
       adsurl = {https://ui.adsabs.harvard.edu/abs/2019ApJ...880..126H}
}

@ARTICLE{Kara2023,
       author = {{Kara}, Erin and {Barth}, Aaron J. and {Cackett}, Edward M. and {Gelbord}, Jonathan and {Montano}, John and {Li}, Yan-Rong and {Santana}, Lisabeth and {Horne}, Keith and {Alston}, William N. and {Buisson}, Douglas and {Chelouche}, Doron and {Du}, Pu and {Fabian}, Andrew C. and {Fian}, Carina and {Gallo}, Luigi and {Goad}, Michael R. and {Grupe}, Dirk and {Gonzalez Buitrago}, Diego H. and {Hernandez Santisteban}, Juan V. and {Kaspi}, Shai and {Hu}, Chen and {Komossa}, S. and {Kriss}, Gerard A. and {Lewin}, Collin and {Lewis}, Tiffany and {Loewenstein}, Michael and {Lohfink}, Anne and {Masterson}, Megan and {McHardy}, Ian M. and {Mehdipour}, Missagh and {Miller}, Jake and {Panagiotou}, Christos and {Parker}, Michael L. and {Pinto}, Ciro and {Remillard}, Ron and {Reynolds}, Christopher and {Rogantini}, Daniele and {Wang}, Jian-Min and {Wang}, Jingyi and {Wilkins}, Dan},
        title = "{UV/Optical disk reverberation lags despite a faint X-ray corona in the AGN Mrk 335}",
      journal = {arXiv e-prints},
         year = 2023,
        month = feb,
          eid = {arXiv:2302.07342},
        pages = {arXiv:2302.07342},
          doi = {10.48550/arXiv.2302.07342},
archivePrefix = {arXiv},
       eprint = {2302.07342},
 primaryClass = {astro-ph.HE},
       adsurl = {https://ui.adsabs.harvard.edu/abs/2023arXiv230207342K}
}

@MISC{pyccf2018,
       author = {{Sun}, Mouyuan and {Grier}, C.~J. and {Peterson}, B.~M.},
        title = "{PyCCF: Python Cross Correlation Function for reverberation mapping studies}",
 howpublished = {Astrophysics Source Code Library, record ascl:1805.032},
         year = 2018,
        month = may,
          eid = {ascl:1805.032},
        pages = {ascl:1805.032},
archivePrefix = {ascl},
       eprint = {1805.032},
       adsurl = {https://ui.adsabs.harvard.edu/abs/2018ascl.soft05032S}
}

@ARTICLE{Fabian2012,
       author = {{Fabian}, A.~C.},
        title = "{Observational Evidence of Active Galactic Nuclei Feedback}",
      journal = {\araa},
         year = 2012,
        month = sep,
       volume = {50},
        pages = {455-489},
          doi = {10.1146/annurev-astro-081811-125521},
archivePrefix = {arXiv},
       eprint = {1204.4114},
 primaryClass = {astro-ph.CO},
       adsurl = {https://ui.adsabs.harvard.edu/abs/2012ARA&A..50..455F}
}

@ARTICLE{Hagen2023b,
       author = {{Hagen}, Scott and {Done}, Chris},
        title = "{Estimating black hole spin from AGN SED fitting: the impact of general-relativistic ray tracing}",
      journal = {\mnras},
         year = 2023,
        month = nov,
       volume = {525},
       number = {3},
        pages = {3455-3467},
          doi = {10.1093/mnras/stad2499},
archivePrefix = {arXiv},
       eprint = {2304.01253},
 primaryClass = {astro-ph.HE},
       adsurl = {https://ui.adsabs.harvard.edu/abs/2023MNRAS.525.3455H}
}

@ARTICLE{MacLeod2010,
       author = {{MacLeod}, C.~L. and {Ivezi{\'c}}, {\v{Z}}. and {Kochanek}, C.~S. and {Koz{\l}owski}, S. and {Kelly}, B. and {Bullock}, E. and {Kimball}, A. and {Sesar}, B. and {Westman}, D. and {Brooks}, K. and {Gibson}, R. and {Becker}, A.~C. and {de Vries}, W.~H.},
        title = "{Modeling the Time Variability of SDSS Stripe 82 Quasars as a Damped Random Walk}",
      journal = {\apj},
         year = 2010,
        month = oct,
       volume = {721},
       number = {2},
        pages = {1014-1033},
          doi = {10.1088/0004-637X/721/2/1014},
archivePrefix = {arXiv},
       eprint = {1004.0276},
 primaryClass = {astro-ph.CO},
       adsurl = {https://ui.adsabs.harvard.edu/abs/2010ApJ...721.1014M}
}

@ARTICLE{kammoun2024,
       author = {{Kammoun}, E. and {Papadakis}, I.~E. and {Dov{\v{c}}iak}, M. and {Panagiotou}, C.},
        title = "{Broadband X-ray/UV/optical time-resolved spectroscopy of NGC 5548: The origin of the UV/optical variability in active galactic nuclei}",
      journal = {arXiv e-prints},
         year = 2024,
        month = mar,
          eid = {arXiv:2403.12208},
        pages = {arXiv:2403.12208},
          doi = {10.48550/arXiv.2403.12208},
archivePrefix = {arXiv},
       eprint = {2403.12208},
 primaryClass = {astro-ph.HE},
       adsurl = {https://ui.adsabs.harvard.edu/abs/2024arXiv240312208K}
}

@ARTICLE{Shen2023,
       author = {{Shen}, Yue and {Grier}, Catherine J. and {Horne}, Keith and {Stone}, Zachary and {Li}, Jennifer I. and {Yang}, Qian and {Homayouni}, Yasaman and {Trump}, Jonathan R. and {Anderson}, Scott F. and {Brandt}, W.~N. and {Hall}, Patrick B. and {Ho}, Luis C. and {Jiang}, Linhua and {Petitjean}, Patrick and {Schneider}, Donald P. and {Tao}, Charling and {Donnan}, Fergus. R. and {AlSayyad}, Yusra and {Bershady}, Matthew A. and {Blanton}, Michael R. and {Bizyaev}, Dmitry and {Bundy}, Kevin and {Chen}, Yuguang and {Davis}, Megan C. and {Dawson}, Kyle and {Fan}, Xiaohui and {Greene}, Jenny E. and {Groller}, Hannes and {Guo}, Yucheng and {Ibarra-Medel}, Hector and {Keenan}, Ryan P. and {Kollmeier}, Juna A. and {Lejoly}, Cassandra and {Li}, Zefeng and {de la Macorra}, Axel and {Moe}, Maxwell and {Nie}, Jundan and {Rossi}, Graziano and {Smith}, Paul S. and {Tee}, Wei Leong and {Weijmans}, Anne-Marie and {Xu}, Jiachuan and {Yue}, Minghao and {Zhou}, Xu and {Zhou}, Zhimin and {Zou}, Hu},
        title = "{The Sloan Digital Sky Survey Reverberation Mapping Project: Key Results}",
      journal = {arXiv e-prints},
         year = 2023,
        month = may,
          eid = {arXiv:2305.01014},
        pages = {arXiv:2305.01014},
          doi = {10.48550/arXiv.2305.01014},
archivePrefix = {arXiv},
       eprint = {2305.01014},
 primaryClass = {astro-ph.GA},
       adsurl = {https://ui.adsabs.harvard.edu/abs/2023arXiv230501014S}
}

@ARTICLE{Brandt:2015,
       author = {{Brandt}, W.~N. and {Alexander}, D.~M.},
        title = "{Cosmic X-ray surveys of distant active galaxies. The demographics, physics, and ecology of growing supermassive black holes}",
      journal = {\aapr},
         year = "2015",
        month = "Jan",
       volume = {23},
          eid = {1},
        pages = {1},
          doi = {10.1007/s00159-014-0081-z},
archivePrefix = {arXiv},
       eprint = {1501.01982},
 primaryClass = {astro-ph.HE},
       adsurl = {https://ui.adsabs.harvard.edu/abs/2015A&ARv..23....1B}
}

@ARTICLE{Collier1999,
       author = {{Collier}, Stefan and {Horne}, Keith and {Wanders}, Ignaz and {Peterson}, Bradley M.},
        title = "{A new direct method for measuring the Hubble constant from reverberating accretion discs in active galaxies}",
      journal = {\mnras},
         year = 1999,
        month = jan,
       volume = {302},
       number = {1},
        pages = {L24-L28},
          doi = {10.1046/j.1365-8711.1999.02250.x},
archivePrefix = {arXiv},
       eprint = {astro-ph/9811278},
 primaryClass = {astro-ph},
       adsurl = {https://ui.adsabs.harvard.edu/abs/1999MNRAS.302L..24C}
}

@ARTICLE{Mudd2018,
       author = {{Mudd}, D. and {Martini}, P. and {Zu}, Y. and {Kochanek}, C. and {Peterson}, B.~M. and {Kessler}, R. and {Davis}, T.~M. and {Hoormann}, J.~K. and {King}, A. and {Lidman}, C. and {Sommer}, N.~E. and {Tucker}, B.~E. and {Asorey}, J. and {Hinton}, S. and {Glazebrook}, K. and {Kuehn}, K. and {Lewis}, G. and {Macaulay}, E. and {Moeller}, A. and {O'Neill}, C. and {Zhang}, B. and {Abbott}, T.~M.~C. and {Abdalla}, F.~B. and {Allam}, S. and {Banerji}, M. and {Benoit-L{\'e}vy}, A. and {Bertin}, E. and {Brooks}, D. and {Carnero Rosell}, A. and {Carollo}, D. and {Carrasco Kind}, M. and {Carretero}, J. and {Cunha}, C.~E. and {D'Andrea}, C.~B. and {da Costa}, L.~N. and {Davis}, C. and {Desai}, S. and {Doel}, P. and {Fosalba}, P. and {Garc{\'\i}a-Bellido}, J. and {Gaztanaga}, E. and {Gerdes}, D.~W. and {Gruen}, D. and {Gruendl}, R.~A. and {Gschwend}, J. and {Gutierrez}, G. and {Hartley}, W.~G. and {Honscheid}, K. and {James}, D.~J. and {Kuhlmann}, S. and {Kuropatkin}, N. and {Lima}, M. and {Maia}, M.~A.~G. and {Marshall}, J.~L. and {McMahon}, R.~G. and {Menanteau}, F. and {Miquel}, R. and {Plazas}, A.~A. and {Romer}, A.~K. and {Sanchez}, E. and {Schindler}, R. and {Schubnell}, M. and {Smith}, M. and {Smith}, R.~C. and {Soares-Santos}, M. and {Sobreira}, F. and {Suchyta}, E. and {Swanson}, M.~E.~C. and {Tarle}, G. and {Thomas}, D. and {Tucker}, D.~L. and {Walker}, A.~R. and {DES Collaboration}},
        title = "{Quasar Accretion Disk Sizes from Continuum Reverberation Mapping from the Dark Energy Survey}",
      journal = {\apj},
         year = 2018,
        month = aug,
       volume = {862},
       number = {2},
          eid = {123},
        pages = {123},
          doi = {10.3847/1538-4357/aac9bb},
archivePrefix = {arXiv},
       eprint = {1711.11588},
 primaryClass = {astro-ph.GA},
       adsurl = {https://ui.adsabs.harvard.edu/abs/2018ApJ...862..123M}
}

@ARTICLE{LyndenBell1969,
       author = {{Lynden-Bell}, D.},
        title = "{Galactic Nuclei as Collapsed Old Quasars}",
      journal = {\nat},
         year = "1969",
        month = "Aug",
       volume = {223},
       number = {5207},
        pages = {690-694},
          doi = {10.1038/223690a0},
       adsurl = {https://ui.adsabs.harvard.edu/abs/1969Natur.223..690L}
}

@ARTICLE{Selsing2016,
       author = {{Selsing}, J. and {Fynbo}, J.~P.~U. and {Christensen}, L. and {Krogager}, J. -K.},
        title = "{An X-Shooter composite of bright 1 < z < 2 quasars from UV to infrared}",
      journal = {\aap},
         year = 2016,
        month = jan,
       volume = {585},
          eid = {A87},
        pages = {A87},
          doi = {10.1051/0004-6361/201527096},
archivePrefix = {arXiv},
       eprint = {1510.08058},
 primaryClass = {astro-ph.GA},
       adsurl = {https://ui.adsabs.harvard.edu/abs/2016A&A...585A..87S}
}

@ARTICLE{Schlegel1998,
       author = {{Schlegel}, David J. and {Finkbeiner}, Douglas P. and {Davis}, Marc},
        title = "{Maps of Dust Infrared Emission for Use in Estimation of Reddening and Cosmic Microwave Background Radiation Foregrounds}",
      journal = {\apj},
         year = 1998,
        month = jun,
       volume = {500},
       number = {2},
        pages = {525-553},
          doi = {10.1086/305772},
archivePrefix = {arXiv},
       eprint = {astro-ph/9710327},
 primaryClass = {astro-ph},
       adsurl = {https://ui.adsabs.harvard.edu/abs/1998ApJ...500..525S}
}

@ARTICLE{Green2018,
       author = {{Green}, Gregory M.},
        title = "{dustmaps: A Python interface for maps of interstellar dust}",
      journal = {The Journal of Open Source Software},
         year = 2018,
        month = jun,
       volume = {3},
       number = {26},
        pages = {695},
          doi = {10.21105/joss.00695},
       adsurl = {https://ui.adsabs.harvard.edu/abs/2018JOSS....3..695G}
}

@ARTICLE{Weaver2022,
       author = {{Weaver}, John R. and {Horne}, Keith},
        title = "{Dust and the intrinsic spectral index of quasar variations: hints of finite stress at the innermost stable circular orbit}",
      journal = {\mnras},
         year = 2022,
        month = may,
       volume = {512},
       number = {1},
        pages = {899-916},
          doi = {10.1093/mnras/stac248},
archivePrefix = {arXiv},
       eprint = {2201.11134},
 primaryClass = {astro-ph.GA},
       adsurl = {https://ui.adsabs.harvard.edu/abs/2022MNRAS.512..899W}
}

@ARTICLE{Wang1999,
       author = {{Wang}, Jian-Min and {Szuszkiewicz}, Ewa and {Lu}, Fang-Jun and {Zhou}, You-Yuan},
        title = "{Emergent Spectra from Slim Accretion Disks in Active Galactic Nuclei}",
      journal = {\apj},
         year = 1999,
        month = sep,
       volume = {522},
       number = {2},
        pages = {839-845},
          doi = {10.1086/307686},
       adsurl = {https://ui.adsabs.harvard.edu/abs/1999ApJ...522..839W}
}

@ARTICLE{Kinemuchi2020,
       author = {{Kinemuchi}, K. and {Hall}, Patrick B. and {McGreer}, Ian and {Kochanek}, C.~S. and {Grier}, Catherine J. and {Trump}, Jonathan and {Shen}, Yue and {Brandt}, W.~N. and {Wood-Vasey}, W.~M. and {Fan}, Xiaohui and {Peterson}, Bradley M. and {Schneider}, Donald P. and {Hern{\'a}ndez Santisteban}, Juan V. and {Horne}, Keith and {Chen}, Yuguang and {Eftekharzadeh}, Sarah and {Guo}, Yucheng and {Jia}, Siyao and {Li}, Feng and {Li}, Zefeng and {Nie}, Jundan and {Ponder}, Kara A. and {Rogerson}, Jesse and {Zhang}, Tianmen and {Zou}, Hu and {Jiang}, Linhua and {Ho}, Luis C. and {Kneib}, Jean-Paul and {Petitjean}, Patrick and {Palanque-Delabrouille}, Nathalie and {Yeche}, Christophe},
        title = "{The Sloan Digital Sky Survey Reverberation Mapping Project: Photometric g and i Light Curves}",
      journal = {\apjs},
         year = 2020,
        month = sep,
       volume = {250},
       number = {1},
          eid = {10},
        pages = {10},
          doi = {10.3847/1538-4365/aba43f},
archivePrefix = {arXiv},
       eprint = {2007.05160},
 primaryClass = {astro-ph.GA},
       adsurl = {https://ui.adsabs.harvard.edu/abs/2020ApJS..250...10K}
}

@ARTICLE{Winkler1992,
       author = {{Winkler}, H. and {Glass}, I.~S. and {van Wyk}, F. and {Marang}, F. and {Jones}, J.~H.~S. and {Buckley}, D.~A.~H. and {Sekiguchi}, K.},
        title = "{Variability studies of seyfert galaxies - I. Broad-band optical photometry .}",
      journal = {\mnras},
         year = 1992,
        month = aug,
       volume = {257},
        pages = {659-676},
          doi = {10.1093/mnras/257.4.659},
       adsurl = {https://ui.adsabs.harvard.edu/abs/1992MNRAS.257..659W}
}

@ARTICLE{Winkler1997,
       author = {{Winkler}, H.},
        title = "{The extinction, flux distribution and luminosity of Seyfert 1 nuclei derived from UBV(RI)\_C aperture photometry}",
      journal = {\mnras},
         year = 1997,
        month = dec,
       volume = {292},
       number = {2},
        pages = {273-288},
          doi = {10.1093/mnras/292.2.273},
       adsurl = {https://ui.adsabs.harvard.edu/abs/1997MNRAS.292..273W}
}

@ARTICLE{Planck2020,
       author = {{Planck Collaboration} and {Aghanim}, N. and {Akrami}, Y. and {Ashdown}, M. and {Aumont}, J. and {Baccigalupi}, C. and {Ballardini}, M. and {Banday}, A.~J. and {Barreiro}, R.~B. and {Bartolo}, N. and {Basak}, S. and {Battye}, R. and {Benabed}, K. and {Bernard}, J. -P. and {Bersanelli}, M. and {Bielewicz}, P. and {Bock}, J.~J. and {Bond}, J.~R. and {Borrill}, J. and {Bouchet}, F.~R. and {Boulanger}, F. and {Bucher}, M. and {Burigana}, C. and {Butler}, R.~C. and {Calabrese}, E. and {Cardoso}, J. -F. and {Carron}, J. and {Challinor}, A. and {Chiang}, H.~C. and {Chluba}, J. and {Colombo}, L.~P.~L. and {Combet}, C. and {Contreras}, D. and {Crill}, B.~P. and {Cuttaia}, F. and {de Bernardis}, P. and {de Zotti}, G. and {Delabrouille}, J. and {Delouis}, J. -M. and {Di Valentino}, E. and {Diego}, J.~M. and {Dor{\'e}}, O. and {Douspis}, M. and {Ducout}, A. and {Dupac}, X. and {Dusini}, S. and {Efstathiou}, G. and {Elsner}, F. and {En{\ss}lin}, T.~A. and {Eriksen}, H.~K. and {Fantaye}, Y. and {Farhang}, M. and {Fergusson}, J. and {Fernandez-Cobos}, R. and {Finelli}, F. and {Forastieri}, F. and {Frailis}, M. and {Fraisse}, A.~A. and {Franceschi}, E. and {Frolov}, A. and {Galeotta}, S. and {Galli}, S. and {Ganga}, K. and {G{\'e}nova-Santos}, R.~T. and {Gerbino}, M. and {Ghosh}, T. and {Gonz{\'a}lez-Nuevo}, J. and {G{\'o}rski}, K.~M. and {Gratton}, S. and {Gruppuso}, A. and {Gudmundsson}, J.~E. and {Hamann}, J. and {Handley}, W. and {Hansen}, F.~K. and {Herranz}, D. and {Hildebrandt}, S.~R. and {Hivon}, E. and {Huang}, Z. and {Jaffe}, A.~H. and {Jones}, W.~C. and {Karakci}, A. and {Keih{\"a}nen}, E. and {Keskitalo}, R. and {Kiiveri}, K. and {Kim}, J. and {Kisner}, T.~S. and {Knox}, L. and {Krachmalnicoff}, N. and {Kunz}, M. and {Kurki-Suonio}, H. and {Lagache}, G. and {Lamarre}, J. -M. and {Lasenby}, A. and {Lattanzi}, M. and {Lawrence}, C.~R. and {Le Jeune}, M. and {Lemos}, P. and {Lesgourgues}, J. and {Levrier}, F. and {Lewis}, A. and {Liguori}, M. and {Lilje}, P.~B. and {Lilley}, M. and {Lindholm}, V. and {L{\'o}pez-Caniego}, M. and {Lubin}, P.~M. and {Ma}, Y. -Z. and {Mac{\'\i}as-P{\'e}rez}, J.~F. and {Maggio}, G. and {Maino}, D. and {Mandolesi}, N. and {Mangilli}, A. and {Marcos-Caballero}, A. and {Maris}, M. and {Martin}, P.~G. and {Martinelli}, M. and {Mart{\'\i}nez-Gonz{\'a}lez}, E. and {Matarrese}, S. and {Mauri}, N. and {McEwen}, J.~D. and {Meinhold}, P.~R. and {Melchiorri}, A. and {Mennella}, A. and {Migliaccio}, M. and {Millea}, M. and {Mitra}, S. and {Miville-Desch{\^e}nes}, M. -A. and {Molinari}, D. and {Montier}, L. and {Morgante}, G. and {Moss}, A. and {Natoli}, P. and {N{\o}rgaard-Nielsen}, H.~U. and {Pagano}, L. and {Paoletti}, D. and {Partridge}, B. and {Patanchon}, G. and {Peiris}, H.~V. and {Perrotta}, F. and {Pettorino}, V. and {Piacentini}, F. and {Polastri}, L. and {Polenta}, G. and {Puget}, J. -L. and {Rachen}, J.~P. and {Reinecke}, M. and {Remazeilles}, M. and {Renzi}, A. and {Rocha}, G. and {Rosset}, C. and {Roudier}, G. and {Rubi{\~n}o-Mart{\'\i}n}, J.~A. and {Ruiz-Granados}, B. and {Salvati}, L. and {Sandri}, M. and {Savelainen}, M. and {Scott}, D. and {Shellard}, E.~P.~S. and {Sirignano}, C. and {Sirri}, G. and {Spencer}, L.~D. and {Sunyaev}, R. and {Suur-Uski}, A. -S. and {Tauber}, J.~A. and {Tavagnacco}, D. and {Tenti}, M. and {Toffolatti}, L. and {Tomasi}, M. and {Trombetti}, T. and {Valenziano}, L. and {Valiviita}, J. and {Van Tent}, B. and {Vibert}, L. and {Vielva}, P. and {Villa}, F. and {Vittorio}, N. and {Wandelt}, B.~D. and {Wehus}, I.~K. and {White}, M. and {White}, S.~D.~M. and {Zacchei}, A. and {Zonca}, A.},
        title = "{Planck 2018 results. VI. Cosmological parameters}",
      journal = {\aap},
         year = 2020,
        month = sep,
       volume = {641},
          eid = {A6},
        pages = {A6},
          doi = {10.1051/0004-6361/201833910},
archivePrefix = {arXiv},
       eprint = {1807.06209},
 primaryClass = {astro-ph.CO},
       adsurl = {https://ui.adsabs.harvard.edu/abs/2020A&A...641A...6P}
}

@ARTICLE{Abramowicz1988,
       author = {{Abramowicz}, M.~A. and {Czerny}, B. and {Lasota}, J.~P. and {Szuszkiewicz}, E.},
        title = "{Slim Accretion Disks}",
      journal = {\apj},
         year = 1988,
        month = sep,
       volume = {332},
        pages = {646},
          doi = {10.1086/166683},
       adsurl = {https://ui.adsabs.harvard.edu/abs/1988ApJ...332..646A}
}

@ARTICLE{Yu2020_DES,
       author = {{Yu}, Zhefu and {Martini}, Paul and {Davis}, T.~M. and {Gruendl}, R.~A. and {Hoormann}, J.~K. and {Kochanek}, C.~S. and {Lidman}, C. and {Mudd}, D. and {Peterson}, B.~M. and {Wester}, W. and {Allam}, S. and {Annis}, J. and {Asorey}, J. and {Avila}, S. and {Banerji}, M. and {Bertin}, E. and {Brooks}, D. and {Buckley-Geer}, E. and {Calcino}, J. and {Rosell}, A. Carnero and {Carollo}, D. and {Kind}, M. Carrasco and {Carretero}, J. and {Cunha}, C.~E. and {D'Andrea}, C.~B. and {Costa}, L.~N. da and {De Vicente}, J. and {Desai}, S. and {Diehl}, H.~T. and {Doel}, P. and {Eifler}, T.~F. and {Flaugher}, B. and {Fosalba}, P. and {Frieman}, J. and {Garc{\'\i}a-Bellido}, J. and {Gaztanaga}, E. and {Glazebrook}, K. and {Gruen}, D. and {Gschwend}, J. and {Gutierrez}, G. and {Hartley}, W.~G. and {Hinton}, S.~R. and {Hollowood}, D.~L. and {Honscheid}, K. and {Hoyle}, B. and {James}, D.~J. and {Kim}, A.~G. and {Krause}, E. and {Kuehn}, K. and {Kuropatkin}, N. and {Lewis}, G.~F. and {Lima}, M. and {Macaulay}, E. and {Maia}, M.~A.~G. and {Marshall}, J.~L. and {Menanteau}, F. and {Miquel}, R. and {M{\"o}ller}, A. and {Plazas}, A.~A. and {Romer}, A.~K. and {Sanchez}, E. and {Scarpine}, V. and {Schubnell}, M. and {Serrano}, S. and {Smith}, M. and {Smith}, R.~C. and {Soares-Santos}, M. and {Sobreira}, F. and {Suchyta}, E. and {Swann}, E. and {Swanson}, M.~E.~C. and {Tarle}, G. and {Tucker}, B.~E. and {Tucker}, D.~L. and {Vikram}, V.},
        title = "{Quasar Accretion Disk Sizes from Continuum Reverberation Mapping in the DES Standard-star Fields}",
      journal = {\apjs},
         year = 2020,
        month = jan,
       volume = {246},
       number = {1},
          eid = {16},
        pages = {16},
          doi = {10.3847/1538-4365/ab5e7a},
archivePrefix = {arXiv},
       eprint = {1811.03638},
 primaryClass = {astro-ph.GA},
       adsurl = {https://ui.adsabs.harvard.edu/abs/2020ApJS..246...16Y}
}

@ARTICLE{Roming:2005,
       author = {{Roming}, Peter W.~A. and {Kennedy}, Thomas E. and {Mason}, Keith O. and
         {Nousek}, John A. and {Ahr}, Lindy and {Bingham}, Richard E. and
         {Broos}, Patrick S. and {Carter}, Mary J. and {Hancock}, Barry K. and
         {Huckle}, Howard E. and {Hunsberger}, S.~D. and {Kawakami}, Hajime and
         {Killough}, Ronnie and {Koch}, T. Scott and {McLelland}, Michael K. and
         {Smith}, Kelly and {Smith}, Philip J. and {Soto}, Juan Carlos and
         {Boyd}, Patricia T. and {Breeveld}, Alice A. and {Holland}, Stephen T. and
         {Ivanushkina}, Mariya and {Pryzby}, Michael S. and {Still}, Martin D. and
         {Stock}, Joseph},
        title = "{The Swift Ultra-Violet/Optical Telescope}",
      journal = {\ssr},
         year = 2005,
        month = oct,
       volume = {120},
       number = {3-4},
        pages = {95-142},
          doi = {10.1007/s11214-005-5095-4},
archivePrefix = {arXiv},
       eprint = {astro-ph/0507413},
 primaryClass = {astro-ph},
       adsurl = {https://ui.adsabs.harvard.edu/abs/2005SSRv..120...95R}
}

@ARTICLE{Gehrels:2004,
       author = {{Gehrels}, N. and {Chincarini}, G. and {Giommi}, P. and {Mason}, K.~O. and
         {Nousek}, J.~A. and {Wells}, A.~A. and {White}, N.~E. and
         {Barthelmy}, S.~D. and {Burrows}, D.~N. and {Cominsky}, L.~R. and
         {Hurley}, K.~C. and {Marshall}, F.~E. and {M{\'e}sz{\'a}ros}, P. and
         {Roming}, P.~W.~A. and {Angelini}, L. and {Barbier}, L.~M. and
         {Belloni}, T. and {Campana}, S. and {Caraveo}, P.~A. and
         {Chester}, M.~M. and {Citterio}, O. and {Cline}, T.~L. and
         {Cropper}, M.~S. and {Cummings}, J.~R. and {Dean}, A.~J. and
         {Feigelson}, E.~D. and {Fenimore}, E.~E. and {Frail}, D.~A. and
         {Fruchter}, A.~S. and {Garmire}, G.~P. and {Gendreau}, K. and
         {Ghisellini}, G. and {Greiner}, J. and {Hill}, J.~E. and
         {Hunsberger}, S.~D. and {Krimm}, H.~A. and {Kulkarni}, S.~R. and
         {Kumar}, P. and {Lebrun}, F. and {Lloyd-Ronning}, N.~M. and
         {Markwardt}, C.~B. and {Mattson}, B.~J. and {Mushotzky}, R.~F. and
         {Norris}, J.~P. and {Osborne}, J. and {Paczynski}, B. and
         {Palmer}, D.~M. and {Park}, H. -S. and {Parsons}, A.~M. and {Paul}, J. and
         {Rees}, M.~J. and {Reynolds}, C.~S. and {Rhoads}, J.~E. and
         {Sasseen}, T.~P. and {Schaefer}, B.~E. and {Short}, A.~T. and
         {Smale}, A.~P. and {Smith}, I.~A. and {Stella}, L. and
         {Tagliaferri}, G. and {Takahashi}, T. and {Tashiro}, M. and
         {Townsley}, L.~K. and {Tueller}, J. and {Turner}, M.~J.~L. and
         {Vietri}, M. and {Voges}, W. and {Ward}, M.~J. and {Willingale}, R. and
         {Zerbi}, F.~M. and {Zhang}, W.~W.},
        title = "{The Swift Gamma-Ray Burst Mission}",
      journal = {\apj},
         year = "2004",
        month = "Aug",
       volume = {611},
       number = {2},
        pages = {1005-1020},
          doi = {10.1086/422091},
archivePrefix = {arXiv},
       eprint = {astro-ph/0405233},
 primaryClass = {astro-ph},
       adsurl = {https://ui.adsabs.harvard.edu/abs/2004ApJ...611.1005G}
}

@ARTICLE{Grier17,
       author = {{Grier}, C.~J. and {Trump}, J.~R. and {Shen}, Yue and {Horne}, Keith and
         {Kinemuchi}, Karen and {McGreer}, Ian D. and {Starkey}, D.~A. and {Brand
        t}, W.~N. and {Hall}, P.~B. and {Kochanek}, C.~S. and {Chen}, Yuguang and
         {Denney}, K.~D. and {Greene}, Jenny E. and {Ho}, L.~C. and
         {Homayouni}, Y. and {I-Hsiu Li}, Jennifer and {Pei}, Liuyi and
         {Peterson}, B.~M. and {Petitjean}, P. and {Schneider}, D.~P. and
         {Sun}, Mouyuan and {AlSayyad}, Yusura and {Bizyaev}, Dmitry and
         {Brinkmann}, Jonathan and {Brownstein}, Joel R. and {Bundy}, Kevin and
         {Dawson}, K.~S. and {Eftekharzadeh}, Sarah and {Fernand
        ez-Trincado}, J.~G. and {Gao}, Yang and {Hutchinson}, Timothy A. and
         {Jia}, Siyao and {Jiang}, Linhua and {Oravetz}, Daniel and
         {Pan}, Kaike and {Paris}, Isabelle and {Ponder}, Kara A. and
         {Peters}, Christina and {Rogerson}, Jesse and {Simmons}, Audrey and
         {Smith}, Robyn and {Wang}, Ran},
        title = "{The Sloan Digital Sky Survey Reverberation Mapping Project: H{\ensuremath{\alpha}} and H{\ensuremath{\beta}} Reverberation Measurements from First-year Spectroscopy and Photometry}",
      journal = {\apj},
         year = "2017",
        month = "Dec",
       volume = {851},
       number = {1},
          eid = {21},
        pages = {21},
          doi = {10.3847/1538-4357/aa98dc},
archivePrefix = {arXiv},
       eprint = {1711.03114},
 primaryClass = {astro-ph.GA},
       adsurl = {https://ui.adsabs.harvard.edu/abs/2017ApJ...851...21G}
}

@ARTICLE{Abbott2021DES,
       author = {{Abbott}, T.~M.~C. and {Adam{\'o}w}, M. and {Aguena}, M. and {Allam}, S. and {Amon}, A. and {Annis}, J. and {Avila}, S. and {Bacon}, D. and {Banerji}, M. and {Bechtol}, K. and {Becker}, M.~R. and {Bernstein}, G.~M. and {Bertin}, E. and {Bhargava}, S. and {Bridle}, S.~L. and {Brooks}, D. and {Burke}, D.~L. and {Carnero Rosell}, A. and {Carrasco Kind}, M. and {Carretero}, J. and {Castander}, F.~J. and {Cawthon}, R. and {Chang}, C. and {Choi}, A. and {Conselice}, C. and {Costanzi}, M. and {Crocce}, M. and {da Costa}, L.~N. and {Davis}, T.~M. and {De Vicente}, J. and {DeRose}, J. and {Desai}, S. and {Diehl}, H.~T. and {Dietrich}, J.~P. and {Drlica-Wagner}, A. and {Eckert}, K. and {Elvin-Poole}, J. and {Everett}, S. and {Evrard}, A.~E. and {Ferrero}, I. and {Fert{\'e}}, A. and {Flaugher}, B. and {Fosalba}, P. and {Friedel}, D. and {Frieman}, J. and {Garc{\'\i}a-Bellido}, J. and {Gaztanaga}, E. and {Gelman}, L. and {Gerdes}, D.~W. and {Giannantonio}, T. and {Gill}, M.~S.~S. and {Gruen}, D. and {Gruendl}, R.~A. and {Gschwend}, J. and {Gutierrez}, G. and {Hartley}, W.~G. and {Hinton}, S.~R. and {Hollowood}, D.~L. and {Honscheid}, K. and {Huterer}, D. and {James}, D.~J. and {Jeltema}, T. and {Johnson}, M.~D. and {Kent}, S. and {Kron}, R. and {Kuehn}, K. and {Kuropatkin}, N. and {Lahav}, O. and {Li}, T.~S. and {Lidman}, C. and {Lin}, H. and {MacCrann}, N. and {Maia}, M.~A.~G. and {Manning}, T.~A. and {Maloney}, J.~D. and {March}, M. and {Marshall}, J.~L. and {Martini}, P. and {Melchior}, P. and {Menanteau}, F. and {Miquel}, R. and {Morgan}, R. and {Myles}, J. and {Neilsen}, E. and {Ogando}, R.~L.~C. and {Palmese}, A. and {Paz-Chinch{\'o}n}, F. and {Petravick}, D. and {Pieres}, A. and {Plazas}, A.~A. and {Pond}, C. and {Rodriguez-Monroy}, M. and {Romer}, A.~K. and {Roodman}, A. and {Rykoff}, E.~S. and {Sako}, M. and {Sanchez}, E. and {Santiago}, B. and {Scarpine}, V. and {Serrano}, S. and {Sevilla-Noarbe}, I. and {Smith}, J. Allyn and {Smith}, M. and {Soares-Santos}, M. and {Suchyta}, E. and {Swanson}, M.~E.~C. and {Tarle}, G. and {Thomas}, D. and {To}, C. and {Tremblay}, P.~E. and {Troxel}, M.~A. and {Tucker}, D.~L. and {Turner}, D.~J. and {Varga}, T.~N. and {Walker}, A.~R. and {Wechsler}, R.~H. and {Weller}, J. and {Wester}, W. and {Wilkinson}, R.~D. and {Yanny}, B. and {Zhang}, Y. and {Nikutta}, R. and {Fitzpatrick}, M. and {Jacques}, A. and {Scott}, A. and {Olsen}, K. and {Huang}, L. and {Herrera}, D. and {Juneau}, S. and {Nidever}, D. and {Weaver}, B.~A. and {Adean}, C. and {Correia}, V. and {de Freitas}, M. and {Freitas}, F.~N. and {Singulani}, C. and {Vila-Verde}, G. and {Linea Science Server}},
        title = "{The Dark Energy Survey Data Release 2}",
      journal = {\apjs},
         year = 2021,
        month = aug,
       volume = {255},
       number = {2},
          eid = {20},
        pages = {20},
          doi = {10.3847/1538-4365/ac00b3},
archivePrefix = {arXiv},
       eprint = {2101.05765},
 primaryClass = {astro-ph.IM},
       adsurl = {https://ui.adsabs.harvard.edu/abs/2021ApJS..255...20A}
}

@ARTICLE{Morganson2018DES,
       author = {{Morganson}, E. and {Gruendl}, R.~A. and {Menanteau}, F. and {Carrasco Kind}, M. and {Chen}, Y. -C. and {Daues}, G. and {Drlica-Wagner}, A. and {Friedel}, D.~N. and {Gower}, M. and {Johnson}, M.~W.~G. and {Johnson}, M.~D. and {Kessler}, R. and {Paz-Chinch{\'o}n}, F. and {Petravick}, D. and {Pond}, C. and {Yanny}, B. and {Allam}, S. and {Armstrong}, R. and {Barkhouse}, W. and {Bechtol}, K. and {Benoit-L{\'e}vy}, A. and {Bernstein}, G.~M. and {Bertin}, E. and {Buckley-Geer}, E. and {Covarrubias}, R. and {Desai}, S. and {Diehl}, H.~T. and {Goldstein}, D.~A. and {Gruen}, D. and {Li}, T.~S. and {Lin}, H. and {Marriner}, J. and {Mohr}, J.~J. and {Neilsen}, E. and {Ngeow}, C. -C. and {Paech}, K. and {Rykoff}, E.~S. and {Sako}, M. and {Sevilla-Noarbe}, I. and {Sheldon}, E. and {Sobreira}, F. and {Tucker}, D.~L. and {Wester}, W. and {DES Collaboration}},
        title = "{The Dark Energy Survey Image Processing Pipeline}",
      journal = {\pasp},
         year = 2018,
        month = jul,
       volume = {130},
       number = {989},
        pages = {074501},
          doi = {10.1088/1538-3873/aab4ef},
archivePrefix = {arXiv},
       eprint = {1801.03177},
 primaryClass = {astro-ph.IM},
       adsurl = {https://ui.adsabs.harvard.edu/abs/2018PASP..130g4501M}
}

@ARTICLE{Flaugher2015DES,
       author = {{Flaugher}, B. and {Diehl}, H.~T. and {Honscheid}, K. and {Abbott}, T.~M.~C. and {Alvarez}, O. and {Angstadt}, R. and {Annis}, J.~T. and {Antonik}, M. and {Ballester}, O. and {Beaufore}, L. and {Bernstein}, G.~M. and {Bernstein}, R.~A. and {Bigelow}, B. and {Bonati}, M. and {Boprie}, D. and {Brooks}, D. and {Buckley-Geer}, E.~J. and {Campa}, J. and {Cardiel-Sas}, L. and {Castander}, F.~J. and {Castilla}, J. and {Cease}, H. and {Cela-Ruiz}, J.~M. and {Chappa}, S. and {Chi}, E. and {Cooper}, C. and {da Costa}, L.~N. and {Dede}, E. and {Derylo}, G. and {DePoy}, D.~L. and {de Vicente}, J. and {Doel}, P. and {Drlica-Wagner}, A. and {Eiting}, J. and {Elliott}, A.~E. and {Emes}, J. and {Estrada}, J. and {Fausti Neto}, A. and {Finley}, D.~A. and {Flores}, R. and {Frieman}, J. and {Gerdes}, D. and {Gladders}, M.~D. and {Gregory}, B. and {Gutierrez}, G.~R. and {Hao}, J. and {Holland}, S.~E. and {Holm}, S. and {Huffman}, D. and {Jackson}, C. and {James}, D.~J. and {Jonas}, M. and {Karcher}, A. and {Karliner}, I. and {Kent}, S. and {Kessler}, R. and {Kozlovsky}, M. and {Kron}, R.~G. and {Kubik}, D. and {Kuehn}, K. and {Kuhlmann}, S. and {Kuk}, K. and {Lahav}, O. and {Lathrop}, A. and {Lee}, J. and {Levi}, M.~E. and {Lewis}, P. and {Li}, T.~S. and {Mandrichenko}, I. and {Marshall}, J.~L. and {Martinez}, G. and {Merritt}, K.~W. and {Miquel}, R. and {Mu{\~n}oz}, F. and {Neilsen}, E.~H. and {Nichol}, R.~C. and {Nord}, B. and {Ogando}, R. and {Olsen}, J. and {Palaio}, N. and {Patton}, K. and {Peoples}, J. and {Plazas}, A.~A. and {Rauch}, J. and {Reil}, K. and {Rheault}, J. -P. and {Roe}, N.~A. and {Rogers}, H. and {Roodman}, A. and {Sanchez}, E. and {Scarpine}, V. and {Schindler}, R.~H. and {Schmidt}, R. and {Schmitt}, R. and {Schubnell}, M. and {Schultz}, K. and {Schurter}, P. and {Scott}, L. and {Serrano}, S. and {Shaw}, T.~M. and {Smith}, R.~C. and {Soares-Santos}, M. and {Stefanik}, A. and {Stuermer}, W. and {Suchyta}, E. and {Sypniewski}, A. and {Tarle}, G. and {Thaler}, J. and {Tighe}, R. and {Tran}, C. and {Tucker}, D. and {Walker}, A.~R. and {Wang}, G. and {Watson}, M. and {Weaverdyck}, C. and {Wester}, W. and {Woods}, R. and {Yanny}, B. and {DES Collaboration}},
        title = "{The Dark Energy Camera}",
      journal = {\aj},
         year = 2015,
        month = nov,
       volume = {150},
       number = {5},
          eid = {150},
        pages = {150},
          doi = {10.1088/0004-6256/150/5/150},
archivePrefix = {arXiv},
       eprint = {1504.02900},
 primaryClass = {astro-ph.IM},
       adsurl = {https://ui.adsabs.harvard.edu/abs/2015AJ....150..150F}
}

@ARTICLE{Arevalo2024,
       author = {{Ar{\'e}valo}, P. and {Churazov}, E. and {Lira}, P. and {S{\'a}nchez-S{\'a}ez}, P. and {Bernal}, S. and {Hern{\'a}ndez-Garc{\'\i}a}, L. and {L{\'o}pez-Navas}, E. and {Patel}, P.},
        title = "{The universal power spectrum of quasars in optical wavelengths. Break timescale scales directly with both black hole mass and the accretion rate}",
      journal = {\aap},
         year = 2024,
        month = apr,
       volume = {684},
          eid = {A133},
        pages = {A133},
          doi = {10.1051/0004-6361/202347080},
archivePrefix = {arXiv},
       eprint = {2306.11099},
 primaryClass = {astro-ph.GA},
       adsurl = {https://ui.adsabs.harvard.edu/abs/2024A&A...684A.133A}
}

@ARTICLE{Uttley2014,
       author = {{Uttley}, P. and {Cackett}, E.~M. and {Fabian}, A.~C. and {Kara}, E. and {Wilkins}, D.~R.},
        title = "{X-ray reverberation around accreting black holes}",
      journal = {\aapr},
         year = 2014,
        month = aug,
       volume = {22},
          eid = {72},
        pages = {72},
          doi = {10.1007/s00159-014-0072-0},
archivePrefix = {arXiv},
       eprint = {1405.6575},
 primaryClass = {astro-ph.HE},
       adsurl = {https://ui.adsabs.harvard.edu/abs/2014A&ARv..22...72U}
}

@ARTICLE{Lewin2023,
       author = {{Lewin}, Collin and {Kara}, Erin and {Cackett}, Edward M. and {Wilkins}, Dan and {Panagiotou}, Christos and {Garc{\'\i}a}, Javier A. and {Gelbord}, Jonathan},
        title = "{X-Ray/UVOIR Frequency-resolved Time Lag Analysis of Mrk 335 Reveals Accretion Disk Reprocessing}",
      journal = {\apj},
         year = 2023,
        month = sep,
       volume = {954},
       number = {1},
          eid = {33},
        pages = {33},
          doi = {10.3847/1538-4357/ace77b},
archivePrefix = {arXiv},
       eprint = {2307.11145},
 primaryClass = {astro-ph.HE},
       adsurl = {https://ui.adsabs.harvard.edu/abs/2023ApJ...954...33L}
}

@ARTICLE{Lewin2024,
       author = {{Lewin}, Collin and {Kara}, Erin and {Barth}, Aaron J. and {Cackett}, Edward M. and {De Rosa}, Gisella and {Homayouni}, Yasaman and {Horne}, Keith and {Kriss}, Gerard A. and {Landt}, Hermine and {Gelbord}, Jonathan and {Montano}, John and {Arav}, Nahum and {Bentz}, Misty C. and {Boizelle}, Benjamin D. and {Dalla Bont{\`a}}, Elena and {Brotherton}, Michael S. and {Dehghanian}, Maryam and {Ferland}, Gary J. and {Fian}, Carina and {Goad}, Michael R. and {Hern{\'a}ndez Santisteban}, Juan V. and {Ili{\'c}}, Dragana and {Kaastra}, Jelle and {Kaspi}, Shai and {Korista}, Kirk T. and {Kosec}, Peter and {Kova{\v{c}}evi{\'c}}, Andjelka and {Mehdipour}, Missagh and {Miller}, Jake A. and {Netzer}, Hagai and {Neustadt}, Jack M.~M. and {Panagiotou}, Christos and {Partington}, Ethan R. and {Popovi{\'c}}, Luka {\v{C}}. and {Sanmartim}, David and {Vestergaard}, Marianne and {Ward}, Martin J. and {Zaidouni}, Fatima},
        title = "{AGN STORM 2. VII. A Frequency-resolved Map of the Accretion Disk in Mrk 817: Simultaneous X-Ray Reverberation and UVOIR Disk Reprocessing Time Lags}",
      journal = {\apj},
         year = 2024,
        month = oct,
       volume = {974},
       number = {2},
          eid = {271},
        pages = {271},
          doi = {10.3847/1538-4357/ad6b08},
archivePrefix = {arXiv},
       eprint = {2409.09115},
 primaryClass = {astro-ph.HE},
       adsurl = {https://ui.adsabs.harvard.edu/abs/2024ApJ...974..271L}
}

@ARTICLE{Done2012D,
       author = {{Done}, Chris and {Davis}, S.~W. and {Jin}, C. and {Blaes}, O. and {Ward}, M.},
        title = "{Intrinsic disc emission and the soft X-ray excess in active galactic nuclei}",
      journal = {\mnras},
         year = 2012,
        month = mar,
       volume = {420},
       number = {3},
        pages = {1848-1860},
          doi = {10.1111/j.1365-2966.2011.19779.x},
archivePrefix = {arXiv},
       eprint = {1107.5429},
 primaryClass = {astro-ph.HE},
       adsurl = {https://ui.adsabs.harvard.edu/abs/2012MNRAS.420.1848D}
}

@ARTICLE{Davis2007,
       author = {{Davis}, Shane W. and {Woo}, Jong-Hak and {Blaes}, Omer M.},
        title = "{The UV Continuum of Quasars: Models and SDSS Spectral Slopes}",
      journal = {\apj},
         year = 2007,
        month = oct,
       volume = {668},
       number = {2},
        pages = {682-698},
          doi = {10.1086/521393},
archivePrefix = {arXiv},
       eprint = {0707.1456},
 primaryClass = {astro-ph},
       adsurl = {https://ui.adsabs.harvard.edu/abs/2007ApJ...668..682D}
}

@ARTICLE{Shen2024,
       author = {{Shen}, Yue and {Grier}, Catherine J. and {Horne}, Keith and {Stone}, Zachary and {Li}, Jennifer I. and {Yang}, Qian and {Homayouni}, Yasaman and {Trump}, Jonathan R. and {Anderson}, Scott F. and {Brandt}, W.~N. and {Hall}, Patrick B. and {Ho}, Luis C. and {Jiang}, Linhua and {Petitjean}, Patrick and {Schneider}, Donald P. and {Tao}, Charling and {Donnan}, Fergus. R. and {AlSayyad}, Yusra and {Bershady}, Matthew A. and {Blanton}, Michael R. and {Bizyaev}, Dmitry and {Bundy}, Kevin and {Chen}, Yuguang and {Davis}, Megan C. and {Dawson}, Kyle and {Fan}, Xiaohui and {Greene}, Jenny E. and {Gr{\"o}ller}, Hannes and {Guo}, Yucheng and {Ibarra-Medel}, H{\'e}ctor and {Jiang}, Yuanzhe and {Keenan}, Ryan P. and {Kollmeier}, Juna A. and {Lejoly}, Cassandra and {Li}, Zefeng and {de la Macorra}, Axel and {Moe}, Maxwell and {Nie}, Jundan and {Rossi}, Graziano and {Smith}, Paul S. and {Tee}, Wei Leong and {Weijmans}, Anne-Marie and {Xu}, Jiachuan and {Yue}, Minghao and {Zhou}, Xu and {Zhou}, Zhimin and {Zou}, Hu},
        title = "{The Sloan Digital Sky Survey Reverberation Mapping Project: Key Results}",
      journal = {\apjs},
         year = 2024,
        month = jun,
       volume = {272},
       number = {2},
          eid = {26},
        pages = {26},
          doi = {10.3847/1538-4365/ad3936},
archivePrefix = {arXiv},
       eprint = {2305.01014},
 primaryClass = {astro-ph.GA},
       adsurl = {https://ui.adsabs.harvard.edu/abs/2024ApJS..272...26S}
}

%\begin{thebibliography}{99}
%\end{thebibliography}

%%%%%%%%%%%%%%%%%%%%%%%%%%%%%%%%%%%%%%%%%%%%%%%%%%

%%%%%%%%%%%%%%%%% APPENDICES %%%%%%%%%%%%%%%%%%%%%

\appendix
\section{Delay measurements}
Here we present the delay measurements for the full sample as measured from PyROA, as well as the properties of each AGN from \citep{Mudd2018} and \citep{Yu2020_DES}.
\renewcommand{\arraystretch}{1.5}
% Remove fits from 1,10,11,14 ,start at 0
\begin{table*}
    \centering
    \begin{tabular}{llccccccccc}
    \hline
    ID&Quasar Name & $z$ & $M_{\bullet}$ & $\tau_g$ & $\tau_r$ & $\tau_i$ & $\tau_z$ & $\tau_Y$ & $\tau_0^{\dagger}$& $\tau_{2500}^{\dagger}$ \\
    &&&$\times10^9$ M$_{\odot}$ & days & days& days& days & days & days& days\\
    \hline
M0 & J025318+000414& 1.56 & 0.49 & $ 0.04_{ 2.78}^{ 2.73}$  & $ 6.88_{ 2.20}^{ 2.25}$  & $ 9.22_{ 2.16}^{ 2.32}$  & $13.61_{ 2.24}^{ 2.36}$ & & $ 3.40\pm 1.15$ & $ 3.67\pm 0.29$\\
M1 & J024753-002137& 1.44 & 1.17 & $-0.01_{ 1.35}^{ 1.37}$  & $ 4.91_{ 1.20}^{ 1.21}$  & $ 7.05_{ 1.17}^{ 1.19}$  & $ 7.88_{ 1.53}^{ 1.49}$ & & $ 2.27\pm 1.29$ & $ 2.42\pm 0.37$\\
M2 & J021500-043007& 1.01 & 0.38 & $-0.02_{ 0.34}^{ 0.33}$  & $ 2.15_{ 0.49}^{ 0.47}$  & $-0.27_{ 0.32}^{ 0.30}$  & $ 3.39_{ 0.63}^{ 0.61}$ & & $ 0.43\pm 6.99$ & $ 0.40\pm33.87$\\
M3 & J022440-043657& 0.91 & 0.2 & $-0.55_{ 1.24}^{ 1.16}$  & $12.23_{ 2.07}^{ 1.92}$  & $99.08_{ 1.55}^{ 0.69}$  & $89.50_{ 5.22}^{ 4.72}$ & & $48.67\pm 1.32$ & $48.69\pm 9.38$\\
M4 & J022436-065912& 1.36 & 0.52 & $ 0.01_{ 0.37}^{ 0.38}$  & $ 0.01_{ 0.65}^{ 0.67}$  & $ 0.64_{ 0.62}^{ 0.63}$  & $ 0.48_{ 1.02}^{ 0.93}$ & & $ 0.20\pm 1.54$ & $-0.04\pm 0.06$\\
M5 & J022108-061753& 1.22 & 0.77 & $-0.05_{ 1.12}^{ 1.16}$  & $ 3.88_{ 1.09}^{ 1.11}$  & $ 7.64_{ 1.29}^{ 1.26}$  & $13.94_{ 1.31}^{ 1.29}$ & & $ 4.25\pm 1.09$ & $ 4.40\pm 0.24$\\
M6 & J022344-064039& 0.98 & 0.06 & $ 0.00_{ 0.23}^{ 0.21}$  & $ 3.64_{ 0.78}^{ 0.62}$  & $ 2.97_{ 3.78}^{ 1.05}$  & $ 4.46_{ 0.52}^{ 0.49}$ & & $ 1.62\pm 1.25$ & $ 1.63\pm 0.20$\\
M7 & J033719-262418& 1.17 & 1.01 & $-0.03_{ 2.19}^{ 2.21}$  & $ 0.04_{ 2.13}^{ 2.19}$  & $-1.34_{ 2.11}^{ 2.20}$  & $ 1.31_{ 2.20}^{ 2.25}$ & & $ 0.24\pm 9.78$ & $ 0.12\pm 2.61$\\
M8 & J024918-001730& 1.24 & 0.49 & $ 0.01_{ 0.47}^{ 0.44}$  & $ 2.06_{ 0.76}^{ 0.78}$  & $-0.47_{ 0.73}^{ 0.69}$  & $ 0.94_{ 0.81}^{ 0.73}$ & & -- & --\\
M9 & J024854+001054& 1.15 & 0.9 & $ 0.01_{ 0.62}^{ 0.59}$  & $ 1.63_{ 0.93}^{ 0.91}$  & $ 4.58_{ 1.08}^{ 1.12}$  & $ 7.55_{ 1.06}^{ 1.10}$ & & $ 2.37\pm 1.11$ & $ 2.43\pm 0.16$\\
M10 & J024133-010724& 1.87 & 0.79 & $-0.02_{ 0.83}^{ 0.85}$  & $ 5.07_{ 1.46}^{ 1.54}$  & $-3.60_{ 1.47}^{ 1.43}$  & $ 8.32_{ 1.42}^{ 1.39}$ & & $ 1.24\pm 3.31$ & $ 1.31\pm 3.22$\\
M11 & J024159-010512& 0.9 & 0.61 & $ 0.08_{ 0.25}^{ 0.80}$  & $ 3.13_{ 1.22}^{ 0.51}$  & $ 2.80_{ 5.17}^{ 0.29}$  & $ 2.85_{ 0.90}^{ 0.93}$ & & $ 1.12\pm 1.41$ & $ 1.13\pm 0.13$\\
M12 & J021514-053321& 0.7 & 0.04 & $-0.03_{ 0.86}^{ 0.89}$  & $ 3.25_{ 1.24}^{ 1.25}$  & $ 7.12_{ 1.05}^{ 1.04}$  & $ 6.35_{ 1.29}^{ 1.28}$ & & $ 3.11\pm 1.32$ & $ 3.05\pm 0.82$\\
M13 & J024357-011330& 0.9 & 0.14 & $-0.03_{ 0.92}^{ 0.91}$  & $ 2.46_{ 1.16}^{ 1.14}$  & $ 3.98_{ 1.31}^{ 1.28}$  & $ 6.18_{ 1.38}^{ 1.30}$ & & $ 2.23\pm 1.05$ & $ 2.24\pm 0.09$\\
M14 & J021952-040919& 0.69 & 0.66 & $-0.01_{ 0.88}^{ 0.89}$  & $11.61_{ 1.10}^{ 1.16}$  & $10.14_{ 0.93}^{ 0.90}$  & $15.29_{ 1.15}^{ 1.15}$ & & -- & --\\
\hline
Y00 & J063037-575610& 0.43 & 1.23 & $-0.04_{ 0.44}^{ 0.45}$  & $ 2.60_{ 0.54}^{ 0.56}$  & $ 2.55_{ 0.62}^{ 0.62}$ & $ 3.00_{ 0.69}^{ 0.69}$  & $ 2.98_{ 0.79}^{ 0.79}$ & $ 1.30\pm 1.42$ & $ 1.27\pm 0.71$\\
Y01 & J063510-585303& 0.22 & 0.23 & $ 0.03_{ 0.50}^{ 0.48}$  & $ 2.05_{ 0.79}^{ 0.78}$  & $ 3.59_{ 0.99}^{ 0.96}$ & $ 4.26_{ 1.26}^{ 1.26}$  &   & $ 2.68\pm 1.13$ & $ 2.49\pm 0.28$\\
Y02 & J063159-590900& 0.73 & 2.0 & $-0.06_{ 0.58}^{ 0.63}$  & $ 3.86_{ 0.76}^{ 0.77}$  & $ 8.86_{ 1.37}^{ 1.40}$ & $17.21_{ 1.97}^{ 2.02}$  &   & $ 5.92\pm 1.15$ & $ 5.84\pm 0.56$\\
Y03 & J062758-582929& 0.49 & 0.2 & $ 0.00_{ 0.43}^{ 0.42}$  & $ 2.62_{ 0.74}^{ 0.71}$  & $ 5.81_{ 0.98}^{ 0.95}$ & $ 6.24_{ 0.88}^{ 0.87}$  &   & $ 3.16\pm 1.15$ & $ 3.02\pm 0.35$\\
Y04 & J033002-273248& 0.53 & 0.8 & $-0.00_{ 0.26}^{ 0.27}$  & $-0.15_{ 0.46}^{ 0.49}$  & $ 0.90_{ 0.69}^{ 0.74}$ & $ 2.21_{ 0.87}^{ 0.87}$  &   & $ 0.72\pm 1.65$ & $ 0.75\pm 0.28$\\
Y05 & J033408-274337& 1.03 & 1.18 & $-0.03_{ 0.71}^{ 0.73}$  & $ 1.74_{ 0.68}^{ 0.70}$  & $ 4.22_{ 0.78}^{ 0.77}$ & $ 4.43_{ 0.95}^{ 0.93}$  &   & $ 1.70\pm 1.22$ & $ 1.72\pm 0.26$\\
Y06 & J034003-264524& 0.49 & 0.44 & $ 0.00_{ 0.54}^{ 0.57}$  & $ 0.83_{ 0.65}^{ 0.65}$  & $ 2.50_{ 0.76}^{ 0.76}$ & $ 3.09_{ 0.93}^{ 0.89}$  &   & $ 1.54\pm 1.14$ & $ 1.49\pm 0.21$\\
Y07 & J032724-274202& 0.76 & 1.71 & $-0.02_{ 0.69}^{ 0.71}$  & $ 3.31_{ 0.84}^{ 0.87}$  & $ 8.02_{ 1.15}^{ 1.12}$ & $ 5.56_{ 1.32}^{ 1.33}$  &   & $ 2.99\pm 1.44$ & $ 2.94\pm 1.06$\\
Y08 & J033545-293216& 0.72 & 0.82 & $-0.02_{ 0.34}^{ 0.34}$  & $ 1.60_{ 0.54}^{ 0.54}$  & $ 2.41_{ 0.80}^{ 0.80}$ & $ 2.16_{ 0.84}^{ 0.82}$  &   & $ 1.11\pm 1.34$ & $ 1.10\pm 0.23$\\
Y09 & J033810-264325& 0.85 & 2.05 & $-0.02_{ 0.79}^{ 0.82}$  & $ 0.07_{ 0.81}^{ 0.80}$  & $-2.73_{ 0.98}^{ 0.96}$ & $-1.57_{ 0.87}^{ 0.91}$  &   & -- & --\\
Y10 & J034001-274036& 1.15 & 1.98 & $ 0.01_{ 0.22}^{ 0.23}$  & $-0.00_{ 0.28}^{ 0.28}$  & $ 0.27_{ 0.55}^{ 0.52}$ & $ 1.42_{ 0.59}^{ 0.57}$  &   & $ 0.31\pm 1.75$ & $ 0.22\pm 0.11$\\
Y11 & J033853-261454& 1.17 & 2.65 & $ 0.00_{ 0.21}^{ 0.20}$  & $ 1.89_{ 0.26}^{ 0.25}$  & $ 3.09_{ 0.45}^{ 0.45}$ & $ 2.99_{ 0.49}^{ 0.51}$  &   & $ 1.17\pm 1.28$ & $ 1.18\pm 0.17$\\
Y12 & J033051-271254& 0.63 & 0.61 & $-0.02_{ 0.41}^{ 0.40}$  & $ 3.46_{ 0.53}^{ 0.53}$  & $ 3.66_{ 0.61}^{ 0.64}$ & $ 4.61_{ 1.08}^{ 1.10}$  &   & $ 2.26\pm 1.37$ & $ 2.20\pm 0.76$\\
Y13 & J032853-281706& 1.0 & 5.65 & $ 0.02_{ 0.97}^{ 0.99}$  & $ 5.00_{ 1.12}^{ 1.13}$  & $ 4.89_{ 1.48}^{ 1.52}$ & $ 4.60_{ 1.81}^{ 1.84}$  &   & $ 1.93\pm 1.72$ & $ 1.95\pm 1.85$\\
Y14 & J032801-273815& 1.59 & 3.28 & $ 0.01_{ 0.54}^{ 0.54}$  & $ 3.61_{ 0.53}^{ 0.54}$  & $ 1.80_{ 0.64}^{ 0.63}$ & $ 7.06_{ 0.67}^{ 0.65}$  &   & $ 1.53\pm 1.59$ & $ 1.63\pm 0.47$\\
Y15 & J033230-284750& 0.86 & 1.49 & $ 0.09_{ 1.83}^{ 1.83}$  & $ 5.42_{ 1.71}^{ 1.73}$  & $ 9.04_{ 1.96}^{ 1.96}$ & $ 7.16_{ 1.84}^{ 1.86}$  &   & $ 2.73\pm 1.60$ & $ 2.72\pm 2.11$\\
Y16 & J033729-294917& 1.35 & 1.95 & $ 0.00_{ 0.31}^{ 0.31}$  & $ 2.00_{ 0.46}^{ 0.46}$  & $ 2.89_{ 0.71}^{ 0.66}$ & $ 3.75_{ 0.95}^{ 0.95}$  &   & $ 1.23\pm 1.16$ & $ 1.27\pm 0.08$\\
Y17 & J033220-285343& 1.27 & 1.37 & $ 0.00_{ 0.19}^{ 0.20}$  & $ 0.25_{ 0.41}^{ 0.41}$  & $ 0.79_{ 0.60}^{ 0.59}$ & $ 0.32_{ 0.86}^{ 0.76}$  &   & $ 0.22\pm 1.55$ & $ 0.03\pm 0.04$\\
Y18 & J033342-285955& 0.55 & 2.05 & $ 0.13_{ 3.89}^{ 4.23}$  & $ 8.13_{ 5.81}^{ 5.46}$  & $15.15_{ 8.52}^{ 7.31}$ & $17.32_{12.01}^{10.74}$  & $15.74_{ 9.87}^{ 9.87}$ & $ 7.26\pm 1.19$ & $ 7.06\pm 1.05$\\
Y19 & J033052-274926& 1.95 & 0.98 & $ 0.01_{ 0.38}^{ 0.38}$  & $ 2.55_{ 0.59}^{ 0.58}$  & $ 3.62_{ 0.76}^{ 0.75}$ & $ 4.15_{ 0.84}^{ 0.81}$  & $ 4.17_{ 0.94}^{ 0.94}$ & $ 0.96\pm 1.20$ & $ 0.94\pm 0.13$\\
Y20 & J033238-273945& 0.84 & 1.33 & $ 0.07_{ 0.85}^{ 0.86}$  & $ 6.01_{ 0.90}^{ 0.95}$  & $ 7.08_{ 1.09}^{ 1.05}$ & $ 9.60_{ 1.38}^{ 1.45}$  & $ 9.78_{ 2.56}^{ 2.56}$ & $ 3.44\pm 1.23$ & $ 3.42\pm 0.63$\\
\hline
    \end{tabular}
    \caption{Interband delay measurements of the DES quasar sample \citep{Mudd2018,Yu2020_DES} using {\sc PyROA}. The best fit parameters as described in Sec.~\ref{sec:discsize}. Delay spectra where no clear trend is observed were omitted from the fit. All fits have 2 degrees of freedom. The lags on all four bands are shown in the observed frame. $^{\dagger}$Both $\tau_0$ and $\tau_{2500}$ are presented in the quasar's rest frame.}
    \label{tab:lags}
\end{table*}

\section{Flux-flux results and SED slope parameters}
\begin{figure}
	\includegraphics[width=1.0\columnwidth,trim=0cm 0.2cm 0cm 0.0cm,clip]{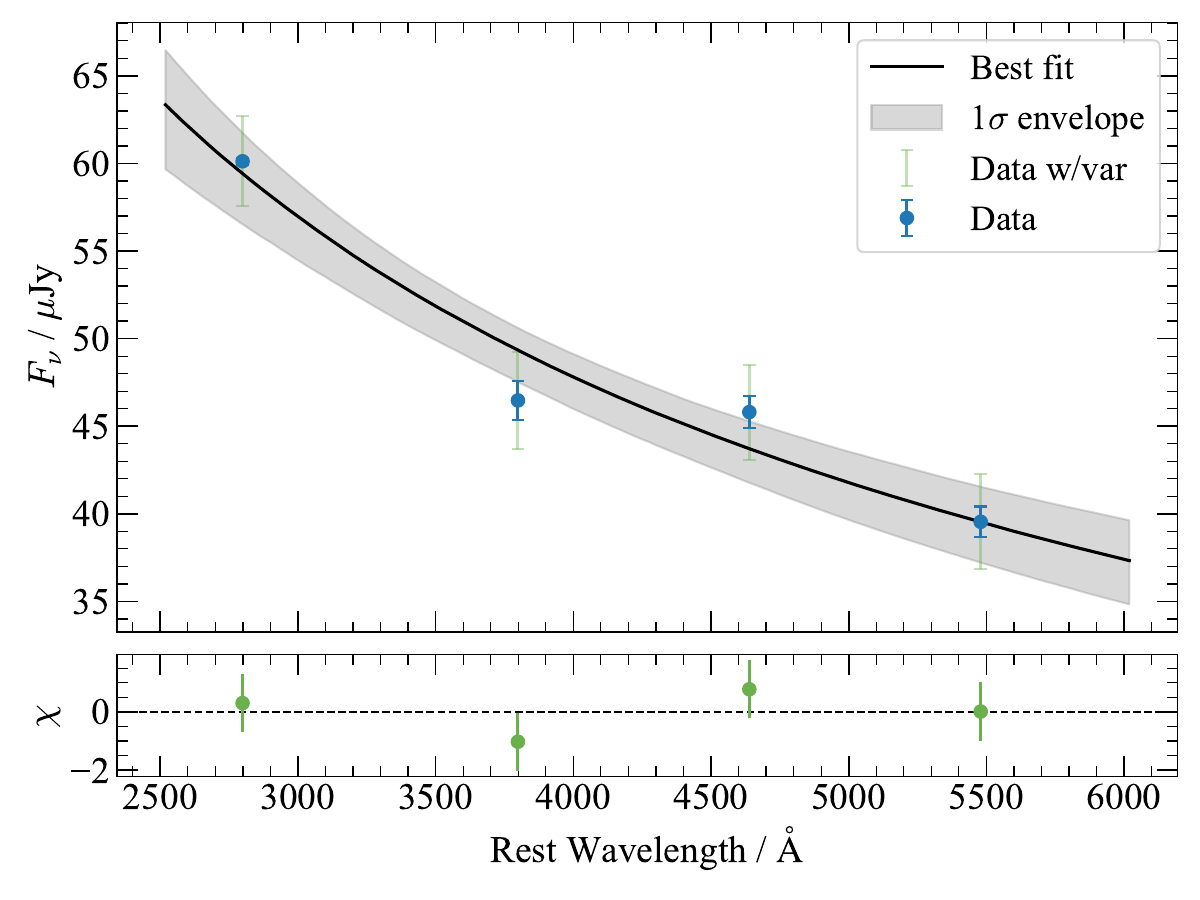}
    \caption{SED fit of M14 as described in Sec.~\ref{sec:sed}. {\it Top:} The best fit to the data (black line) and its 68\% confidence interval region (shaded grey band). The nominal errors (blue) and after the additional variance is added (green) are shown. {\it Bottom:} Normalised residuals using the total variance, $\sigma_i^2 + \sigma_{\rm int}^2$.}
    \label{fig:sed_app14}
\end{figure}

Here we present in Fig.~\ref{fig:sed_app14} an example of the SED fit described in Sec.~\ref{sec:sed}. We also present the flux-flux analysis results for all DES quasars, as well as the power-law best fit parameters in Tables~\ref{tab:sed_fits_mudd} and \ref{tab:sed_fits_yu}. 

\begin{landscape}
\begin{table}
    \centering
    \caption{Results of the SED fits from the flux-flux decomposition for the \citep{Mudd2018} sample.}
    \vspace{0.1em}
    \begin{tabular}{ccccccccccccccccc}
    \hline
    ID & $E(B-V)$ & \multicolumn{4}{|c|}{$F_{\rm AGN, \,  faint}$ }& \multicolumn{4}{|c|}{$F_{\rm  AGN, \, bright}$ } & $\beta$ & $\sigma_{\rm int}$\\
    & & $g$&$r$&$i$&$z$& $g$&$r$&$i$&$z$ & & \\
    & &  $\mu$Jy& $\mu$Jy& $\mu$Jy& $\mu$Jy& $\mu$Jy& $\mu$Jy& $\mu$Jy& $\mu$Jy & &$\mu$Jy\\
    \hline
M00 & 0.058  & $ 38.5\pm  0.4$    & $ 46.7\pm  2.3$    & $ 46.2\pm  2.3$    & $ 41.8\pm  2.0$    & $ 53.9\pm  0.3$    & $ 65.3\pm  2.4$    & $ 64.7\pm  2.3$    & $ 58.5\pm  2.1$  &$ 0.16\pm 0.26$ & $ 6.54\pm 1.81$\\
M01 & 0.037  & $ 36.4\pm  0.2$    & $ 42.6\pm  1.5$    & $ 45.4\pm  1.5$    & $ 49.6\pm  1.9$    & $ 49.1\pm  0.2$    & $ 57.4\pm  1.6$    & $ 61.2\pm  1.6$    & $ 66.9\pm  2.1$  &$ 0.45\pm 0.05$ & $ 0.68\pm 2.16$\\
M02 & 0.019  & $ 36.4\pm  0.1$    & $ 36.4\pm  0.8$    & $ 30.9\pm  0.8$    & $ 26.2\pm  0.8$    & $ 51.6\pm  0.1$    & $ 51.5\pm  1.0$    & $ 43.7\pm  1.0$    & $ 37.0\pm  1.1$  &$-0.44\pm 0.24$ & $ 4.86\pm 1.70$\\
M03 & 0.020  & $ 15.4\pm  0.3$    & $ 12.9\pm  0.6$    & $ 12.3\pm  1.5$    & $ 10.7\pm  0.9$    & $ 31.3\pm  0.1$    & $ 26.2\pm  1.2$    & $ 24.9\pm  3.2$    & $ 21.6\pm  1.5$  &$-0.55\pm 0.12$ & $ 0.66\pm 2.11$\\
M04 & 0.028  & $ 10.8\pm  0.1$    & $ 10.7\pm  0.3$    & $  9.2\pm  0.2$    & $ 10.3\pm  0.3$    & $ 24.7\pm  0.1$    & $ 24.4\pm  0.4$    & $ 20.9\pm  0.3$    & $ 23.4\pm  0.4$  &$-0.15\pm 0.19$ & $ 1.99\pm 1.69$\\
M05 & 0.021  & $ 16.6\pm  0.1$    & $ 16.1\pm  0.4$    & $ 16.5\pm  0.4$    & $ 14.2\pm  0.4$    & $ 27.6\pm  0.1$    & $ 26.7\pm  0.5$    & $ 27.4\pm  0.5$    & $ 23.6\pm  0.4$  &$-0.17\pm 0.14$ & $ 1.74\pm 1.72$\\
M06 & 0.025  & $ 35.5\pm  0.1$    & $ 33.4\pm  1.0$    & $ 32.3\pm  1.0$    & $ 31.9\pm  0.8$    & $ 52.8\pm  0.1$    & $ 49.7\pm  1.5$    & $ 48.0\pm  1.4$    & $ 47.4\pm  1.0$  &$-0.17\pm 0.04$ & $ 0.61\pm 2.01$\\
M07 & 0.011  & $ 14.0\pm  0.2$    & $ 13.9\pm  0.9$    & $ 10.4\pm  0.6$    & $ 10.5\pm  0.7$    & $ 23.4\pm  0.2$    & $ 23.2\pm  1.1$    & $ 17.3\pm  0.8$    & $ 17.5\pm  0.8$  &$-0.47\pm 0.21$ & $ 1.84\pm 2.43$\\
M08 & 0.046  & $ 31.2\pm  0.2$    & $ 27.5\pm  0.8$    & $ 21.8\pm  0.6$    & $ 22.9\pm  0.7$    & $ 47.7\pm  0.2$    & $ 42.0\pm  1.1$    & $ 33.4\pm  0.8$    & $ 35.0\pm  0.9$  &$-0.54\pm 0.17$ & $ 3.00\pm 1.79$\\
M09 & 0.041  & $ 20.1\pm  0.2$    & $ 18.0\pm  0.4$    & $ 15.4\pm  0.4$    & $ 15.2\pm  0.4$    & $ 40.3\pm  0.1$    & $ 36.2\pm  0.6$    & $ 31.0\pm  0.4$    & $ 30.4\pm  0.5$  &$-0.45\pm 0.07$ & $ 1.13\pm 1.94$\\
M10 & 0.026  & $ 21.3\pm  0.2$    & $ 23.5\pm  0.8$    & $ 18.9\pm  0.7$    & $ 18.4\pm  0.7$    & $ 36.3\pm  0.2$    & $ 40.1\pm  1.1$    & $ 32.3\pm  1.0$    & $ 31.3\pm  0.9$  &$-0.24\pm 0.27$ & $ 4.26\pm 1.72$\\
M11 & 0.027  & $ 51.2\pm 12.2$    & $ 74.7\pm 97.2$    & $ 70.8\pm 94.0$    & $ 73.8\pm104.6$    & $ 88.9\pm  9.0$    & $114.0\pm 98.6$    & $107.0\pm 95.9$    & $110.3\pm108.2$  &$-0.47\pm 2.47$ & $ 5.03\pm 8.37$\\
M12 & 0.018  & $ 36.2\pm  0.2$    & $ 27.0\pm  0.5$    & $ 28.3\pm  0.5$    & $ 25.2\pm  0.5$    & $ 66.9\pm  0.1$    & $ 50.0\pm  0.7$    & $ 52.3\pm  0.6$    & $ 46.6\pm  0.7$  &$-0.53\pm 0.22$ & $ 5.31\pm 1.65$\\
M13 & 0.026  & $ 27.7\pm  0.1$    & $ 23.7\pm  0.6$    & $ 27.9\pm  0.7$    & $ 25.2\pm  0.7$    & $ 42.4\pm  0.1$    & $ 36.3\pm  0.7$    & $ 42.7\pm  0.9$    & $ 38.6\pm  0.8$  &$-0.07\pm 0.25$ & $ 4.39\pm 1.68$\\
M14 & 0.017  & $ 30.3\pm  0.4$    & $ 23.4\pm  0.9$    & $ 23.1\pm  0.8$    & $ 19.9\pm  0.7$    & $ 60.1\pm  0.2$    & $ 46.5\pm  1.1$    & $ 45.8\pm  0.9$    & $ 39.5\pm  0.9$  &$-0.61\pm 0.12$ & $ 2.56\pm 2.01$\\
\hline
\end{tabular}
    \label{tab:sed_fits_mudd}
\end{table}
\end{landscape}

\begin{landscape}
\begin{table}
    \centering
    \caption{Results of the SED fits from the flux-flux decomposition for the \citep{Yu2020_DES} sample.}
    \vspace{0.1em}
    \begin{tabular}{ccccccccccccccccccc}
    \hline
    ID & $E(B-V)$ & \multicolumn{5}{|c|}{$F_{\rm AGN, \,  faint}$ }& \multicolumn{5}{|c|}{$F_{\rm  AGN, \, bright}$ } & $\beta$ & $\sigma_{\rm int}$\\
    & & $g$&$r$&$i$&$z$&$Y$ & $g$&$r$&$i$&$z$&$Y$ & & \\
    & &  $\mu$Jy& $\mu$Jy& $\mu$Jy& $\mu$Jy& $\mu$Jy& $\mu$Jy& $\mu$Jy& $\mu$Jy& $\mu$Jy& $\mu$Jy & &$\mu$Jy\\
    \hline
Y00 & 0.114  & $260.7\pm  1.0$    & $273.0\pm  3.8$    & $241.8\pm  3.6$    & $215.8\pm  4.1$    & $225.3\pm  4.9$    & $399.0\pm  0.9$    & $417.8\pm  4.2$    & $369.9\pm  4.4$    & $330.2\pm  5.3$    & $344.7\pm  6.7$  &$-0.25\pm 0.13$ & $27.13\pm 1.49$\\
Y01 & 0.084  & $ 47.9\pm  0.1$    & $ 39.5\pm  0.5$    & $ 31.9\pm  0.6$    & $ 31.5\pm  0.6$  &   & $ 65.4\pm  0.1$    & $ 53.8\pm  0.7$    & $ 43.5\pm  0.7$    & $ 42.9\pm  0.7$  &&$-0.68\pm 0.11$ & $ 2.61\pm 1.81$\\
Y02 & 0.073  & $ 28.7\pm  0.1$    & $ 23.5\pm  0.4$    & $ 24.4\pm  0.6$    & $ 23.9\pm  0.9$  &   & $ 42.6\pm  0.1$    & $ 34.9\pm  0.5$    & $ 36.1\pm  0.9$    & $ 35.5\pm  1.2$  &&$-0.29\pm 0.18$ & $ 2.88\pm 1.69$\\
Y03 & 0.070  & $ 28.3\pm  0.1$    & $ 28.0\pm  0.5$    & $ 24.8\pm  0.5$    & $ 23.8\pm  0.5$  &   & $ 44.3\pm  0.2$    & $ 43.9\pm  0.7$    & $ 38.9\pm  0.6$    & $ 37.3\pm  0.7$  &&$-0.26\pm 0.11$ & $ 1.92\pm 1.83$\\
Y04 & 0.008  & $ 26.2\pm  0.2$    & $ 21.4\pm  0.6$    & $ 19.4\pm  0.9$    & $ 20.3\pm  1.1$  &   & $ 38.8\pm  0.2$    & $ 31.6\pm  0.9$    & $ 28.7\pm  1.0$    & $ 30.1\pm  1.6$  &&$-0.51\pm 0.11$ & $ 1.47\pm 2.98$\\
Y05 & 0.007  & $ 33.8\pm  0.4$    & $ 34.8\pm  1.0$    & $ 36.6\pm  1.0$    & $ 31.4\pm  1.1$  &   & $ 50.4\pm  0.3$    & $ 52.0\pm  1.0$    & $ 54.6\pm  1.1$    & $ 46.8\pm  1.2$  &&$-0.04\pm 0.19$ & $ 4.32\pm 1.74$\\
Y06 & 0.012  & $  7.5\pm  0.0$    & $  6.1\pm  0.1$    & $  5.5\pm  0.1$    & $  5.4\pm  0.1$  &   & $ 15.5\pm  0.1$    & $ 12.6\pm  0.1$    & $ 11.4\pm  0.2$    & $ 11.1\pm  0.2$  &&$-0.54\pm 0.09$ & $ 0.52\pm 1.68$\\
Y07 & 0.007  & $  9.3\pm  0.0$    & $  9.8\pm  0.2$    & $ 10.7\pm  0.2$    & $  9.2\pm  0.2$  &   & $ 13.6\pm  0.0$    & $ 14.2\pm  0.2$    & $ 15.5\pm  0.3$    & $ 13.4\pm  0.3$  &&$ 0.06\pm 0.20$ & $ 1.33\pm 1.72$\\
Y08 & 0.010  & $ 24.7\pm  0.1$    & $ 25.5\pm  0.3$    & $ 23.1\pm  0.4$    & $ 20.8\pm  0.4$  &   & $ 38.9\pm  0.1$    & $ 40.2\pm  0.4$    & $ 36.4\pm  0.6$    & $ 32.8\pm  0.6$  &&$-0.23\pm 0.17$ & $ 2.96\pm 1.70$\\
Y09 & 0.011  & $ 21.0\pm  0.0$    & $ 23.4\pm  0.4$    & $ 18.7\pm  0.3$    & $ 18.9\pm  0.4$  &   & $ 29.1\pm  0.1$    & $ 32.3\pm  0.4$    & $ 26.0\pm  0.4$    & $ 26.1\pm  0.4$  &&$-0.20\pm 0.29$ & $ 3.67\pm 1.67$\\
Y10 & 0.012  & $ 29.0\pm  0.2$    & $ 22.9\pm  0.2$    & $ 19.8\pm  0.2$    & $ 18.3\pm  0.3$  &   & $ 61.0\pm  0.1$    & $ 48.1\pm  0.3$    & $ 41.6\pm  0.4$    & $ 38.6\pm  0.6$  &&$-0.72\pm 0.04$ & $ 0.93\pm 2.05$\\
Y11 & 0.011  & $ 13.6\pm  0.0$    & $ 13.9\pm  0.1$    & $ 11.7\pm  0.1$    & $ 10.5\pm  0.2$  &   & $ 24.8\pm  0.1$    & $ 25.4\pm  0.2$    & $ 21.3\pm  0.2$    & $ 19.1\pm  0.3$  &&$-0.38\pm 0.23$ & $ 2.33\pm 1.66$\\
Y12 & 0.009  & $ 21.6\pm  0.1$    & $ 21.1\pm  0.3$    & $ 18.5\pm  0.3$    & $ 16.0\pm  0.4$  &   & $ 33.9\pm  0.1$    & $ 33.0\pm  0.4$    & $ 28.9\pm  0.4$    & $ 25.1\pm  0.6$  &&$-0.40\pm 0.18$ & $ 2.37\pm 1.70$\\
Y13 & 0.007  & $ 11.1\pm  0.1$    & $ 12.8\pm  0.4$    & $ 13.8\pm  0.5$    & $ 11.8\pm  0.5$  &   & $ 16.2\pm  0.1$    & $ 18.7\pm  0.4$    & $ 20.1\pm  0.6$    & $ 17.2\pm  0.7$  &&$ 0.16\pm 0.27$ & $ 2.07\pm 1.76$\\
Y14 & 0.010  & $ 43.4\pm  0.1$    & $ 45.5\pm  0.4$    & $ 35.7\pm  0.4$    & $ 36.6\pm  0.4$  &   & $ 68.8\pm  0.1$    & $ 72.1\pm  0.5$    & $ 56.6\pm  0.5$    & $ 58.0\pm  0.5$  &&$-0.31\pm 0.27$ & $ 7.53\pm 1.62$\\
Y15 & 0.007  & $ 61.1\pm  0.2$    & $ 71.9\pm  1.6$    & $ 72.9\pm  1.8$    & $ 61.8\pm  1.5$  &   & $ 86.6\pm  0.3$    & $101.9\pm  1.8$    & $103.2\pm  2.0$    & $ 87.5\pm  1.6$  &&$ 0.07\pm 0.31$ & $13.35\pm 1.69$\\
Y16 & 0.011  & $ 10.7\pm  0.1$    & $  9.4\pm  0.1$    & $  8.3\pm  0.2$    & $  7.5\pm  0.3$  &   & $ 23.7\pm  0.1$    & $ 20.7\pm  0.2$    & $ 18.2\pm  0.3$    & $ 16.6\pm  0.5$  &&$-0.51\pm 0.06$ & $ 0.42\pm 1.56$\\
Y17 & 0.007  & $ 44.7\pm  0.2$    & $ 42.1\pm  0.9$    & $ 39.6\pm  1.3$    & $ 42.7\pm  1.8$  &   & $ 62.8\pm  0.2$    & $ 59.1\pm  1.2$    & $ 55.6\pm  1.8$    & $ 59.9\pm  2.4$  &&$-0.15\pm 0.07$ & $ 1.18\pm 3.00$\\
Y18 & 0.008  & $ 13.9\pm  0.2$    & $ 16.8\pm  1.2$    & $ 21.0\pm  1.9$    & $ 23.0\pm  2.4$    & $ 31.4\pm  3.5$    & $ 23.2\pm  0.5$    & $ 27.9\pm  1.7$    & $ 35.0\pm  2.7$    & $ 38.3\pm  3.6$    & $ 52.4\pm  5.6$  &$ 0.88\pm 0.14$ & $ 1.03\pm 2.77$\\
Y19 & 0.008  & $ 35.7\pm  0.3$    & $ 31.2\pm  1.0$    & $ 29.8\pm  1.3$    & $ 41.1\pm  2.0$    & $ 38.0\pm  2.2$    & $ 73.5\pm  0.6$    & $ 64.2\pm  1.8$    & $ 61.3\pm  2.2$    & $ 84.5\pm  3.7$    & $ 78.2\pm  4.2$  &$ 0.17\pm 0.30$ & $11.25\pm 1.55$\\
Y20 & 0.008  & $ 66.2\pm  0.7$    & $ 65.0\pm  1.9$    & $ 60.0\pm  2.0$    & $ 55.2\pm  2.3$    & $ 51.0\pm  3.4$    & $ 92.3\pm  0.9$    & $ 90.7\pm  2.4$    & $ 83.7\pm  2.5$    & $ 77.0\pm  2.8$    & $ 71.2\pm  4.2$  &$-0.26\pm 0.08$ & $ 1.98\pm 3.94$ \\
\hline
\end{tabular}
    \label{tab:sed_fits_yu}
\end{table}
\end{landscape}
% \begin{table}
%     \centering
%     \caption{Results of the SED fits from the flux-flux decomposition for the \citep{Yu2020_DES} sample.}
%     \begin{tabular}{ccc}
%     \hline
%     Quasar Name & $\beta$ & $\sigma_{\rm int}$\\
%     & & mJy\\
%     \hline
% Y00  & $-0.25\pm 0.13$& $27.13\pm 1.49$\\
% Y01  & $-0.68\pm 0.11$& $ 2.61\pm 1.81$\\
% Y02  & $-0.29\pm 0.18$& $ 2.88\pm 1.69$\\
% Y03  & $-0.26\pm 0.11$& $ 1.92\pm 1.83$\\
% Y04  & $-0.51\pm 0.11$& $ 1.47\pm 2.98$\\
% Y05  & $-0.04\pm 0.19$& $ 4.32\pm 1.74$\\
% Y06  & $-0.54\pm 0.09$& $ 0.52\pm 1.68$\\
% Y07  & $ 0.06\pm 0.20$& $ 1.33\pm 1.72$\\
% Y08  & $-0.23\pm 0.17$& $ 2.96\pm 1.70$\\
% Y09  & $-0.20\pm 0.29$& $ 3.67\pm 1.67$\\
% Y10  & $-0.72\pm 0.04$& $ 0.93\pm 2.05$\\
% Y11  & $-0.38\pm 0.23$& $ 2.33\pm 1.66$\\
% Y12  & $-0.40\pm 0.18$& $ 2.37\pm 1.70$\\
% Y13  & $ 0.16\pm 0.27$& $ 2.07\pm 1.76$\\
% Y14  & $-0.31\pm 0.27$& $ 7.53\pm 1.62$\\
% Y15  & $ 0.07\pm 0.31$& $13.35\pm 1.69$\\
% Y16  & $-0.51\pm 0.06$& $ 0.42\pm 1.56$\\
% Y17  & $-0.15\pm 0.07$& $ 1.18\pm 3.00$\\
% Y18  & $ 0.88\pm 0.14$& $ 1.03\pm 2.77$\\
% Y19  & $ 0.17\pm 0.30$& $11.25\pm 1.55$\\
% Y20  & $-0.26\pm 0.08$& $ 1.98\pm 3.94$\\
% \hline
%     \end{tabular}

%     \label{tab:sed_fitsYu}
% \end{table}
\bsp	% typesetting comment
\label{lastpage}
\end{document}